\documentclass[pdflatex,sn-nature]{sn-jnl}

\usepackage{graphicx}
\usepackage{amsmath,amssymb,amsfonts}
\usepackage{booktabs,tabularx,threeparttable,longtable,array,multirow}
\usepackage{textcomp}
\usepackage{url}

\newcommand{\Ha}{\ensuremath{\rm H\alpha}}
\newcommand{\Hb}{\ensuremath{\rm H\beta}}
\newcommand{\ion}[2]{\text{#1\,\textsc{\lowercase{#2}}}}
\newcommand{\Oiii}{[\ion{O}{III}]}
\newcommand{\Oiirec}{\ion{O}{II}}
\newcommand{\Oii}{[\ion{O}{II}]}

\newcommand{\Nii}{[\ion{N}{II}]}
\newcommand{\Sii}{[\ion{S}{II}]}
\newcommand{\Siii}{[\ion{S}{III}]}
\newcommand{\Ariii}{[\ion{Ar}{III}]}
\newcommand{\Hii}{\ion{H}{II}}
\newcommand{\Hi}{\ion{H}{I}}

\newcommand{\secpoint}{\mbox{$''\mskip-7.6mu.\,$}}

\newcommand{\kms}{\ensuremath{\mathrm{km}/\mathrm{s}}}
\newcommand{\ebv}{\ensuremath{\mathrm{E}(\mathrm{B}-\mathrm{V})}}
\newcommand{\Op}{\ensuremath{\mathrm{O}^+}}
\newcommand{\Opp}{\ensuremath{\mathrm{O}^{++}}}
\newcommand{\Hep}{\ensuremath{\mathrm{He}^+}}
\newcommand{\Hepp}{\ensuremath{\mathrm{He}^{++}}}
\newcommand{\Hp}{\ensuremath{\mathrm{H}^+}}
\newcommand{\Np}{\ensuremath{\mathrm{N}^+}}
\newcommand{\Npp}{\ensuremath{\mathrm{N}^{++}}}
\newcommand{\Sp}{\ensuremath{\mathrm{S}^+}}
\newcommand{\Spp}{\ensuremath{\mathrm{S}^{++}}}
\newcommand{\Arpp}{\ensuremath{\mathrm{Ar}^{++}}}
\newcommand*{\diff}{\mathop{}\!\mathrm{d}}
\newcommand{\aterm}[5]{\ensuremath{#1}\,${}^{#2}\mathrm{#3}\ifx#5o^{\circ}\fi_{#4}$}
\newcommand{\farcs}{\mbox{\ensuremath{.\!\!^{\prime\prime}}}}

\begin{document}

\title[Continuum pumping revises metallicity]{Continuum Pumping Revises Metallicity in the Orion Nebula}

\author*[1]{\fnm{Yuguang} \sur{Chen}}\email{yuguangchen@cuhk.edu.hk}
\author[2]{\fnm{Tucker} \sur{Jones}}
\author[3]{\fnm{Ryan} \sur{Sanders}}
\author[2]{\fnm{Paige} \sur{Kelly}}
\author[1]{\fnm{Zhuo} \sur{Cheng}}
\author[4,5]{\fnm{Xuan} \sur{Fang}}
\author[2]{\fnm{Sunny} \sur{Rhoades}}
\author[6]{\fnm{Keerthi} \sur{Vasan GC}}
\author[7]{\fnm{Fabio} \sur{Bresolin}}
\author[8,9]{\fnm{Rodrigo} \sur{Herrera-Camus}}
\author[2]{\fnm{Erin} \sur{Huntzinger}}
\author[10]{\fnm{Xiao-Wei} \sur{Liu}}
\author[6]{\fnm{Peter} \sur{Senchyna}}
\author[11]{\fnm{Charles} \sur{Steidel}}
\author[1]{\fnm{Renbin} \sur{Yan}}
\author[12]{\fnm{Danielle} \sur{Berg}}
\author[13]{\fnm{Dario} \sur{Fadda}}
\author[13]{\fnm{Ryan} \sur{Rickards Vaught}}
\author[14]{\fnm{Guido} \sur{Roberts-Borsani}}
\author[15]{\fnm{Daniel} \sur{Stark}}
\author[16]{\fnm{Jessica} \sur{Sutter}}
\author[17]{\fnm{Bejamin} \sur{Weiner}}

\affil[1]{\orgdiv{Department of Physics}, \orgname{The Chinese University of Hong Kong}, \orgaddress{\city{Hong Kong}, \country{China}}}
\affil[2]{\orgdiv{Department of Physics \& Astronomy}, \orgname{University of California, Davis}, \orgaddress{\city{Davis}, \postcode{95616}, \country{USA}}}
\affil[3]{\orgdiv{Department of Physics and Astronomy}, \orgname{University of Kentucky}, \orgaddress{\city{Lexington}, \postcode{40506}, \country{USA}}}
\affil[4]{\orgname{National Astronomical Observatories, Chinese Academy of Sciences}, \orgaddress{\city{Beijing}, \postcode{100101}, \country{China}}}
\affil[5]{\orgdiv{School of Astronomy and Space Sciences}, \orgname{University of Chinese Academy of Science}, \orgaddress{\city{Beijing}, \postcode{100049}, \country{China}}}
\affil[6]{\orgname{The Observatories of the Carnegie Institution for Science}, \orgaddress{\city{Pasadena}, \postcode{91101}, \country{USA}}}
\affil[7]{\orgname{Institute for Astronomy, University of Hawaii}, \orgaddress{\city{Honolulu}, \postcode{96822}, \country{USA}}}
\affil[8]{\orgdiv{Departamento de Astronom\'ia}, \orgname{Universidad de Concepci\'on}, \orgaddress{\city{Concepci\'on}, \country{Chile}}}
\affil[9]{\orgname{Millennium Nucleus for Galaxies}, \orgaddress{\city{Concepci\'on}, \country{Chile}}}
\affil[10]{\orgname{South-Western Institute for Astronomy Research, Yunnan University}, \orgaddress{\city{Kunming}, \postcode{650500}, \country{China}}}
\affil[11]{\orgname{Cahill Center for Astronomy and Astrophysics, California Institute of Technology}, \orgaddress{\city{Pasadena}, \postcode{91125}, \country{USA}}}
\affil[12]{\orgdiv{Department of Astronomy}, \orgname{The University of Texas at Austin}, \orgaddress{\city{Austin}, \postcode{78712}, \country{USA}}}
\affil[13]{\orgname{Space Telescope Science Institute}, \orgaddress{\city{Baltimore}, \postcode{21218}, \country{USA}}}
\affil[14]{\orgname{University College London}, \orgaddress{\city{London}, \postcode{WC1E 6AE}, \country{UK}}}
\affil[15]{\orgdiv{Department of Astronomy}, \orgname{University of California, Berkeley}, \orgaddress{\city{Berkeley}, \postcode{94720}, \country{USA}}}
\affil[16]{\orgname{Whitman College}, \orgaddress{\city{Walla Walla}, \postcode{99362}, \country{USA}}}
\affil[17]{\orgname{MMT/Steward Observatory, University of Arizona}, \orgaddress{\city{Tucson}, \postcode{85721}, \country{USA}}}

\abstract{
The enrichment of heavy elements in the interstellar medium is central to understanding galaxy evolution. Oxygen, the most abundant element after H and He, provides the principal tracer of gas-phase metallicity, typically measured using nebular emission lines in star-forming regions. However, a long-standing issue—the abundance discrepancy factor (ADF)—produces a $\sim 0.25$ dex offset between metallicities derived from recombination lines (RLs) and collisionally excited lines (CELs). We present deep spectroscopic observations of the Orion Nebula (M42), and use them to assess the contribution of the often neglected continuum pumping process. We find that the continuum pumping effect is at least a factor of two stronger than previously expected. After accounting for the flux contribution from continuum pumping, the RL metallicity is consistent with the CEL metallicity, thus explaining the bulk of the ADF. Reassessing O/H measurements in five literature star-forming regions using the same method, we similarly find a systematic reduction of RL-based metallicities. These results indicate that RL metallicities, which have previously been considered more reliable than CELs, are subject to potentially large bias from continuum pumping. This study provides a foundation to correct this bias and establish accurate element abundance scales throughout cosmic time.

}

\maketitle

\noindent
Oxygen is the third most abundant element in the present-day Universe \cite{asplund09}, after H and He produced in the Big Bang. The O/H abundance (hereafter ``metallicity'') is therefore the most widely adopted tracer of ionized-gas metallicity in galaxies. Measurements of the O/H abundance have been used to understand galaxy formation and chemical enrichment over the past 13 billion years of cosmic history, reaching $z\gtrsim10$ \cite{sanders24, nakajima2023, jones2020, harikane25}. Two ``direct'' approaches are commonly used to infer metallicity from nebular emission in photoionized gas \cite{osterbrock06}: measurements based on strong collisionally excited lines (CELs; e.g., \Oiii\ $\lambda\lambda 4363,5007$) and those based on faint recombination lines (RLs; for example, the \Oiirec\ V1 multiplet near 4650~\AA). In star-forming \Hii\ regions, these methods yield systematically discrepant abundances, with RL-based metallicities typically exceeding CEL-based values by $\simeq 0.2$--0.3 dex (a factor of $\sim 2$) \cite{liu03, liu06, garcia-rojas07b, esteban09, peimbert17}. This offset is known as the abundance discrepancy factor (ADF).

There are at least three physical effects proposed to explain the ADF. The most commonly invoked is temperature fluctuations \cite{peimbert67}. 
Because CEL emissivities depend strongly on the electron temperature ($T_e$), temperature fluctuations within the nebula can cause CEL diagnostics to preferentially sample high-temperature gas, leading to an overestimate of the emissivity and thus an underestimate of the abundance based on CELs. By contrast, RLs are comparatively insensitive to $T_e$ and are therefore regarded as a more reliable method for abundance and metallicity measurements. 
Alternatively, the presence of high-metallicity, hydrogen-poor inclusions embedded within the ionized gas \cite{liu00, stasinska07, chen24} is often invoked to explain the high ADF found in planetary nebulae (PNe). In this scenario, a substantial fraction of oxygen and other heavy elements resides in cold clumps that contribute negligibly to CEL emission. A third explanation that has been studied in PNe, but often overlooked in \Hii\ regions, is that RLs could be contaminated by the continuum pumping process (also known as fluorescence), whereby continuum photons from the ionizing source excite ions from the ground state, contaminating the recombination signal \cite{grandi75a, escalante12}. This continuum pumping effect is known to influence recombination lines arising from ions such as \ion{N}{ii}, \ion{O}{i} and \ion{C}{ii}. For more widely used \Oiirec\ lines, however, this effect has generally been considered minor. 

\begin{figure}[htbp!]
\centering
\includegraphics[height=2.9in]{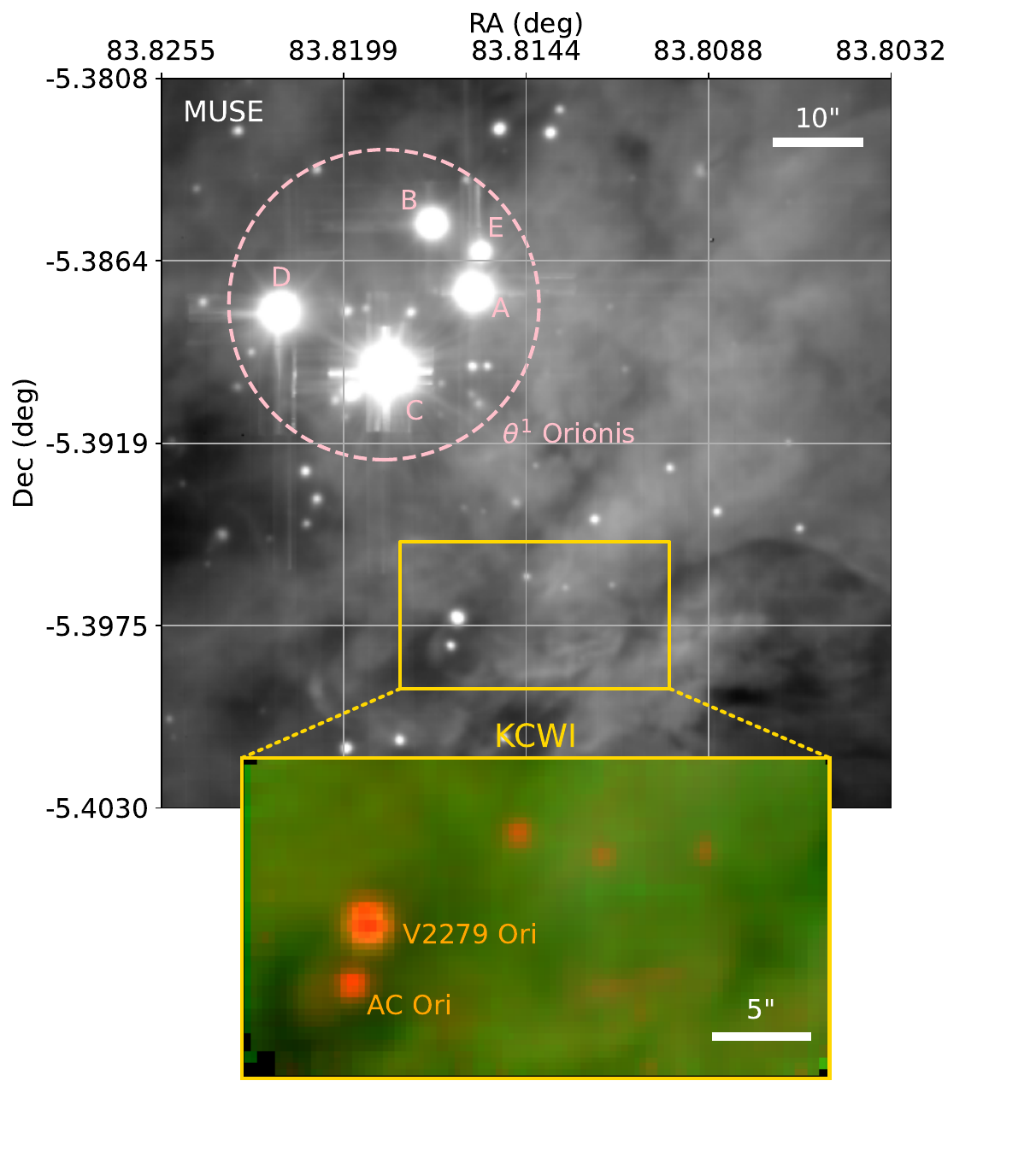}
\includegraphics[height=2.8in]{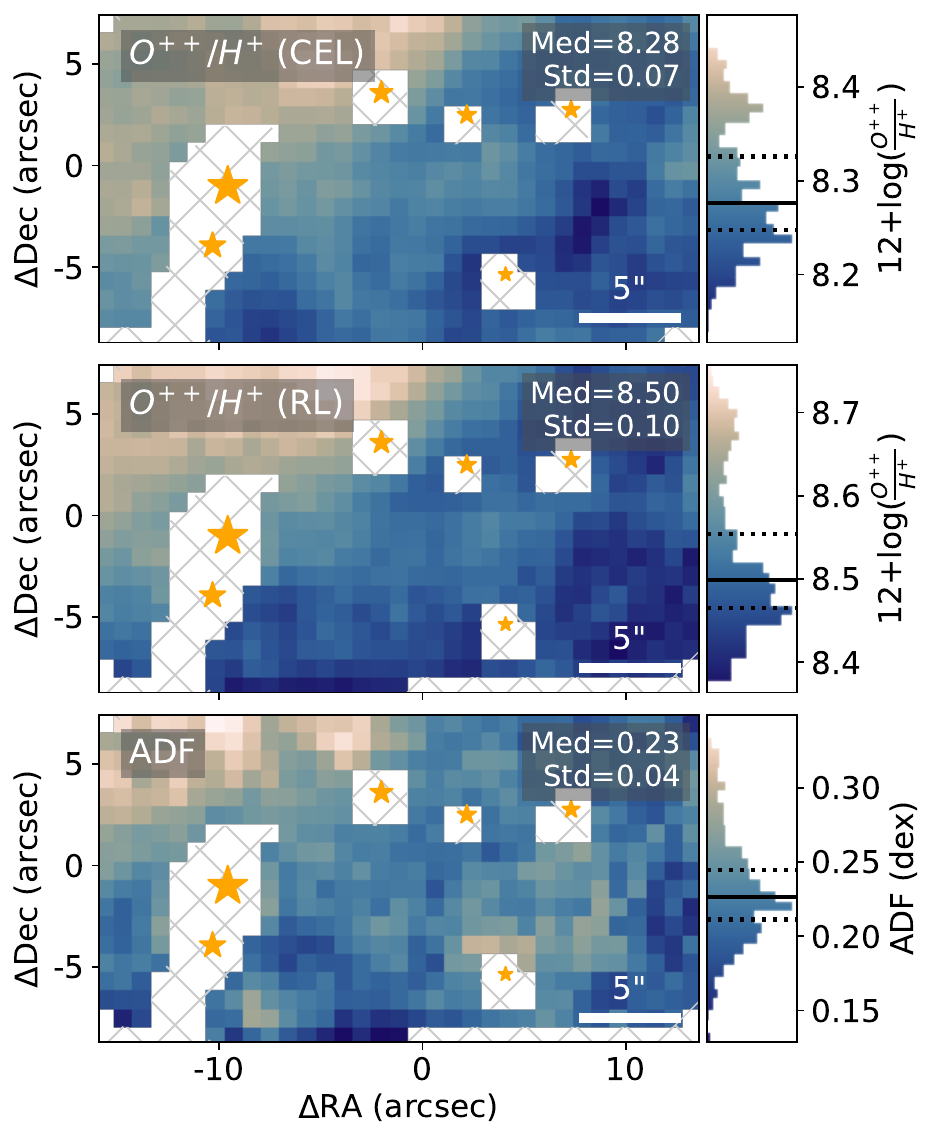}
\caption{   \textbf{Overview of the KCWI IFU observations.} \textbf{Left:} Location of the KCWI field of view (FoV) within the Orion Nebula. The background white-light image is reconstructed from a shallower IFU dataset obtained with MUSE on the VLT \cite{welbacher15}, showing the diffuse nebular emission and bright stars. The pink dashed circle marks the Trapezium ($\theta^1$ Ori) cluster, the dominant ionizing source. The yellow rectangle outlines the KCWI field of view. The inset below shows a pseudo-colour image constructed from the KCWI cube with blue = (3580, 4928]~\AA, green = (4928, 6939]~\AA, and red = (6939, 9770]~\AA. Red points are stars, and the green background is the diffuse nebular emission dominated by the \Oiii\ and H$\beta$ emission. 
\textbf{Right:} Spatial distributions of the $\Opp/\Hp$ abundance derived from \Oiii\ CELs and \Oiirec\ RLs, and the ADF (defined as the logarithmic difference between the RL and CEL abundances). Stars and their associated scattered light are masked, with orange stars in the maps marking the location of the stars and the size of symbols roughly scaled with the stars' brightness. The panel adjacent to each map shows the associated histogram and color bar. Solid horizontal lines denote the median values, and dotted lines indicate the 32nd and 68th percentiles. 
The lower-left panel is shown with $0\farcs3 \times 0\farcs3$ sampling, while the right panels and all subsequent figures are shown with $0\farcs9 \times 0\farcs9$ spaxel sampling which is approximately matched to the seeing resolution of the data.
}
\label{fig:fov}
\end{figure}

We obtained deep observations of the Orion Nebula (M42) using the Keck Cosmic Web Imager (KCWI) \cite{morrissey18}, an integral field unit (IFU) spectrograph at W. M. Keck Observatory, on UT 2024-12-29 as part of the ongoing Comprehensive IFU Research on Accurate Abundance Studies (CIRAAS) survey. Observational details and data reduction procedures are described in Methods. The data cover a field of view (FoV) of $30\farcs5 \times 16\farcs8$, centered $26\farcs1$ south and $16\farcs6$ west of $\theta^1$ Ori C (Fig.~\ref{fig:fov}, left), selected to overlap with the position of a previously published high-resolution long-slit spectrum \cite{esteban04}. The observations provide nearly continuous spectral coverage from $\simeq 3575$ to $9775~\mathrm{\AA}$ at a resolution of $R \sim 2000$. For the \Oiirec\ RLs analysed here, we additionally used a higher-resolution configuration ($R \sim 4000$), covering the wavelength range 3980--4860~\AA\ to fully resolve the multiplet structure. 

The \Oiirec\ RL V1 multiplet, which is commonly used for \Opp/\Hp\ abundance measurements, is well detected in all spaxels with a median S/N ratio of $\simeq 55$. In each spaxel we calculate the RL-based abundance using the V1 multiplet, as well as the CEL-based abundance using the standard ``direct-$T_e$'' method (see Methods). The difference in abundance between these two methods is the ADF, which we find to have a median value of 0.22 dex with a standard deviation of 0.04 dex (Fig.~\ref{fig:fov}, right) for individual $0\farcs9 \times 0\farcs9$ spaxels, corresponding to a physical size of $\simeq 373 \times 373~\mathrm{AU}^{2}$. These ADF values are well within the typical range found for \Hii\ regions \cite{esteban04}. We can thus conclude that the spatially resolved ionized gas in our field of view is typical of \Hii\ regions, and that the physical driver of the ADF acts on a projected physical scale $< 400~\mathrm{AU}$. 

Beyond the V1 multiplet, many other \Oiirec\ RLs are much fainter. To increase the S/N ratio, we create stacked spectra by combining spaxels in the Fov. The ionized gas across the FoV shows clear bimodality in its ionization state as traced by the strong line ratios \Nii/\Ha\ and \Oiii/\Hb\ (see Methods). Therefore, the three stacked spectra are from (1) the entire FoV, and in the (2) high and (3) low ionization regions. 
The stacked spectra allow us to relibly measure the fluxes of the ultra-faint \Oiirec\ RLs while excluding transitions affected by possible blending. Ultimately we are able to measure the fluxes of 28 \Oiirec\ RLs (including 8 from the V1 multiplet). Each \Oiirec\ RL/H$\beta$ ratio is then independently converted to an apparent \Opp/\Hp\ abundance. The typical apparent \Opp/\Hp\ abundance spread within the same spectrum across different lines can exceed 1 dex, far more than the typical measurement uncertainty of $\sim 0.03$~dex, indicating large physical or systematic effects whose strength varies among the recombination lines. 

\begin{figure}[htbp!]
\centering
\includegraphics[height=2.7in]{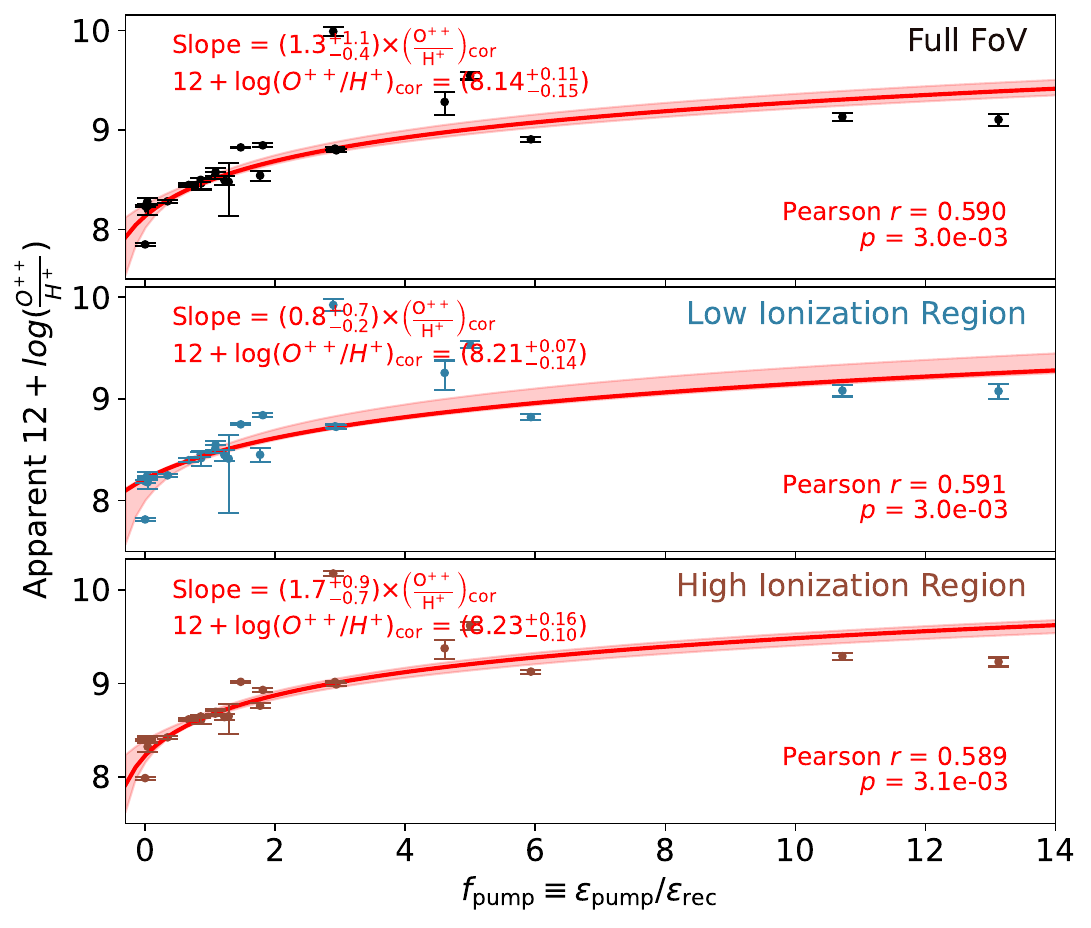} 
\includegraphics[height=1.8in]{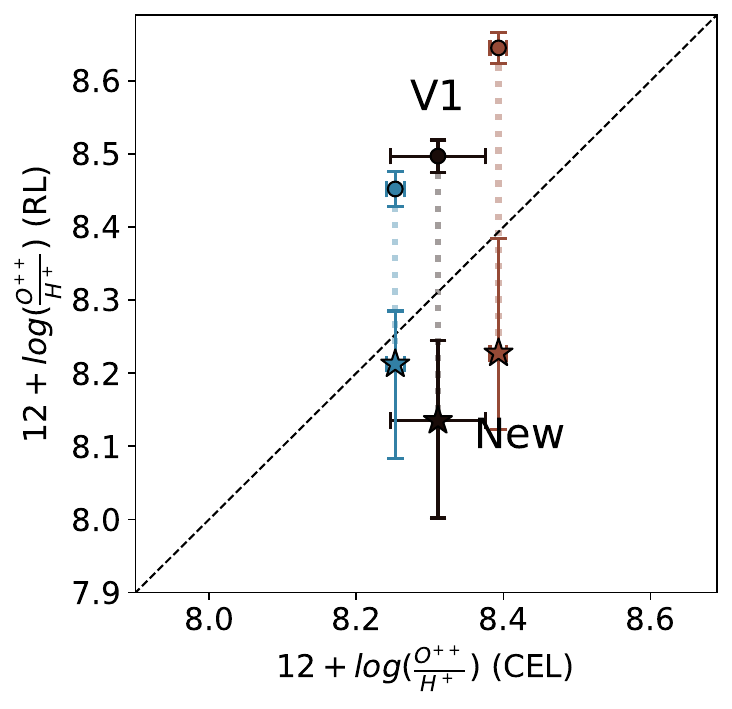} 
\caption{ \textbf{Continuum pumping explains the O \textsc{II} RL emission.} \textbf{Left:} Apparent $\Opp/\Hp$ abundances as a function of the continuum pumping ratio, $f_\mathrm{pump}$, derived from photoionization models (see Methods). Strong correlations are evident for all three stacked spectra, including in independent ionization regions within the field of view (see Methods), indicating that continuum pumping indeed contributes significantly to the \Oiirec\ RL fluxes. \textbf{Right:} After correcting for the continuum pumping effect, the newly calculated \Oiirec\ RL abundances (bottom stars) are significantly reduced compared to the metalllicities based on the V1 multiplet (top circles) by $\simeq 0.25$~dex, sufficient to be consistent with the CEL abundances shown on the abscissa. 
The ADF in M42 is therefore fully explained by continuum pumping effects.
} 
\label{fig:o2rl_fluo}
\end{figure}

To examine the extent to which continuum pumping can explain the large spread in apparent \Opp/\Hp\ abundances, we construct a photoionization model following the framework developed for the PN IC418 \cite{escalante12}, but with the properties of ionizing radiation and gas tailored to M42. For each detected \Oiirec\ transition, the ratio of continuum pumping to pure recombination emissivity $f_\mathrm{pump} = \varepsilon_\mathrm{pump}/\varepsilon_\mathrm{rec}$ is calculated from the photoionization model (see Methods). 
If continuum pumping is important, its contribution will increase the observed RL fluxes and hence the derived abundance, such that we expect a linear relation between the apparent \Opp/\Hp\ and $f_\mathrm{pump}$. 
Indeed, we find strong correlations between the apparent $\Opp/\Hp$ abundance and $f_\mathrm{pump}$ ($r \simeq 0.59$, $p \simeq 3 \times 10^{-3}$) in the two independent ionization regions and across the full FoV.
We perform a linear fit between the apparent \Opp/\Hp\ and $f_\mathrm{pump}$. As expected, a linear function can well describe ($\mathrm{RMS} < 0.03$~dex after excluding one outlier) the relation in all three stacked spectra, with only a few outliers likely arising from residual blending or unidentified contamination (Fig.~\ref{fig:o2rl_fluo}). The best-fitting slope also suggests that the photoionization model can predict an adequet amount of continuum pumping as observed. As a result, continuum pumping contributes to $\gtrsim 50\%$ of the flux even for the V1 multiplet, exceeding the previous expectations that its contribution is negligible.

Ultimately we seek to determine the true elemental abundance, which is best measured using lines that are negligibly affected by continuum pumping ($f_\mathrm{pump} \simeq 0$). However, the \Oiirec\ RLs with $f_\mathrm{pump} = 0$ are fainter compared to the V1 multiplet. To minimize contamination from unknown sources, we instead estimate the ``pumping free'' \Oiirec\ RL abundance from the intercept at $f_\mathrm{pump}=0$ with a linear best-fit relation between the apparent $\Opp/\Hp$ abundance and $f_\mathrm{pump}$. For all stacked spectra, the corrected RL abundances are lower than the V1 abundances by $> 0.2$~dex (or $>3.5\sigma$), but consistent with the corresponding CEL abundances within $1\sigma$. For example, for the full-FoV spectrum, the corrected abundance $12 + \log(\Opp/\Hp)_\mathrm{cor} = 8.14_{-0.15}^{+0.11}$ is consistent with the CEL-based abundance of $12 + \log(\Opp/\Hp) = 8.31 \pm 0.06$ within $1.5\sigma$. In our photoionization model, five \Oiirec\ RLs have $f_\mathrm{pump} < 0.1$ (4085.11, 4092.93, 4185.44, 4189.79, and 4590.97), with median apparent $12 + \log(\Opp/\Hp)_\mathrm{cor} = 8.21$ in the full-FoV spectrum, indicating that these lines may also be used to estimate the ``pumping free'' abundance. The abundance discrepancy in M42 can therefore be explained entirely by accouting for continuum pumping. 

The spatial resolution of the KCWI IFU observations additionally allows us to examine the more commonly invoked explanations of the ADF (see Methods). In the temperature fluctuation scenario, the observed ADF can be explained with a standard deviation in temperature $T_e$ of $\simeq 1500$~K. By contrast, we measure a standard deviation among individual spaxels that is an order of magnitude smaller across the FoV [$\simeq 120~\mathrm{K}$; see (see Methods) for details]. While temperature fluctuations on scales below the spatial resolution of 373~AU cannot be ruled out, the short thermal relaxation time of only a few years places stringent constraints on the characteristic scales over which such fluctuations could be maintained to account for the ADF. 

We can also probe temperature variations along the line of sight. A common interpretation is that interionic temperature differences [e.g., $T_e(\Oiii) - T_e(\Nii)$] are indicative of temperature fluctuations \cite{mendez-delgado23, garcia-rojas07b}. Although we find significant temperature differences between gas associated with different ionization states within individual spaxels, we generally do not detect a clear positive correlation between these differences and the ADF, indicating that interionic temperature differences are not the primary driver of the ADF. Previous high-spectral-resolution studies have also revealed multiple kinematic components along several sightlines within our FoV \cite{garcia-diaz08, blagrave06, mendez-delgado21a, mendez-delgado21b}. We find that the \Oiirec\ RLs in all spectrally resolved components are well reproduced by our continuum pumping prescriptions, and that the ADF remains resolved within the individual kinematic components of M42.

Alternatively, high-metallicity inclusions also cause the ADF. In this scenario, the \Oiirec\ RLs would be expected to originate from colder gas that may also be more obscured by dust. We find no evidence for either colder gas or additional dust obscuration (see Methods). First, temperatures derived from purely RL-based diagnostics (\Oiirec\ $\lambda\lambda 4649/4089$ and \Oiirec\ $\lambda\lambda 4649/4087$) \cite{storey17} are high ($T_e > 10,000$~K) and may even be inconsistent with the ratios predicted by current atomic data. Second, if the RLs originated in more dust-obscured gas, transitions at longer wavelengths would experience less extinction and therefore yield systematically larger apparent \Opp/\Hp\ abundances. Instead, we find no significant dependence on wavelength. Thus, although both effects may be present at some level, neither temperature fluctuations nor high-metallicity inclusions can account for the observed ADF in M42 and are therefore ruled out as its dominant origin.

Why, then, does our continuum pumping model predict substantial contamination, whereas previous models \cite{grandi76} concluded that it is negligible? We find that the absolute pumping strength is sensitive to the assumed ionizing spectrum, particularly at $\simeq 390$--430~\AA, where the relevant pumping transitions are located, as well as to the ionization parameter ($U$). A direct comparison of our predicted $f_\mathrm{pump}$ with previous models indicates that the underlying results are broadly consistent. However, under more realistic nebular conditions, our models predict values of $f_\mathrm{pump}$ that can exceed earlier estimates by several orders of magnitude (see Methods). Furthermore, although the absolute value of $f_\mathrm{pump}$ depends strongly on the adopted ionizing spectrum, the relative pumping fractions among different RLs are remarkably insensitive to this assumption, ensuring that the resulting corrected abundance, $12 + \log(\Opp/\Hp)_\mathrm{cor}$ at $f_\mathrm{pump}=0$, remains robust.

\begin{figure}[htbp!]
\centering
\includegraphics[width=0.9\textwidth]{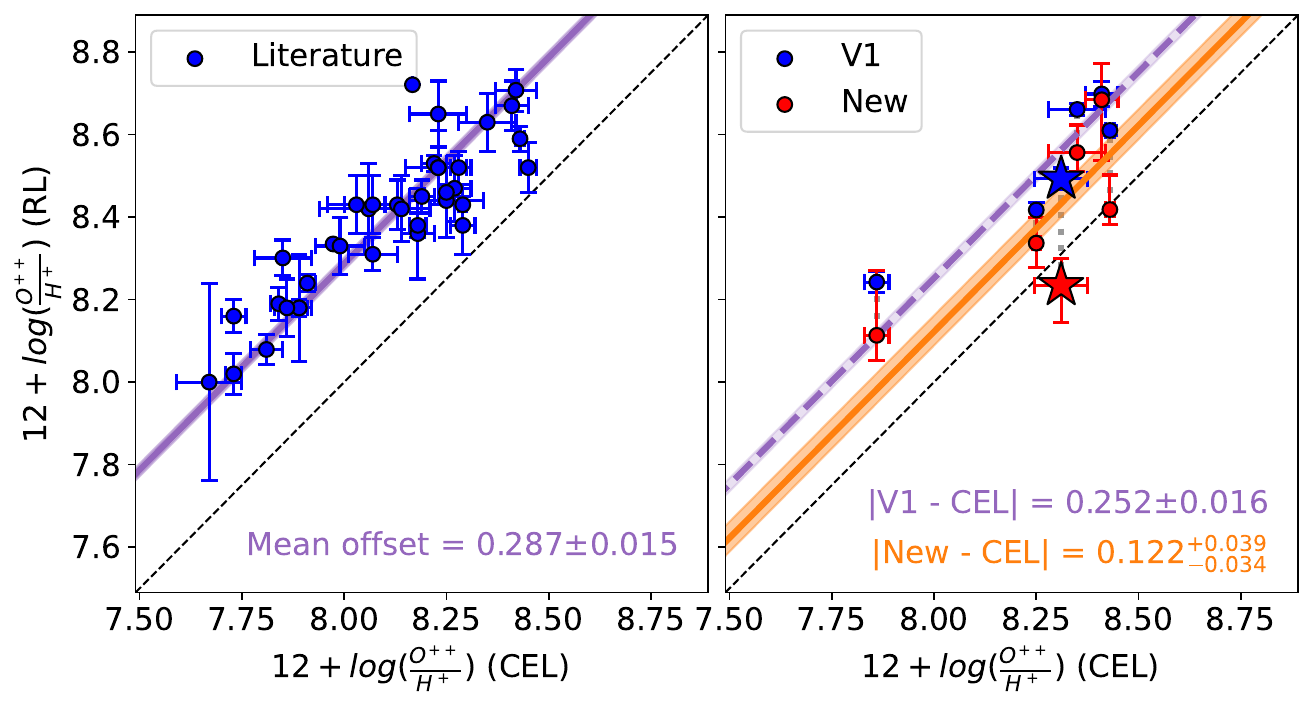}
\caption{ \textbf{Considering the effect continuum pumping reduces the ADF by $\simeq 50\%$.} \textbf{Left:} Abundance discrepancy factors compiled from literature \Hii\ regions \cite{esteban09, toribio-san-cipriano16, toribio-san-cipriano17, esteban14, pena-guerrero12, esteban04, garcia-rojas04, garcia-rojas05, garcia-rojas06, garcia-rojas07, peimbert03, tsamis03, esteban17}. The \Oiirec\ RL abundances (y-axis) are systematically higher than the direct-$T_e$ CEL abundances (x-axis), with a mean offset of $0.287 \pm 0.015$ dex. \textbf{Right:} The five literature \Hii\ regions with sufficient \Oiirec\ detections for continuum-pumping corrections (circles), together with the KCWI measurement of M42 from this work (star). The blue points show recalculated \Oiirec\ V1 abundances derived exclusively from V1 multiplet lines. The resulting ADF for these six objects is $0.252 \pm 0.016$, consistent with the larger literature sample and indicating that this subset is representative. After applying the pumping correction (see Fig.~\ref{fig:recalculate_met}), the mean ADF decreases to $0.122_{-0.034}^{+0.039}$, corresponding to a substantial reduction of $\gtrsim 50\%$. } 
\label{fig:new_adf}
\end{figure}

Having established that fluorescence in \Oiirec\ RLs is non-negligible in M42, we apply the same approach to test whether published \Oiirec\ RL metallicities have been systematically overestimated. Robust linear fits require measurements of multiple \Oiirec\ transitions spanning a range of $f_\mathrm{pump}$. We therefore compile five literature \Hii\ regions with sufficient reliable \Oiirec\ RL detections: M42 (a long-slit spectrum obtained near our FoV) \cite{esteban04}, NGC~3576 \cite{garcia-rojas04}, M8 and M17 \cite{garcia-rojas07}, and 30~Dor \cite{peimbert03}. For each object, we tailor the photoionization model to the published physical conditions and recompute $f_\mathrm{pump}$ for the measured lines (see Methods). In all five regions, the best-fitting slopes between the apparent $\Opp/\Hp$ abundances and $f_\mathrm{pump}$ are positive, implying that continuum pumping is a generic contributor to \Oiirec\ RL intensities. The strength of the effect, however, varies substantially. The literature M42 spectrum again requires a large correction of $\sim 0.2$ dex, sufficient to remove the ADF, whereas the remaining targets show more modest offsets (Fig.~\ref{fig:recalculate_met}). Averaged over the sample, the inferred downward revision of RL-based metallicities is 0.13 dex, accounting for $\gtrsim 50\%$ of the ADF inferred from the V1 multiplet in these same objects (Fig.~\ref{fig:new_adf}).

Although M42 may represent an extreme case with unusually strong continuum pumping, the pumping contribution in other objects may also be systematically underestimated, particularly in spectra of lower signal-to-noise. The ultrafaint \Oiirec\ transitions with $f_\mathrm{pump} \sim 0$ are especially susceptible to blending or contamination from unrelated features, which would artificially elevate the intercept in the linear fit. 
As a result, the inferred corrections in these objects are statistically weak, even if continuum pumping contributes non-negligibly to the RL fluxes. 

Finally, although we show that temperature fluctuations do not drive the ADF for \Opp, this does not imply that temperature variations are absent or unimportant for other ions. On the contrary, within our FoV the standard deviations of $T_e$(\Sp) and $T_e$(\Spp) both exceed $1000~\mathrm{K}$ (Fig.~\ref{fig:te_maps}). Such fluctuations can introduce substantial biases in CEL-based abundance determinations for these species. Temperature structure must therefore be assessed on an ion-by-ion basis, rather than assumed to affect all abundance diagnostics uniformly \cite{esteban09, esteban14}. We also emphasize that our analysis is restricted to star-forming \Hii\ regions. Although planetary nebulae (PNe) likewise exhibit significant ADFs, the largest ADFs are closely associated with binary central stars \cite{wesson18}, indicating a physical origin distinct from that in \Hii\ regions. We therefore exclude PNe from the present analysis.

Our results highlight the substantial impact of continuum pumping on recombination-line diagnostics. Owing to this effect, gas-phase O/H abundances derived from \Oiirec\ RLs—often regarded as the ``gold standard'' of metallicity measurements—can be overestimated by up to 0.25~dex in \Hii\ regions. Studies that directly compare observed gas-phase metallicities with stars \cite{simon-diaz11}, numerical simulations \cite{schaye15, pillepich18}, star-formation and feedback models \cite{muratov15, ma16}, dust depletion estimates \cite{pena-guerrero12}, or empirical strong-line calibrations \cite{pettini04, maiolino08, curti20} should therefore be revisited in light of this bias. Moreover, as direct-$T_e$ metallicities are increasingly measured at high redshift \cite{sanders24, rogers26}, it is essential to quantify both temperature-structure effects and continuum pumping across a range of physical environments. A systematic approach requires deep, spatially resolved spectroscopy of nearby targets spanning diverse conditions, combined with consistent analyses using multiple abundance diagnostics and tailored photoionization modeling. We are pursuing this strategy through the ongoing CIRAAS program.

\clearpage

\section*{Methods}
\subsection*{\textbf{Observations and Data Reduction}}

The KCWI observations were obtained on UT 2024-12-29 under clear, photometric conditions, with a typical seeing of $\simeq 0\farcs8$. KCWI comprises two independent channels (blue and red), each equipped with a dedicated spectrograph that can operate simultaneously. We employed two observational configurations. A ``snapshot'' setup with short integrations was designed to avoid saturation in the brightest nebular lines, while a ``deep'' setup with longer exposures enabled the detection of faint auroral transitions and the \Oiirec\ RLs. The snapshot configuration used the ``BL'' grating centered at 4500~\AA\ in the blue channel and the ``RL'' grating centered at 8000~\AA\ in the red channel, both providing a spectral resolution of $R \sim 2{,}000$. The deep configuration employed the ``BM'' grating centered at 4400~\AA\ in the blue channel ($R \sim 4{,}000$) and the RL grating centered at 7250~\AA\ in the red channel ($R \sim 2{,}000$). Together, the $R \sim 2{,}000$ settings (BL4500, RL7250 and RL8000) deliver nearly continuous wavelength coverage from 3575 to 9775~\AA. Examples of the reduced spectra are shown in Fig.~\ref{fig:full_spec}. The higher-resolution blue setup ($R \sim 4{,}000$) covers 3980--4860~\AA, sufficient to resolve the \Oiirec\ multiplet structure and measure individual components. The spectroscopic configuration adopted for each strong emission line measurement in Table~\ref{tab:strong_line_fluxes} is listed therein.

Both configurations were used in combination with the ``Medium'' image slicer, yielding a field of view of $\simeq 16" \times 20"$ per pointing and a native spatial sampling of $0\farcs68 \times 0\farcs29$. Our final mosaic comprises two adjacent pointings arranged along the west–east direction. For each pointing, we obtained a sequence of individual exposures separated by $90^\circ$ increments in position angle (PA) to improve spatial symmetry and mitigate instrumental systematics in flat-fielding and background subtraction. The science sequence was accompanied by offset blank-sky exposures for sky subtraction. The full observing log is provided in Table~\ref{tab:kcwi_observing_log}. Despite the 1-s integrations in the snapshot configuration, the bright \Oiii\ $\lambda 5007$ line is saturated and therefore excluded from the analysis. We instead use the unsaturated \Oiii\ $\lambda 4959$ transition for all subsequent measurements.

Data reduction followed Ref.~\cite{chen24}, using the \textsc{Python} implementation of the KCWI Data Reduction Pipeline (DRP) (\url{https://github.com/Keck-DataReductionPipelines/KCWI_DRP}), with substantial modifications adapted from \textsc{KCWIKit} \cite{prusinski24}. Key adjustments are summarized here. For cube construction and rectification, we adopted linear interpolation to conserve flux and to minimize artificial ripples around strong emission features. Cosmic-ray rejection in the red channel was performed via sigma clipping across adjacent exposures, masking discrepant pixels prior to feeding in the DRP. Sky subtraction was achieved by removing the DRP-generated sky model derived from dedicated blank-sky frames. Although the red channel encompasses numerous time-variable OH emission lines from the sky, the intrinsic surface brightness of M42 dominates over the sky background, rendering this approach sufficient. Flux calibration was based on spectrophotometric standard stars (Feige~110 and Feige~34) observed with identical instrumental configurations during evening and morning twilight. Telluric absorption corrections were applied using \textsc{KSkyWizard} (\url{https://github.com/zhuyunz/KSkyWizard}) and \textsc{PypeIt} \cite{prrochaska20}, adopting the same flux standards.

The final data cubes were generated by resampling individual exposures onto a common $0\farcs3 \times 0\farcs3$ grid using \textsc{KCWIKit} and \textsc{Montage} (\url{http://montage.ipac.caltech.edu/}), after which the two adjacent pointings were mosaicked into a single cube. Astrometric alignment was refined by cross-correlating a pseudo-white-light image constructed from the KCWI cube with the MUSE pseudo-white-light reference shown in Fig.~\ref{fig:fov}. KCWI is seeing-limited; during our observations the spatial resolution was $\sim 0\secpoint8$. To improve the signal-to-noise ratio while preserving spatial information, we rebinned the data into $0\secpoint9 \times 0\secpoint9$ spaxels, corresponding to physical scales of $373 \times 373$ AU at a distance of 414~pc \cite{menten07}. Stars and regions affected by stray light within the field were masked in subsequent analyses.

For strongly detected emission lines, uncertainties in flux ratios may be dominated by relative flux-calibration errors. We quantify this effect by comparing the calibrated spectra of the standard stars in individual exposures with their reference spectra from the \textsc{CALSPEC} database \cite{bohlin14}, measuring relative flux differences over a range of wavelength intervals. As expected, over narrow separations the discrepancy is minimal, with an RMS of $\sim 0.5\%$ at $\Delta\lambda = 15$~\AA. The relative differences increase gradually with wavelength separation and plateau at $\Delta\lambda \sim 100$~\AA, reaching an RMS of $\sim 1\%$. We therefore adopt a conservative 1\% systematic uncertainty for all individual flux measurements, corresponding to a 1.4\% uncertainty in flux ratios.

\subsection*{\textbf{Spatially Resolved Nebular Analysis}}

For each rebinned spaxel, nebular emission-line fluxes are measured by fitting single Gaussian profiles, with the exception of H$\alpha$. The H$\alpha$ profile is modeled with two Gaussian components—a narrow and a broad component sharing a common velocity centroid—to represent the nebular and outflow contributions. The broad component contributes $2.0\% \pm 0.3\%$ of the narrow-component flux. Owing to its substantially larger velocity dispersion ($\sigma \simeq 16~\mathrm{\AA}$ versus $\sigma \lesssim 2~\mathrm{\AA}$ for the narrow component), the total flux difference between single- and two-component fits is only $\simeq 0.5\%$ across all spaxels. For emission lines in close wavelength proximity (for example, \Nii\ + H$\alpha$, \Oiii\ + H$\beta$, and the \Oii\ $\lambda\lambda 3726,29$ doublet), we perform simultaneous multi-line fitting within the relevant spectral window. The local continuum is modeled with a low-order polynomial using manually selected line-free regions, and all lines in the window are fit with tied velocity offsets and dispersions. When a line is covered by multiple spectral configurations, we preferentially adopt measurements from the deeper BM4400 and RL7250 setups unless saturation occurs. The \Oiii\ $\lambda 5007$ transition is saturated in a subset of spaxels even in the shallow BL4500 exposure and is therefore excluded from direct analysis. When required, its flux is inferred as 2.98 times the \Oiii\ $\lambda 4959$ flux, following the theoretical ratio from atomic transition probabilities \cite{storey00}. For lines robustly detected in multiple configurations, flux measurements agree within $0.5\%$, below the adopted 1\% systematic uncertainty.

Within the FoV, the ionization structure exhibits a clear bimodality, traced by the dust-insensitive $\mathrm{O3N2} \equiv \log(\Oiii~\lambda 5007/\mathrm{H}\beta) - \log(\Nii~\lambda 6583/\mathrm{H}\alpha)$ ratio (Fig.~\ref{fig:o3n2}). The north-eastern portion of the FoV, located closer to the Trapezium cluster, corresponds to a high-ionization zone with $\mathrm{O3N2} \simeq 1.4$, whereas the south-western side is characterized by lower ionization with $\mathrm{O3N2} \simeq 1.0$. On the O3–N2 BPT diagram \cite{baldwin81}, the spaxels occupy loci above the canonical star-forming ridge line. The high-ionization region lies near the Kauffmann demarcation \cite{kauffmann03}, consistent with previous measurements of M42 \cite{esteban04}, while the low-ionization region falls between the Kauffmann and Kewley \cite{kewley01} boundaries. Based on the O3N2 distribution, we split the FoV into three regions: a high-ionization zone with $\mathrm{O3N2} > 1.30$ (comprising $\simeq 30\%$ of the FoV), a low-ionization zone with $\mathrm{O3N2} < 1.17$ ($\simeq 60\%$), and an intermediate transition zone occupying the remaining $\simeq 10\%$. For the high- and low-ionization regions, we construct integrated 1D spectra by co-adding the corresponding spaxels; the combined spectrum of the full FoV is denoted the ``full-FoV'' spectrum. These three stacked spectra are shown in Fig.~\ref{fig:full_spec}. Emission-line measurements and nebular diagnostics are performed identically for both individual spaxels and the stacked spectra. A summary of the measured fluxes is provided in Table~\ref{tab:strong_line_fluxes}.

Dust attenuation was corrected on a spaxel-by-spaxel basis. We adopt the mean \ebv\ derived from the H$\gamma$/H$\beta$ and H$\beta$/H$\alpha$ Balmer decrements, assuming $R_V = 5.5$—as previously reported for M42 \cite{costero70, blagrave07}—and the Milky Way extinction law of Ref. \cite{cardelli89}. Because stellar sources are masked within the FoV, the contribution from scattered stellar absorption to the Balmer lines is negligible. In the deep $R \sim 4{,}000$ spectra, we detect no broad absorption wings around higher-order Balmer transitions or \ion{He}{I} absorption lines for typical OB stars. The resulting \ebv\ map is shown in Fig.~\ref{fig:ebv_map}.

Determination of metallicity via the direct-$T_e$ method requires measurements of both the electron density ($n_e$) and temperature ($T_e$), obtained from line ratios with different sensitivities to these quantities. For \Sp, we derive $n_e$ from the density-sensitive \Sii\ $\lambda\lambda 6716/6731$ ratio and $T_e$ from the temperature-sensitive \Sii\ $\lambda\lambda 4068/(6716+6731)$ ratio, solving simultaneously for $(n_e, T_e)$ in each spaxel. The weaker \Sii\ $\lambda 4076$ auroral transition is excluded owing to concerns that it may be contaminated by nearby \Oiirec\ recombination emission \cite{liu04}. For \Op, we apply a similar procedure, deriving $n_e$ from the \Oii\ $\lambda\lambda 3726/3729$ ratio and $T_e$ from the \Oii\ $\lambda\lambda (3726+3729)/(7319+7330)$ ratio, again solving simultaneously for consistency between the two diagnostics.

The $n_e(\Op)$ measurements are reported for completeness but are not adopted in the subsequent abundance analysis; instead, we use $n_e(\Sp)$ throughout. In M42, the relatively high densities render the \Oiii\ $\lambda 4959$ emissivity mildly density dependent, such that the inferred abundance depends on the adopted $n_e$. We adopt $n_e(\Sp)$ to maintain consistency with previous determinations of the \Op\ abundance in M42 \cite{esteban04}. Additionally, the \Oii\ $\lambda\lambda 3726,3729$ density-sensitive doublet is partially blended at $R \sim 2{,}000$ in the BL4500 setup, introducing additional uncertainty. The $n_e(\Op)$ values derived from \Oii\ are systematically lower than $n_e(\Sp)$ by 10\%--40\%. Adopting $n_e(\Op)$ would decrease the inferred \Op\ and total oxygen abundances by $\simeq 0.15$~dex. This choice does not affect the derived temperatures or abundances of other ions, as their relevant transitions have critical densities typically exceeding $10^5~\mathrm{cm}^{-3}$. Nor does it influence our conclusions regarding the \Oiirec\ RLs and the ADF, given the comparably weak dependence of the \Oiirec\ emissivities on $n_e$.

Adopting $n_e(\Sp)$ for each spaxel, we recalculate $T_e(\Op)$, and determine $T_e(\Opp)$ from the dust-corrected \Oiii\ $\lambda\lambda 4363/4959$ ratio. The temperature of \Np\ is derived from the \Nii\ $\lambda\lambda 5755/6583$ ratio. These ionic temperatures are then used to compute the emissivities of the corresponding nebular transitions, including \Oii\ $\lambda\lambda 3726+3729$, \Oiii\ $\lambda 4959$, and \Nii\ $\lambda 6583$. To facilitate comparison with empirical relations linking the ADF or temperature fluctuations to $T_e(\Oiii) - T_e(\ion{H}{I})$, we also derive the hydrogen recombination temperature from the Paschen Jump (PJ). As shown in Fig.~\ref{fig:PJ}, the PJ is measured as the median continuum flux-density difference between the relatively line-free intervals [8135, 8179]~\AA\ and [8558, 8575]~\AA. The PJ/Pa11 ratio is modeled using \textsc{PyNeb} \cite{luridiana15}, assuming $\Hep/\Hp = 0.0874$ from Ref.~\cite{esteban04}. No \ion{He}{II} emission is detected within the FoV, and we therefore adopt $\Hepp/\Hp = 0$. We do not derive a Balmer Jump temperature, as the BL4500 configuration lacks sufficient depth and spectral resolution to identify uncontaminated continuum windows in the relevant wavelength range.

Dust-corrected flux ratios of nebular lines relative to H$\alpha$ or H$\beta$—chosen according to proximity in wavelength—are used to derive ionic abundances. The resulting \Opp\ abundance maps are presented in Fig.~\ref{fig:fov}. For references, the $n_e$ distributions are shown in Fig.~\ref{fig:ne_maps}. The $T_e$ maps derived from \Oiii, \Sii, the Paschen Jump, \Oii, \Nii, \Siii\ and \Ariii\ are shown in Fig.~\ref{fig:te_maps}. Ionic abundances of \Sp, \Op, \Np, \Spp\ and \Arpp\ are displayed in Fig.~\ref{fig:ion_maps}, while the total abundances of O, N and S are shown in Fig.~\ref{fig:totmet_maps}. The detailed methodology is summarized in Table~\ref{tab:phys_method}, and the derived $T_e$, $n_e$ and abundances for the three stacked spectra are listed in Table~\ref{tab:phys_properties}.

We estimate uncertainties via Monte Carlo (MC) propagation. For each spaxel, we generate 2000 realizations of the measured line fluxes, perturbing each flux according to its uncertainty, and repeat the full analysis chain—deriving \ebv, $n_e$, $T_e$ and ionic abundances—for every realization. This procedure naturally captures non-linear error propagation and covariances among derived quantities. Running this analysis over the full set of $>600$ spaxels is computationally expensive. We therefore apply the MC framework to a representative subset of seven spaxels randomly selected across the FoV; their locations are indicated in Fig.~\ref{fig:o3n2}.

\subsection*{\textbf{The O~{\sc II}} Recombination Lines}

The \Oiirec\ recombination lines are central to this study and are therefore measured and analysed with particular care. The most widely used transitions for abundance determinations is the V1 multiplet, arising from the \Op\ $3p~^4D^\circ \rightarrow 3s~^4P$ transitions (Fig.~\ref{fig:v1_grotrian}). The multiplet comprises the lines at 4638.86, 4641.81, 4649.13, 4650.84, 4661.63, 4673.73, 4676.23 and 4696.35~\AA. We measure the V1 fluxes in each spaxel by jointly fitting Gaussian profiles with a common velocity centroid and dispersion, while allowing the line intensities to vary independently. All components are robustly detected in the spatially resolved spectra except the 4696.35~\AA\ transition, which is excluded from the V1 abundance analysis. The median signal-to-noise ratio of the total V1 flux per spaxel is $\simeq 55$. Assuming $T_e = T_e(\Oiii)$ and $n_e = n_e(\Sii)$, and adopting Case~B recombination coefficients from Ref.~\cite{storey17}, we derive the \Opp\ abundance from the dust-corrected \Oiirec\ V1/H$\beta$ ratio. The RL-based abundance is compared with the CEL-based value to compute the abundance discrepancy factor, $\mathrm{ADF} = \log[\Opp/\Hp(\mathrm{RL})] - \log[\Opp/\Hp(\mathrm{CEL})]$. The resulting maps of $\Opp/\Hp(\mathrm{RL})$ and the ADF are presented in Fig.~\ref{fig:fov}.

\subsection*{\textbf{Examining Temperature Variations across the FoV}}

Temperature fluctuations are the most commonly adopted explanation of the ADF. However, both the origin of the ADF and the physical drivers of the putative temperature fluctuations remain debated. Long-slit observations have established that the ADF persists down to sub-pc scales \cite{odell03}. However, classical photoionization models of chemically homogeneous \Hii\ regions predict temperature variations that are generally too small to reproduce the required $t^2$, unless several additional heating sources—such as turbulence, shocks or cosmic rays—are invoked simultaneously \cite{kingdon95, giammanco05, garcia-rojas07b}. Attempts to directly measure temperature fluctuations often lead to contradictory results \cite{tsamis03, croxall13, stasinska13, chen23, kelly25, huntzinger26}. 

In this and the following sections, we examine whether temperature variations within the observation window can explain the ADF in M42. Following the temperature-fluctuation formalism of Ref.~\cite{peimbert67}, the dimensionless fluctuation parameter is defined as,
\begin{equation}
    t^2 = \frac{\int n_e n(\Opp) (T_e - T_0)^2 \diff V}{\int n_e n(\Opp) T_0^2 \diff V },
\end{equation}
where $n(\Opp)$ is the number density of \Opp, $T_0$ is the mean temperature, and $\int \diff V$ denotes a volume integral. In the classical treatment, $n_e$ and $n(\Opp)$ are assumed constant and the $T_e$ distribution is taken to be Gaussian. Following Refs.~\cite{chen23, chen24}, we perform a numerical integration of the temperature distribution by jointly fitting the \Oiii\ $\lambda\lambda 4363,4959$ lines and the \Oiirec\ V1 fluxes, with $T_0$ and $t^2$ as free parameters. The corresponding temperature dispersion is expressed as, 
\begin{equation}
    \sigma_T(t^2) = \sqrt{t^2} T_0.
\end{equation}
Maps of $t^2$, $\sigma_T(t^2)$ and $T_0$ are presented in Fig.~\ref{fig:t2_maps}. Compared with the traditional first-order Taylor expansion \cite{peimbert67}, which approximates a linear relation between the apparent $T_e$ and $T_0$, the numerical approach is more accurate and typically yields larger $t^2$ and $\sigma_T(t^2)$. For a representative median $T_0 = 7400$~K and $t^2 = 0.041$, the first-order approximation would instead give $t^2 = 0.03$, reducing $\sigma_T(t^2)$ by $\simeq 10\%$. We therefore adopt the numerical integration throughout.

Although the inferred median temperature variation is modest at $\sigma_T = 1510~\mathrm{K}$, the spatially resolved $T_e$(\Opp) map across the FoV (Fig. \ref{fig:te_maps}) is remarkably uniform, with a standard deviation of only 120~K—far too small to account for the required temperature fluctuations. To further test the spatial effect, we progressively bin adjacent spaxels and recompute the ADF (Fig.~\ref{fig:t2_cog}). From the native spatial scale of 373~AU to the full FoV ($\gtrsim 7000$~AU), the $\sigma_T$ does not exhibit a systematic increase with bin size. This absence of scale dependence indicates that temperature variations across the FoV are not the primary driver of the ADF, and instead constrains the underlying mechanism to operate on spatial scales smaller than 373~AU. This result is consistent with previous spatially resolved measurements of $T_e$(\Oiii) based on Hubble Space Telescope narrowband imaging at finer spatial resolution \cite{odell03}, which found a scatter of only 495~K over a larger FoV.  

Assuming a turbulent velocity of 10~\kms, the thermal relaxation time ($\lesssim$ few years) is orders of magnitude shorter than the turbulent crossing time ($\sim180$ yr at 373 AU), implying that temperature perturbations are rapidly damped and the gas behaves close to isothermal on these scales. Consequently, subsonic to transonic turbulence is unlikely to sustain the observed $\sigma_T \simeq 1500~\mathrm{K}$ without additional localized heating or multiphase structure. Our ADF map thus provides the strongest spatial-resolution constraints to date on the physical origin of the ADF in \Hii\ regions.

\subsection*{\textbf{Examining Interionic Temperature Variations along the Line of Sight}}

The analysis above constrains the physical origin of the ADF across the plane of the sky. However, the three-dimensional structure of M42 is highly complex \cite{odell17}, and multiple gas phases may remain unresolved along the line of sight, potentially broadening the effective $T_e$(\Oiii) distribution within individual spaxels. Indeed, each spaxel contains emission from ions spanning a wide range of ionization states (Fig.~\ref{fig:te_maps}), reflecting the multiphase nature of the nebula. Differences in $T_e$ derived from different ions—for example, $T_e$(O$^{++}$) versus $T_e$(N$^+$)—have therefore been proposed as evidence for temperature inhomogeneities in \Hii\ regions \cite{mendez-delgado23, garcia-rojas07b}.

We compare the spatially resolved $\sigma_T(t^2)$ with $\Delta T_e \equiv T_e(\Oiii) - T_e(X)$ for all available diagnostics (Fig. \ref{fig:tsig_to_dt}). Two empirical relations are compared with our analysis. The first links $t^2$ to $T_e(\Oiii) - T_e(\Nii)$ \cite{mendez-delgado23}; we convert $t^2$ to $\sigma_T(t^2)$ assuming $T_0 = 7400~\mathrm{K}$, representative of our FoV. The second connects the ADF with $T_e(\Oiii) - T_e(\Hi)$, where $T_e(\Hi)$ is inferred from the Balmer jump \cite{garcia-rojas07b}. Although our spaxels generally lie near the empirical relations, correlations established between different \Hii\ regions do not persist in the spatially resolved analysis within a single object. Of the six $\Delta T_e$ comparisons examined, two ($T_e(\Oiii) - T_e(\Nii)$ and $T_e(\Oiii) - T_e(\Siii)$) exhibit statistically significant positive correlations ($p<0.01$), one shows a significant negative correlation ($T_e(\Oiii) - T_e(\Sii)$), and three show no significant correlation despite sufficiently small uncertainties to detect one if present ($T_e(\Oiii) - T_e(\mathrm{PJ})$, $T_e(\Oiii) - T_e(\Oii)$, and $T_e(\Oiii) - T_e(\Ariii)$). Even for the statistically significant cases, the distributions reveal distinct behaviors in the low- and high-ionization regions, rather than a single continuous trend. 

The breakdown of the $\sigma_T$-$\Delta T_e$ relation in our spatially resolved analysis can be understood if the physical mechanism responsible for the ADF is distinct from that producing the interionic temperature differences. When an \Hii\ region is considered as a whole, the interionic $\Delta T_e$, particularly between ions with different ionization potentials (e.g., \Oiii\ and \Nii), primarily reflects the ionization structure established by photoionization. Such large-scale temperature gradients do not necessarily translate into temperature fluctuations within the \Opp\ zone itself, unless gas spanning different ionization phases is substantially mixed on spatial scales smaller than our physical resolution of 373~AU.

We therefore conclude that, although M42 as a whole follows the empirical ADF-$\Delta T_e$ relation established from object-to-object comparisons, interionic temperature differences are not the physical driver of the ADF on the spatial scales probed here. Rather, the observed $\Delta T_e$ primarily traces the nebular ionization structure, whereas the mechanism responsible for the ADF appears to operate independently of these large-scale interionic temperature differences. Below (Continuum Pumping in the Line-of-Sight structures), we examine the kinematically resolved \Opp\ structures near our FoV separately and determines that they too follow the prediction of continuum pumping.

\subsection*{\textbf{Examining Cool Metal Inclusions and Differential Dust Attenuation}}

In planetary nebulae (PNe) with extreme abundance discrepancy factors, the classical temperature-fluctuation framework would require implausibly large $t^2$ values. This tension has motivated chemically inhomogeneous ``bi-abundance'' models, in which cold, hydrogen-poor, metal-rich inclusions are embedded within the ionized gas \cite{liu00, liu04, wesson03}. These inclusions may contain dust and experience internal extinction. As more recently, two-phase models coupled with non-uniform dust obscuration have been invoked to reconcile discrepancies between optical CELs and far-infrared fine-structure lines \cite{chen24, harikane25, usui25}. Note that the cause of ADF in PNe has been historically viewed as different to \Hii\ regions, especially high-ADF PNe (ADF$\sim 1$) that have been found to be associated with the binarity of the central objects. This analysis is only to examine whether this explanation is suitable for M42. The result of this analysis should not be applied broadly to PNe. 

Our deep KCWI data enable a direct test of this scenario using the ultrafaint \Oiirec\ RL diagnostics for $T_e$ and $n_e$, originally developed for PNe \cite{mcnabb13, storey17}. The recombination coefficients depend on temperature, while density-dependent collisional processes redistribute the populations of the \Opp\ parent fine-structure levels, thereby modifying the subsequent radiative cascades. The \Oiirec\ $\lambda\lambda 4649.13/4661.63$ ratio, which arises within the V1 multiplet, is primarily sensitive to $n_e$. In contrast, the \Oiirec\ $\lambda\lambda 4649.13/4089.29$ ratio, which connects the V1 and V48 multiplets, has been proposed as a diagnostic of $T_e$. We measure the weak \Oiirec\ $\lambda 4089.29$ line using the same joint-fitting procedure applied to the V1 multiplet, achieving a median signal-to-noise ratio of 8.2 per spaxel.

Previous studies have suggested that the \Oiirec\ $\lambda 4089.29$ line may be contaminated by several sources. One possibility is blending with \ion{Si}{IV} $\lambda 4088.86$. In particular, Ref.~\cite{peimbert13} suggested that the neighboring $\lambda 4116.10$ feature detected by Ref.~\cite{esteban04} is also due to \ion{Si}{IV}, consistent with the theoretical \ion{Si}{IV} $\lambda\lambda 4088.86/4116.10$ intensity ratio of 2.4. This interpretation, however, does not appear applicable to our data. We find that the observed $4089.29/4116.10$ ratio differs substantially from the theoretical 2.4:1 expectation. In our high-ionization spectrum (Fig.~\ref{fig:full_spec}), the $\lambda 4116.10$ feature is only marginally detected, whereas both the $\lambda 4116.10$ feature and the adjacent [\ion{Fe}{II}] $\lambda 4114.48$ line are clearly detected with comparable strengths in Ref.~\cite{esteban04}. Moreover, even if the \Oiirec\ $\lambda 4089.29$ line were partially contaminated by \ion{Si}{IV} $\lambda 4088.86$, removing this contribution would increase the intrinsic \Oiirec\ $\lambda\lambda 4649.13/4089.29$ ratio, thereby exacerbating, rather than alleviating, the discrepancy between the observed and theoretically predicted ratios (see below).

A second possibility is that the \Oiirec\ $\lambda 4089.29$ line is affected by scattered \ion{Si}{IV} $\lambda 4088.86$ absorption originating in the spectra of O stars. Indeed, our continuum spectra exhibit prominent scattered stellar features. The scattered light is dominated by $\theta^1$ Ori C, as indicated by the distinctive \ion{He}{II} $\lambda\lambda 4542/4686$ equivalent-width ratio of $\simeq 1.5$. This ratio is typically $<1$ for normal O stars, but $\theta^1$ Ori C is an O7Vp star with peculiar stellar wind and magnetic field properties, and its high-resolution spectrum exhibits a similarly elevated \ion{He}{II} $\lambda\lambda 4542/4686$ ratio \cite{simon-diaz06}. Following the methodology of Ref.~\cite{liu95}, we therefore construct a simple continuum model by linearly combining the spectrum of $\theta^1$ Ori C (downloaded from the ESO archive; observed with VLT/UVES on UT2016-02-22; Program ID: 194.C-0833; PI: N. Cox) and the Balmer continuum predicted by \textsc{PyNeb} over $\simeq 4000$--5000~\AA. The relative normalization is determined by matching the observed \ion{He}{II} $\lambda 4542$ equivalent width of $\simeq -0.22$~\AA\ in each spaxel. After subtracting this modeled stellar continuum, the measured \Oiirec\ $\lambda 4089.29$ flux increases by approximately 30\%.

After correcting for scattered stellar light, the resulting spatially resolved \Oiirec\ RL diagnostics for $n_e$ and $T_e$, based on the \Oiirec\ $\lambda\lambda 4649.13/4661.63$ and $\lambda\lambda 4649.13/4089.29$ ratios, are shown in Fig.~\ref{fig:o2rl_tene} (top). To assess the robustness of this result, we also replace the potentially contaminated \Oiirec\ $\lambda 4089.29$ line with the \Oiirec\ $\lambda 4087.15$ transition (Fig.~\ref{fig:o2rl_tene}, bottom). Like the $\lambda 4089.29$ line, the $\lambda 4087.15$ transition corresponds to a \aterm{4f}{}{}{}{} $\rightarrow$ \aterm{3d}{}{}{}{} cascade, and the \Oiirec\ $\lambda\lambda 4649.13/4087.15$ ratio is likewise sensitive to both $T_e$ and $n_e$. Importantly, the spectrum of $\theta^1$ Ori C exhibits no stellar feature near $\lambda 4087.15$. Owing to the faintness of this line, however, its flux can only be measured reliably from the spatially binned spectra. Nevertheless, both diagnostic diagrams yield the same conclusion: the observed \Oiirec\ $\lambda\lambda 4649.13/4089.29$ and $\lambda\lambda 4649.13/4087.15$ ratios are systematically high, implying $T_e > 10{,}000~\mathrm{K}$ across most of the FoV. These temperatures are inconsistent with the expectation that the RL emission originates from a substantially cooler gas component.

The apparently unphysical \Oiirec\ line ratios could still be reconciled with a distinct cool component if the RL-emitting gas were significantly more obscured by dust, artificially enhancing the observed $\lambda\lambda 4649/4089$ ratio.
To investigate the possibility of the scenario, owing to their higher signal-to-noise ratios, the stacked spectra enable secure measurements of additional \Oiirec\ transitions beyond those accessible in the spatially resolved analysis. The line fluxes are measured following the same procedure as for other emission lines: transitions in close wavelength proximity are grouped and modeled jointly with a common velocity centroid and dispersion. In the deep BM4400 configuration, we identify 33 \Oiirec\ RLs. Of these, five are potentially affected by blending with other faint features or suffer from wavelength ambiguities, and are therefore excluded from subsequent quantitative analysis. The remaining 28 lines are considered secure. Representative examples of the detected \Oiirec\ RLs in the stacked spectra are shown in Fig.~\ref{fig:full_spec}, and the measured fluxes are listed in Table~\ref{tab:oii_rl}. For each detected \Oiirec\ transition, we compute its apparent abundance, $12 + \log(\mathrm{O}^{++}/\mathrm{H}^{+})$, following the same procedure adopted for the V1 multiplet. Specifically, we use the corresponding $T_e(\Oiii)$ and $n_e(\Sii)$ values listed in Table~\ref{tab:phys_properties} to calculate the recombination emissivity of each \Oiirec\ line with \textsc{PyNeb}. The dust-corrected \Oiirec/H$\beta$ flux ratio is then converted into an ionic abundance. The resulting apparent abundances for the individual RLs are provided in Table~\ref{tab:oii_rl}.

If the anomalous \Oiirec\ $\lambda\lambda 4649/4089$ and $\lambda\lambda 4946/4087$ ratios were produced by a cool, dust-embedded component, the ratio of observed to theoretically predicted recombination flux would exhibit a systematic dependence on wavelength. We find no such trend (Fig.~\ref{fig:o2rl_wave}, bottom). The apparent $\Opp/\Hp$ abundance inferred from individual \Oiirec\ RLs shows no significant correlation with wavelength (Pearson $r = 0.179$, $p = 0.36$). A linear fit with $3\sigma$ clipping yields a slope of $(1.8 \pm 1.6) \times 10^{-7}$, with uncertainties comparable to the slope itself and substantial scatter among the data points. These results disfavor the cool clumps or differential dust obscuration as the origin of the elevated $\lambda\lambda 4649/4089$ and $\lambda\lambda 4649/4087$ ratios. 

\subsection*{\textbf{The Continuum Pumping Model and Its Contribution}}

Another possible reason for the elevated \Oiirec\ $\lambda\lambda 4649/4089$ ratio is fluorescent continuum pumping \cite{grandi75a, grandi75b, grandi76, escalante05}. Continuum photons from the ionizing source, at wavelengths coincident with resonant transitions, can be absorbed by ions and overpopulate excited levels relative to the pure recombination case, thereby enhancing the RL intensity. This mechanism is known to dominate RL emission in some PNe—for example, in \ion{N}{ii}, \ion{O}{i} and \ion{C}{ii}, where pumping can account for $>99\%$ of the total flux \cite{escalante12}. 

For \Oiirec\ transitions, continuum pumping arises when far-UV photons from the ionizing source are absorbed by \Op\ ions through resonant transitions. In the case of the widely used \Oiirec\ V1 multiplet, the relevant pumping channels involve resonant transitions to the \aterm{5s}{4}{P}{}{} levels at $\simeq 32$~eV (388~\AA), the \aterm{4s}{4}{P}{}{} levels at $\simeq 30$~eV (420~\AA), and the \aterm{3d}{4}{D}{}{} levels at $\simeq 29$~eV (430~\AA), as illustrated in Fig.~\ref{fig:v1_grotrian}. These far-UV transitions link the ground configuration to excited states that feed the V1 multiplet through cascade. However, existing analytical model \cite{grandi76} predicts a negligible ($<0.1\%$) amount of contribution for the V1 multiplet from continuum pumping. Past photoionization models for PN IC418 also predict a modest contribution, typically $\lesssim 30\%$ for the V1 multiplet \cite{escalante12}. 

We construct our own continuum pumping models using the photoionization code \textsc{Cloudy} (version c22.01) \cite{ferland17}. The default \textsc{Cloudy} treatment simulates only 15 energy levels and computes the V1 multiplet as a composite feature and does not explicitly predict several of the additional \Oiirec\ transitions detected in our spectra. We therefore enable the full \Op\ atomic model (240 levels) with ``induced processes'' to capture fluorescent excitation. However, activating the full level-resolved simulation disables the recombination cascade calculations to the level populations. To circumvent this issue, we extract from \textsc{Cloudy} the spatial profiles of $T_e$, $n_e$ and $n(\Opp)$ across the model and recompute the \Oiirec\ recombination emissivities zone by zone using \textsc{PyNeb}. The resulting total V1 recombination flux agrees with the default \textsc{Cloudy} prediction (computed with a 15-level \Op\ atom and pure recombination) to within $<10\%$. We calculate the total line emissivity as the sum of the recombination and pumping contributions integrated in the simulated nebula,
\begin{equation}
\label{eq:emissivity_approx}
\varepsilon_\mathrm{tot} = \varepsilon_\mathrm{rec} + \varepsilon_\mathrm{pump},
\end{equation}
and use this decomposition to quantify the relative importance of continuum pumping for each transition.

For our fiducial model presented in Figs.~\ref{fig:o2rl_fluo} and \ref{fig:hh_oh_fpump}, the ionizing spectrum is synthesized for the Trapezium cluster using the TLUSTY OSTAR2002 stellar atmosphere models \cite{lanz03}, adopting the stellar parameters and luminosities derived by Ref.~\cite{simon-diaz06}. The synthesis includes $\theta^1$ Ori A, C, and D, while the $\theta^2$ Ori stars are omitted because of their larger projected distances from the observed nebular field. The ionizing continuum is overwhelmingly dominated by $\theta^1$ Ori C. We note, however, that this synthesized Trapezium spectrum is likely imperfect because it approximates the peculiar O7Vp star $\theta^1$ Ori C with a standard O7V TLUSTY model, despite the former's unusual stellar wind and magnetic field properties.

The ionization parameter is set to $\log U = -2.5$, and the initial hydrogen density to $n(\mathrm{H}) = 10^4~\mathrm{cm}^{-3}$, in agreement with the typical $n_e(\Sii)$ measurements. The gas-phase abundance pattern follows the default \textsc{Cloudy} Orion composition, while the $12 + \log(\mathrm{O}/\mathrm{H})$ ratio is adopted as 8.51 using the CEL-based measurement from Ref. \cite{esteban04}, and dust grains are excluded to isolate radiative effects from depletion and extinction. We adopt a microturbulent velocity of $10~\kms$ and assume spherical geometry with an inner radius of $10^{19}~\mathrm{cm}$. Simulations are iterated to convergence. The recombination ($\varepsilon_\mathrm{rec}$) and pumping ($\varepsilon_\mathrm{pump}$) emissivities are obtained by spherically integrating the zone-by-zone emissivity profiles (Fig.~\ref{fig:pumping_profile}). 
Among the 28 securely measured \Oiirec\ RLs, five transitions lack the required Einstein coefficients in the current \textsc{Cloudy} atomic dataset. These lines (4083.90, 4087.15, 4089.29, 4107.09, and 4609.44; Table~\ref{tab:oii_rl}) all originate from the \aterm{4f}{}{}{}{} levels and have previously been assumed to be immune to continuum pumping \cite{grandi76, escalante12}. However, our examination of the \Op\ atomic data reveals significant discrepancies between the level structure adopted by \textsc{Cloudy} and that reported by Ref.~\cite{nahar10}. In particular, additional pumping channels may exist through the higher \aterm{5d}{}{}{}{} and \aterm{6d}{}{}{}{} levels, which can be populated by continuum excitation from both the ground state, \aterm{2p^3}{4}{S}{3/2}{\circ}, and the metastable \aterm{2p^3}{2}{D}{}{\circ}\ levels. We therefore conservatively exclude these five transitions from the subsequent analysis rather than assume that they receive no pumping contribution.
For the remaining transitions, the median continuum pumping emissivity is 1.3 times the recombination emissivity, corresponding to an enhancement of 0.36~dex. The relative pumping strength, defined as $f_\mathrm{pump} \equiv \varepsilon_\mathrm{pump}/\varepsilon_\mathrm{rec}$, is listed for each transition in Table~\ref{tab:oii_rl}.

If continuum pumping contributes non-negligibly to the \Oiirec\ emissivities, the apparent abundance should scale linearly with $f_\mathrm{pump}$ as,
\begin{equation}
    \left(\frac{\Opp}{\Hp}\right)_\mathrm{ap} = (1 + f_\mathrm{pump}) \left( \frac{\Opp}{\Hp} \right)_\mathrm{cor},
\end{equation}
where $(\Opp/\Hp)_\mathrm{ap}$ and $(\Opp/\Hp)_\mathrm{cor}$ denote the apparent and intrinsic abundances, respectively. We indeed observe a strong positive correlation between $f_\mathrm{pump}$ and the measured $12 + \log(\Opp/\Hp)$ (Pearson $r = 0.59$, $p = 3.0 \times 10^{-3}$; Fig.~\ref{fig:o2rl_fluo}). We fit a linear relation between $(\Opp/\Hp)_\mathrm{ap}$ and $f_\mathrm{pump}$ with iterative $3\sigma$ clipping to exclude outliers. Fitting uncertainties are estimated via MC simulations in which individual $(\Opp/\Hp)$ values are perturbed according to their measurement errors. The best-fitting slope is $1.3_{-0.4}^{+1.1} \times (\Opp/\Hp)_\mathrm{cor}$, demonstrating that the predicted continuum pumping strengths are consistent with the observations. The resulting corrected abundance is $12 + \log(\Opp/\Hp) = 8.14_{-0.15}^{+0.11}$, in agreement within $1.5\sigma$ with the direct-$T_e$ CEL value of $8.311 \pm 0.065$. 
Historically, the V25 multiplet, arising from the \aterm{3p}{2}{D}{}{\circ} levels, has been regarded as particularly susceptible to continuum pumping because of its direct connection to the metastable \aterm{2p^3}{2}{D}{}{\circ} state. To assess its influence, we repeat the analysis after excluding the two V25 transitions. The inferred intrinsic abundance is essentially unchanged, yielding $12 + \log(\Opp/\Hp) = 8.10_{-0.23}^{+0.14}$ for the full FoV.

Using $f_\mathrm{pump}$ from our \textsc{Cloudy} model, we repeat the linear fitting separately for the low- and high-ionization stacked spectra (Fig.~\ref{fig:o2rl_fluo}; Table~\ref{tab:oii_rl}). In both regimes, the apparent $\Opp/\Hp$ abundance exhibits a strong positive correlation with $f_\mathrm{pump}$. After correction, the RL-based metallicities are consistent with the corresponding CEL-based abundances for each ionization zone within $1\sigma$.

We find that the predicted continuum pumping fraction depends sensitively on both the adopted ionizing spectrum and the ionization parameter. In addition to our fiducial Trapezium synthesis model, we consider three alternative ionizing spectra: (1) a blackbody with $T = 37{,}500$~K; (2) a Starburst99 \cite{leitherer99} stellar population synthesis model with $\log(Z_*) = -1.85$, assuming an instantaneous burst 1~Myr ago and a Salpeter initial mass function (IMF); and (3) a BPASS v2.2.1 \cite{stanway18} stellar population synthesis model with $\log(Z_*) = -1.85$, assuming an instantaneous burst 1~Myr ago and a Kroupa-like broken power-law IMF. Fig. \ref{fig:fpump_from_sed} compares the predicted $f_\mathrm{pump}$ values from these alternative ionizing spectra with those from the fiducial Trapezium model. Although the absolute pumping fractions differ, the predictions are tightly correlated and follow nearly linear relations. Consequently, the extrapolated intrinsic abundance, $12 + \log(\Opp/\Hp)$ at $f_\mathrm{pump}=0$, is nearly independent of the assumed ionizing spectrum. The absolute values of $f_\mathrm{pump}$, however, vary substantially even when all other model parameters are held fixed. For example, the BPASS model predicts pumping fractions only $\simeq 25\%$ of those in the fiducial model.

\subsection*{\textbf{Comparison to Previous Continuum Pumping Models}}

We also compare our predicted pumping fractions ($f_\mathrm{pump}$) with those from the PNe model of IC418 presented by Ref.~\cite{escalante12} (hereafter E12). As shown in Fig.~\ref{fig:model_comparison}, the two sets of predictions remain well correlated and follow an approximately linear relation. However, because the E12 model is specifically tailored to a planetary nebula rather than an \Hii\ region, its predicted $f_\mathrm{pump}$ values exhibit noticeably larger scatter about the best-fitting relation than the alternative \Hii-region models considered above. The corresponding corrected abundance is $12 + \log(\mathrm{O}^{++}/\mathrm{H}^+) = 8.14_{-0.20}^{+0.07}$, almost the same as our fiducial result.

In contrast to the analytical model of Ref.~\cite{grandi76} (hereafter G76), our simulations predict continuum pumping fractions that are several orders of magnitude larger. In the G76 model, continuum pumping is negligible for nearly all \Oiirec\ RLs, with the sole exception of the \Oiirec\ $\lambda 4153.30$ transition, for which $f_\mathrm{pump} \simeq 0.1$. Following the formalism of G76, the emissivity produced by continuum pumping can be written as,
\begin{equation}
j_{\Oiirec}^{\rm pump} = \sum_k J_\nu(\nu_k)
\left[ 
\frac{n_{s_k}}{g_{s_k}} \frac{\nu_{\Oiirec}}{\nu_k} \frac{\pi e^2}{m_e c} (gf)_k P_k \beta_{\Oiirec}
\right],
\label{eq:grandi_analytical}
\end{equation}
where the sum is over all pumping transitions $k$ that can populate an upper level which subsequently cascades into the corresponding \Oiirec\ transition. $J_\nu(\nu_k)$ is the mean intensity of the ionizing radiation at the frequency $\nu_k$ of the pumping transition. The factor $n_{s_k}/g_{s_k}$ is the population of the lower source level $s_k$, normalized by its statistical weight. $(gf)_k$ is the weighted oscillator strength of the pumping transition. $P_k$ indicates the cascade probability from the pumped upper level to the upper level of the given \Oiirec\ transition, including any intermediate branching ratios, and can be expressed as, 
\begin{equation}
P_k = \prod_i \frac{A_{i \rightarrow i-1}}{\sum_m A_{i \rightarrow m}},
\end{equation}
where $A$ is the Einstein coefficient. Finally, $\beta_{\Oiirec}$ is the branching ratio for emission through the given \Oiirec\ RL line once its upper level is populated, i.e., 
\begin{equation}
\beta_{\Oiirec} = \frac{A_{u\ell}}{\sum_m A_{um}}.
\end{equation}

For the \Oiirec\ $\lambda 4153$ transition, the dominant pumping channel proceeds directly from the ground \aterm{2p^3}{4}{S}{}{\circ} level to the upper \aterm{3d}{4}{P}{}{} level via absorption at a vacuum wavelength of 430.17~\AA. For this transition, the relevant atomic parameters are $\log(gf)_k = 0.254$, $P_k = 1$, and $\beta_{\Oiirec} = 3.01 \times 10^{-4}$. In G76, the pumping emissivity was parameterized in terms of an effective pumping coefficient, $(JB)^\mathrm{eff}$ (their Equation~7), which encapsulates the dependence on the incident radiation field $J_\nu(\nu_k)$. Adopting their predicted pumping fraction of $f_\mathrm{pump}(4153) = 0.1$, we infer $(JB)^\mathrm{eff} = 6.17 \times 10^{-10}~\mathrm{s}^{-1}$. This value is approximately three orders magnitude smaller than that obtained from Equation~\ref{eq:grandi_analytical} under our fiducial \textsc{Cloudy} model, which yields $(JB)^\mathrm{eff} = 2.25 \times 10^{-7}~\mathrm{s}^{-1}$. Because $(JB)^\mathrm{eff}$ scales directly with the intensity of the ionizing radiation, it is effectively normalized by the ionization parameter, $\log U$. The value adopted by G76 therefore corresponds to an ionization parameter of approximately $\log U \simeq -5.1$, substantially lower than the values inferred for M42 in the central region from modern photoionization studies \cite{odell17b}.

A second key difference between our photoionization model and the analytical treatment of G76 is that G76 assumes the pumping rate is proportional to the oscillator strength, $(gf)_k$, of the upward pumping transition. This approximation is valid only in the optically thin limit. Under the physical conditions of our \textsc{Cloudy} models, however, the dominant pumping transitions are strongly optically thick. For example, the 430.2~\AA\ transition, which is the primary pumping channel for the \Oiirec\ $\lambda 4153$ line, reaches an optical depth of $\tau(430.2~\mathrm{\AA}) \simeq 748$. Likewise, the dominant pumping channels for the \Oiirec\ $\lambda 4649$ V1 transition have optical depths of $\tau(418.6~\mathrm{\AA}) \simeq 29$ and $\tau(485.1~\mathrm{\AA}) \simeq 11$. Consequently, the pumping rate no longer scales linearly with $(gf)_k$, as line saturation substantially suppresses the increase in pumping efficiency. This effect is particularly evident for the \Oiirec\ $\lambda 4153$ transition. Although the inferred $(JB)^\mathrm{eff}$ under our fiducial \textsc{Cloudy} conditions is three orders of magnitude larger than the value adopted by G76, the corresponding pumping fraction increases by only a factor of $\sim$15 in the full photoionization simulation. Similarly, the predicted ratio $f_\mathrm{pump}(4153)/f_\mathrm{pump}(4649)$ is only $\simeq 2.4$ in the \textsc{Cloudy} models, compared to $\simeq 110$ from the G76-style analytical calculation. This discrepancy arises because the 430.2~\AA\ pumping transition saturates much more rapidly than the 418.6~\AA\ and 485.1~\AA\ transitions that dominate the excitation of the V1 multiplet.

In summary, our photoionization calculations are fundamentally consistent with the frameworks of both E12 and G76. The apparent discrepancies arise primarily from differences in the adopted physical conditions and the treatment of radiative transfer, rather than from the underlying pumping mechanism itself. When realistic conditions appropriate for M42 are adopted, our models predict continuum pumping fractions that closely reproduce the observed \Oiirec\ RLs, providing strong empirical support for the inferred pumping strengths.

\subsection*{\textbf{Continuum Pumping in the Line-of-Sight Structures}}

The Orion Nebula has been extensively observed with high-resolution ($R > 20{,}000$) slit spectroscopy, revealing multiple kinematic components along individual sightlines \cite{garcia-diaz08}. A major contributor to this complexity is the presence of Herbig-Haro (HH) objects, small-scale emission nebulae produced when collimated outflows from young stellar objects interact with the surrounding interstellar medium. The kinematically resolved nebular emission associated with HH objects near our FoV has been studied in detail \cite{blagrave06, mendez-delgado21a, mendez-delgado21b}. Their emission-line spectra are generally consistent with photoionization, and the $T_e$(\Oiii) values derived from CELs differ from those of the surrounding nebula by only a few hundred kelvin, comparable to the spatial variations in $T_e$(\Oiii) measured across our FoV. Our field contains one known HH object, HH~529 I near AC Ori \cite{mendez-delgado21a}, as well as several filamentary structures identified in previous imaging studies \cite{odell03}.

It is therefore essential to establish whether continuum pumping also dominates the ADF within these kinematically distinct structures along the line of sight. In Fig.~\ref{fig:hh_oh_fpump}, we apply the same analysis to the literature measurements of \Oiirec\ RLs in kinematically resolved HH objects and their surrounding nebular components. In every case, the apparent \Opp/\Hp\ abundance exhibits a clear positive correlation with the predicted $f_\mathrm{pump}$ from our fiducial M42 photoionization model, providing strong evidence that continuum pumping also operates within these line-of-sight structures. Consequently, the continuum-pumping-corrected RL abundances are systematically lower than those derived from the V1 multiplet alone. Figure~\ref{fig:hh_adf} summarizes the resulting changes in the ADF for the kinematically resolved components before and after applying the correction. The literature \Oiirec\ RL fluxes, predicted $f_\mathrm{pump}$ values, and the corrected \Opp/\Hp\ abundances are listed in Table~\ref{tab:hh_oii_rl}--\ref{tab:hh_oii_rl_continued}.

Accounting for continuum pumping resolves the bulk of the ADF in both the kinematically resolved HH objects and their surrounding nebular components in M42. This result is expected because the HH objects are predominantly photoionized by the same radiation field as the ambient nebula, and their CEL-based $T_e$(\Oiii) values differ only marginally from those of the surrounding gas. The similar continuum pumping fractions and electron temperatures among the individual velocity components also explain why our KCWI observations, despite not resolving the kinematic substructure, remain amenable to the same pumping correction. The unresolved spectra effectively combine components that are affected in a similar manner, allowing the correction derived from the integrated emission to remain valid.

\subsection*{\textbf{Recalculating the Literature ADF}}

To assess whether the non-negligible continuum-pumping contribution to \Oiirec\ RLs is universal, we compile a sample of five \Hii\ regions from the literature in which multiple \Oiirec\ transitions from different series have been measured with high-resolution spectroscopy. The sample comprises a long-slit pointing of M42 \cite{esteban04}, Galactic \Hii\ regions: NGC~3576 \cite{garcia-rojas04}, M8 and M17 \cite{garcia-rojas07}, and Large Magellanic Cloud \Hii\ region, 30~Dor \cite{peimbert03}. For each object, we adopt the published dust-corrected \Oiirec\ fluxes and compute the apparent $\Opp/\Hp$ abundances for individual transitions using the reported physical conditions. We then compare these abundances with the corresponding $f_\mathrm{pump}$ values predicted by tailored photoionization models.

To compute $f_\mathrm{pump}$ for each literature object, we tailor the photoionization models to approximate the reported physical conditions. Unlike M42, however, the ionizing sources of these objects are generally not characterized in comparable detail. Moreover, because variations in the ionizing spectrum primarily rescale the absolute values of $f_\mathrm{pump}$ while preserving the relative line-to-line pumping strengths, this uncertainty is largely absorbed into the fitted slope of the apparent \Opp/\Hp\ versus $f_\mathrm{pump}$ relation. We therefore adopt BPASS stellar population synthesis models as a uniform choice for the ionizing spectrum. The stellar population is assumed to have an age of 1~Myr, and we retain the same ionization parameter ($\log U = -2.5$), spherical geometry, and inner radius of $10^{19}~\mathrm{cm}$ as in our fiducial M42 model.
The gas-phase O/H metallicity is fixed to the direct-$T_e$ CEL metallicity reported in the corresponding reference. The stellar metallicity is scaled with the gas-phase oxygen abundance, assuming $\log(Z_*/O) = 1.64$. The hydrogen density, $\log(n_H)$, is set according to the $n_e(\Oii)$ values reported in each study. For the calculation of the apparent RL $\Opp/\Hp$, we adopt the literature $T_e(\Oiii)$ values. A summary of the adopted model parameters, together with a comparison to the fiducial M42 simulation, is provided in Table~\ref{tab:model_assumptions_new_adf}.

Tables~\ref{tab:new_lit_m42}--\ref{tab:new_lit_30dor} list the dust-corrected \Oiirec\ fluxes, the corresponding $f_\mathrm{pump}$ values, and the apparent $12 + \log(\Opp/\Hp)$ abundances, following the format of Table~\ref{tab:oii_rl}. In all five \Hii\ regions, we find a positive slope between the apparent abundance and $f_\mathrm{pump}$ (Fig.~\ref{fig:recalculate_met}), indicating that continuum pumping is a generic contributor to the \Oiirec\ emissivity. The strength and statistical significance of the correlation, however, vary among objects. Except for M42, the inferred slopes are generally less significant. This could reflects the limited number of securely measured \Oiirec\ transitions in the literature datasets. Particularly, transitions with $f_\mathrm{pump} \simeq 0$ are often intrinsically faint and therefore more susceptible to blending or contamination by unrelated weak features, resulting in bias in the inferred relation toward smaller pumping corrections and thus lead to an underestimation of the true continuum contribution.

Despite these limitations, comparison of the traditional V1-based \Oiirec\ abundances with the continuum-pumping–corrected values (Fig.~\ref{fig:new_adf}) shows that the mean ADF decreases by $\simeq 0.13$~dex--corresponding to $>50\%$ of the original discrepancy. This systematic reduction indicates that continuum pumping is a non-negligible contributor to the \Oiirec\ emissivity in all examined \Hii\ regions and may be stronger than previously assumed. Further progress will require refined pumping models and deeper spectroscopic measurements targeting the faintest \Oiirec\ transitions to better constrain the high-$f_\mathrm{pump}$ regime.

\section*{Data availability}
The raw Keck Cosmic Web Imager data in this work are available in the Keck Observatory Archive: \url{https://www2.keck.hawaii.edu/koa/ public/koa.php}. All data analysis softwares are publicly available and have been referenced along the text.

\section*{Acknowledgements}
The authors would like to thank Robert Minchin, Karin Sandstrom, and Justin Spilker for their contribution to the planning of the project. Y.C. appreciates the helpful discussions with Xihan Ji and Tsang Keung Chan. We would like to thank Peter Weilbacher for timely providing the MUSE data, enabling some initial assessment related to this work. This work is based on data obtained at the W. M. Keck Observatory, which is operated as a scientific partnership among the California Institute of Technology, the University of California, and the National Aeronautics and Space Administration. The Observatory was made possible by the generous financial support of the W. M. Keck Foundation. The authors wish to recognize and acknowledge the very significant cultural role and reverence that the summit of Maunakea has always had within the indigenous Hawaiian community.  We are most fortunate to have the opportunity to conduct observations from this mountain.

\section*{Funding}
Y.C. and T.J. gratefully acknowledge support for this work from the National Aeronautics and Space Administration (NASA) under grant 80NSSC23K1132. Y.C. is supported by the Direct Grant for Research (C0010-4053720, C001-4053782) from the Faculty of Science, The Chinese University of Hong Kong. This work is supported by the Hong Kong Jockey Club Charities Trust through the project, JC STEM Lab of Astronomical Instrumentation.

\section*{Author contributions}
Y.C. led the design, observation, data reduction, analysis, and interpretation of the project. T.J. and R.S. conceived the project. P.K., Z.C., and F.X. contributed to the analysis of the dataset. S.R. and K.V.G.C. conducted the observation. F.B., R.H.-C.,  E.H., X.-W.L., P.S., C.S., and R.Y. contributed to the preparation of manuscript. All authors contributed to the planning of the observations, the overall interpretation of the results, and various aspects of the manuscript.

\section*{Competing interests}
There are no competing interests to declare.

\clearpage
\section*{Supplementary Information}
\renewcommand{\thefigure}{S\arabic{figure}}
\renewcommand{\thetable}{S\arabic{table}}
\renewcommand{\theequation}{S\arabic{equation}}
\renewcommand{\theHfigure}{S\arabic{figure}}
\renewcommand{\theHtable}{S\arabic{table}}
\setcounter{figure}{0}
\setcounter{table}{0}
\setcounter{equation}{0}
\begin{figure}[htbp!]
\centering
\includegraphics[width=0.8\textwidth]{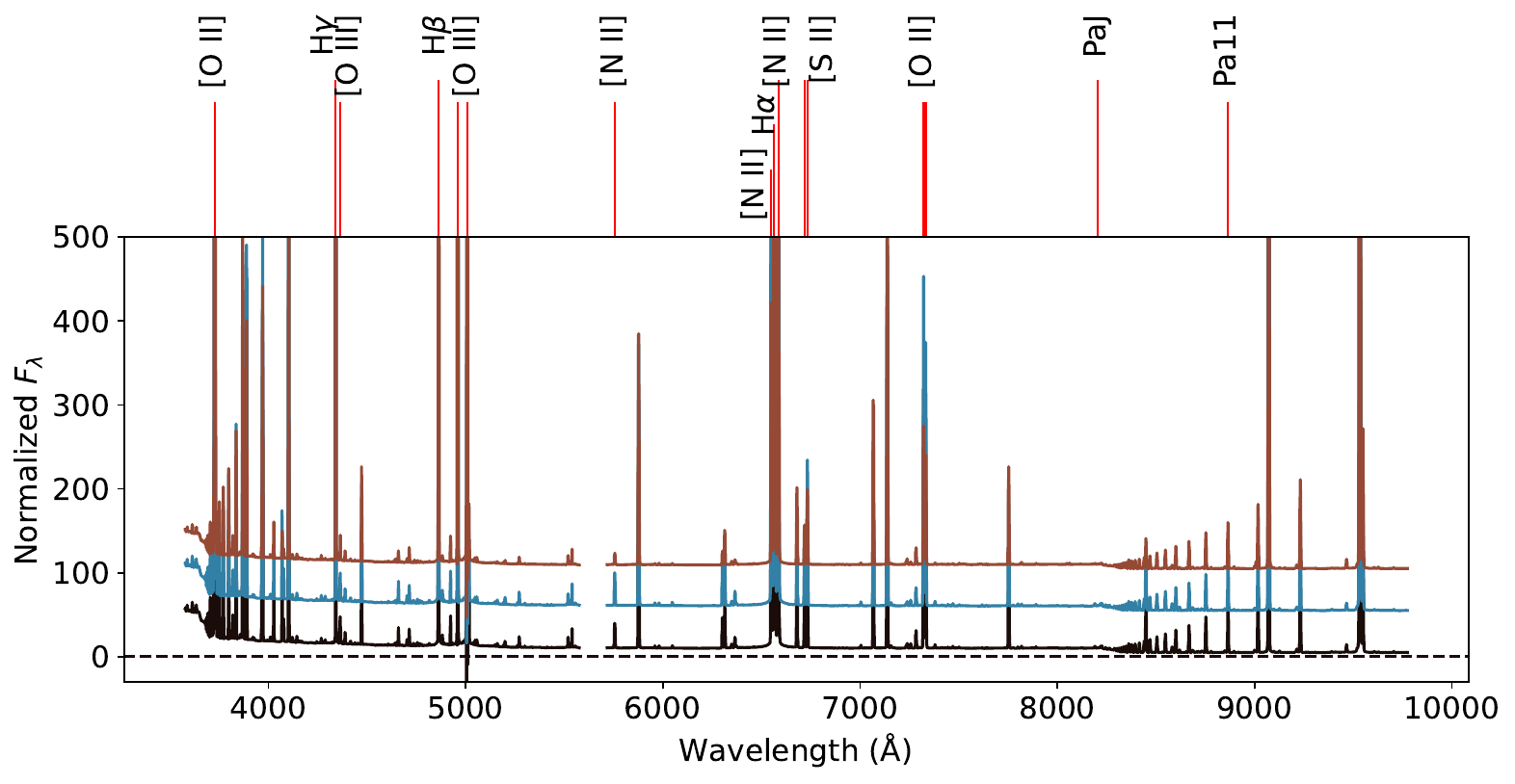}\\
\includegraphics[width=0.8\textwidth]{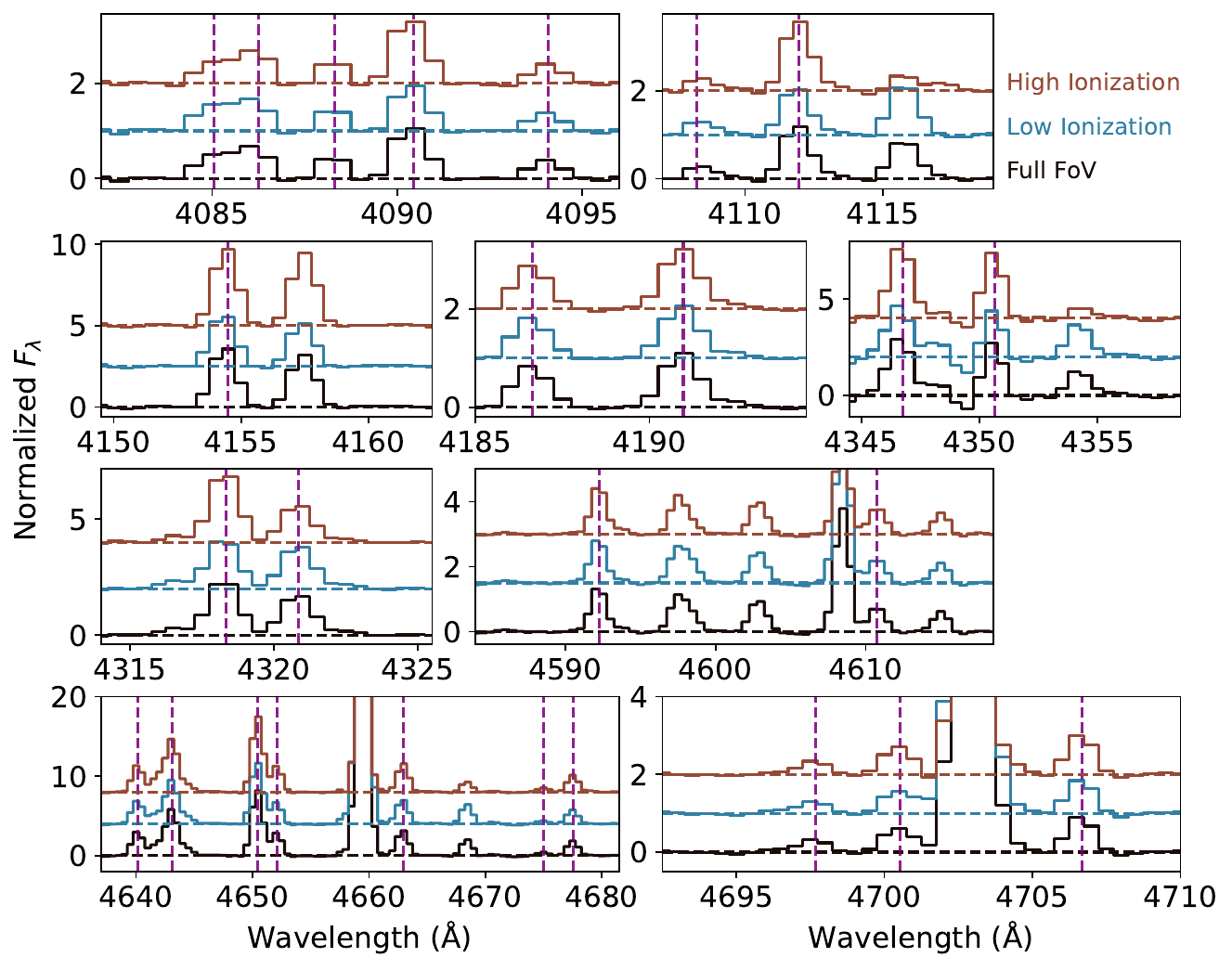}
\caption{  \textbf{Spectra extracted from the KCWI observations.} \textbf{Top:} Stacked spectra from the high-ionization region (red), low-ionization region (blue) and the full FoV (black). Prominent emission features are highlighted in red. Complete line flux measurements are provided in Tables~\ref{tab:strong_line_fluxes} and \ref{tab:oii_rl}. \textbf{Bottom panels:} Enlarged views of the \Oiirec\ recombination lines. Vertical purple dashed lines mark transitions with reliably measured fluxes. For clarity, the high- and low-ionization spectra (red and blue) are vertically offset in all panels. Note that the spectra are calibrated to vacuum wavelengths. } 
\label{fig:full_spec}
\end{figure} 

\begin{figure}[htbp]
\centering
\includegraphics[width=0.9\textwidth]{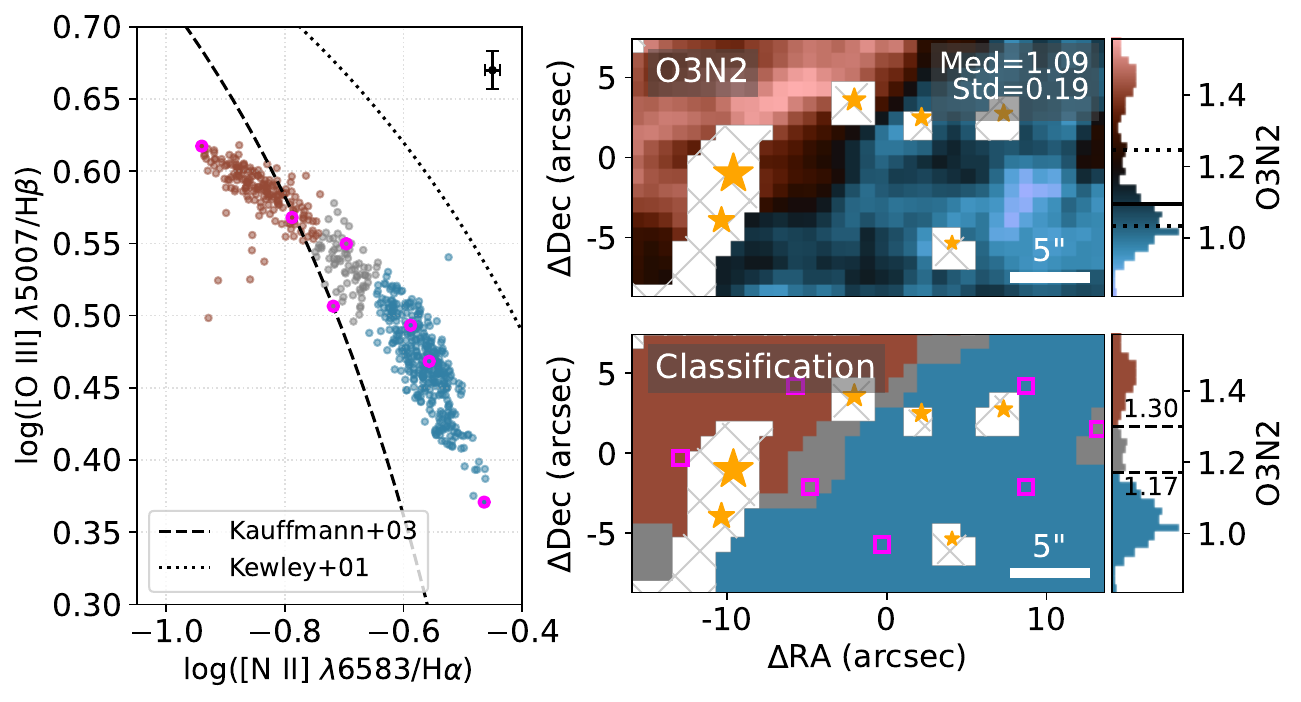}
\caption{  \textbf{Classification of high- and low-ionization regions within the KCWI FoV.} \textbf{Left:} Distribution of spaxels on the O3–N2 BPT diagram. The dashed and dotted curves denote the Kauffmann and Kewley demarcation lines \cite{kauffmann03, kewley01}, respectively, separating \Hii\ regions from active galactic nuclei. \textbf{Right:} Spatial map of $\mathrm{O3N2} \equiv \log(\Oiii~\lambda 5007/\mathrm{H}\beta - \Nii~\lambda 6583/\mathrm{H}\alpha)$. The O3N2 values exhibit a clear bimodal distribution driven by distinct ionization conditions (top), motivating our classification into high- and low-ionization regions for spectral stacking (bottom). Magenta circles and squares mark representative spaxels selected for subsequent Monte-Carlo uncertainty analysis. } 
\label{fig:o3n2}
\end{figure} 

\begin{figure}[htbp]
\centering
\includegraphics[width=0.8\textwidth]{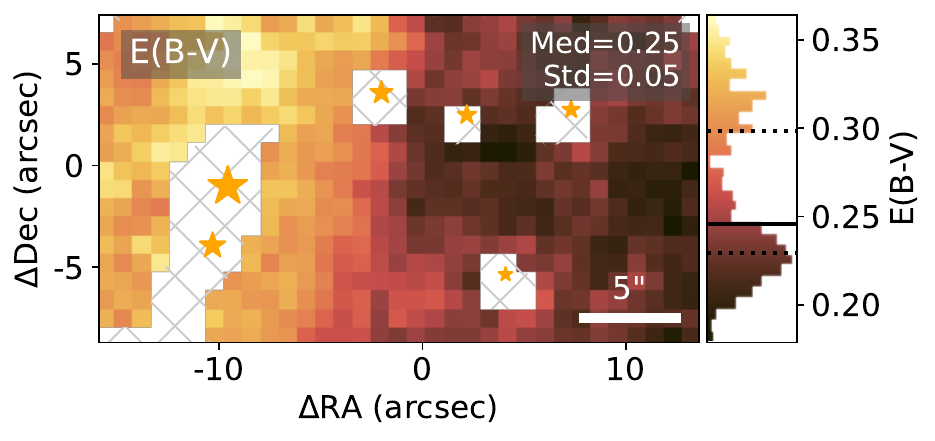}\\
\caption{   \textbf{Spatial distribution of \ebv.} The \ebv\ is derived as the average of the values inferred from the H$\beta$/H$\alpha$ and H$\gamma$/H$\beta$ Balmer decrements.}
\label{fig:ebv_map}
\end{figure}

\begin{figure}[htbp]
\centering
\includegraphics[width=0.6\textwidth]{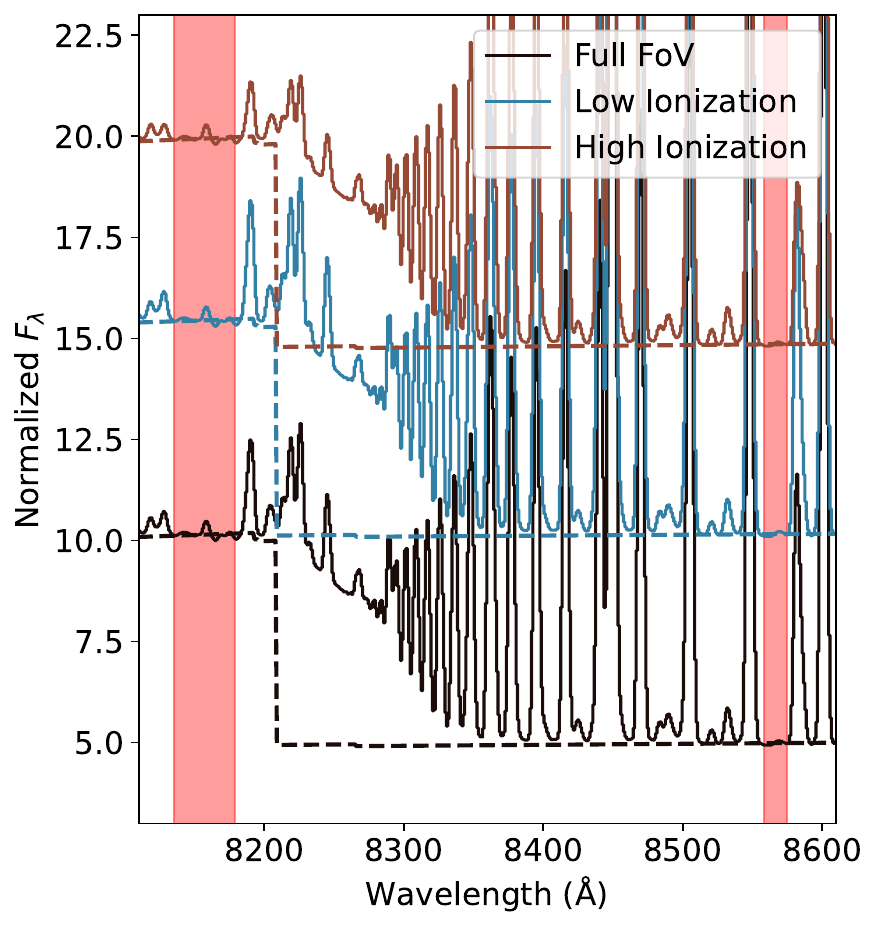}\\
\caption{   \textbf{Examples of Paschen Jump temperature measurements from the stacked spectra.} Solid curves show the observed spectra, with colours corresponding to the different ionization-level bins. The dashed curves represent the best-fitting Paschen Jump models. The red shaded regions, [8135, 8179]~\AA\ and [8558, 8575]~\AA, indicate the wavelength intervals used to compute the mean continuum flux densities for the Paschen Jump determination.}
\label{fig:PJ}
\end{figure}

\begin{figure}[htbp!]
\centering
\includegraphics[width=0.99\textwidth]{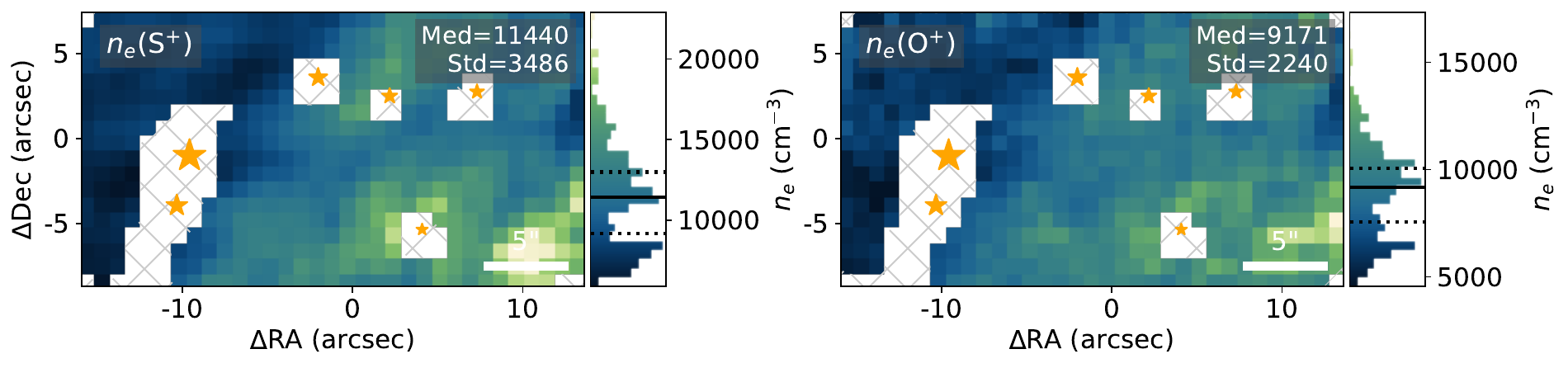}\\
\caption{   \textbf{Spatial distributions of $n_e$ derived from [S \textsc{II}] and [O \textsc{II}].} The $n_e(\Oii)$ values are shown for reference only; $n_e(\Sii)$ is adopted in all subsequent analyses.}
\label{fig:ne_maps}
\end{figure}

\begin{figure}[htbp!]
\centering
\includegraphics[width=0.99\textwidth]{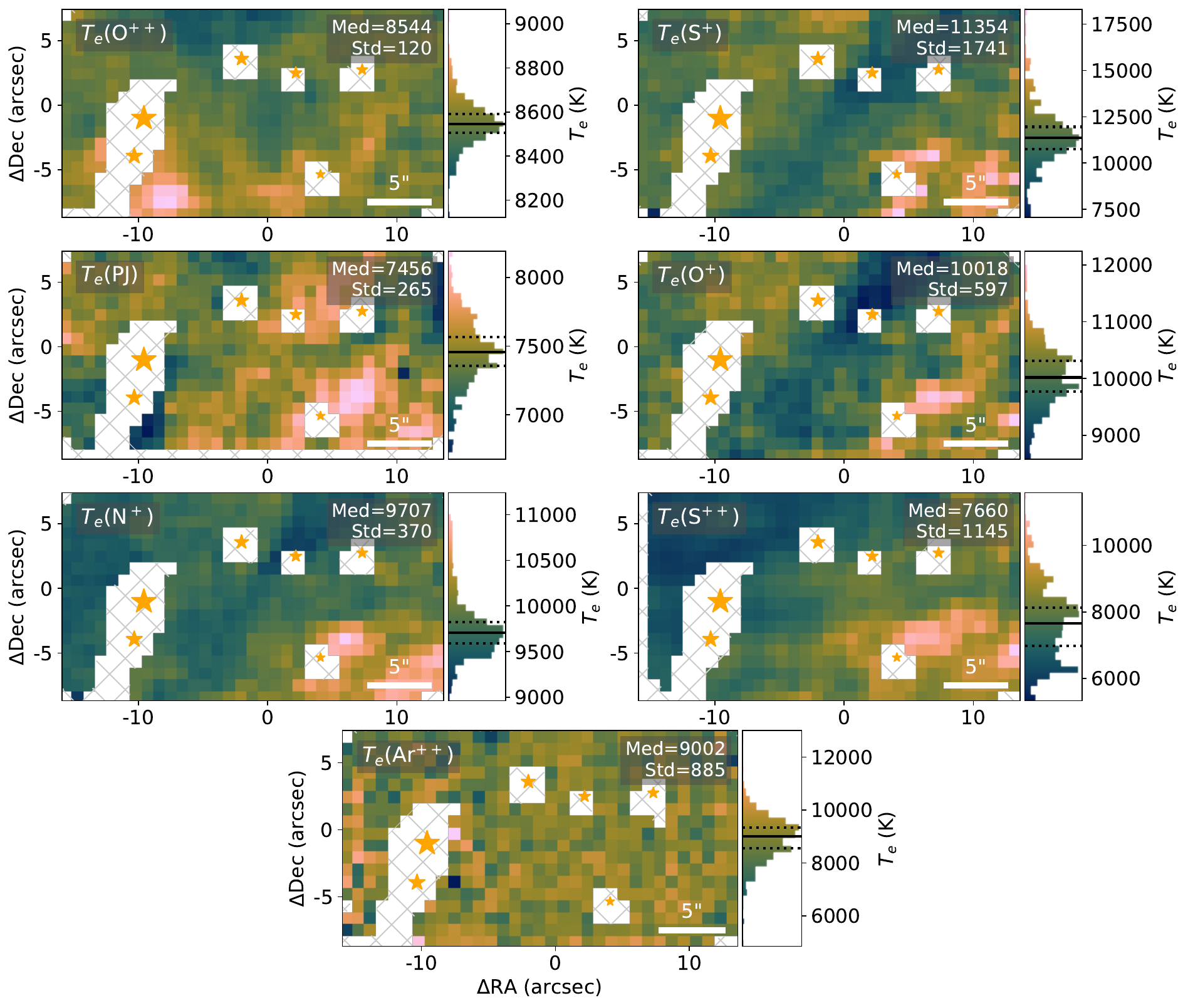}\\
\caption{    \textbf{Maps of electron temperature ($T_e$) derived from multiple diagnostics.} We derive $T_e$ from \Oiii, \Sii, the Paschen Jump (PJ), \Oii, \Nii, \Siii\ and \Ariii. Except for PJ, temperatures are calculated from the auroral-to-nebular ratios listed in Table~\ref{tab:phys_method}, adopting $n_e(\Sii)$ as the electron density.}
\label{fig:te_maps}
\end{figure}

\begin{figure}[htbp]
\centering
\includegraphics[width=0.99\textwidth]{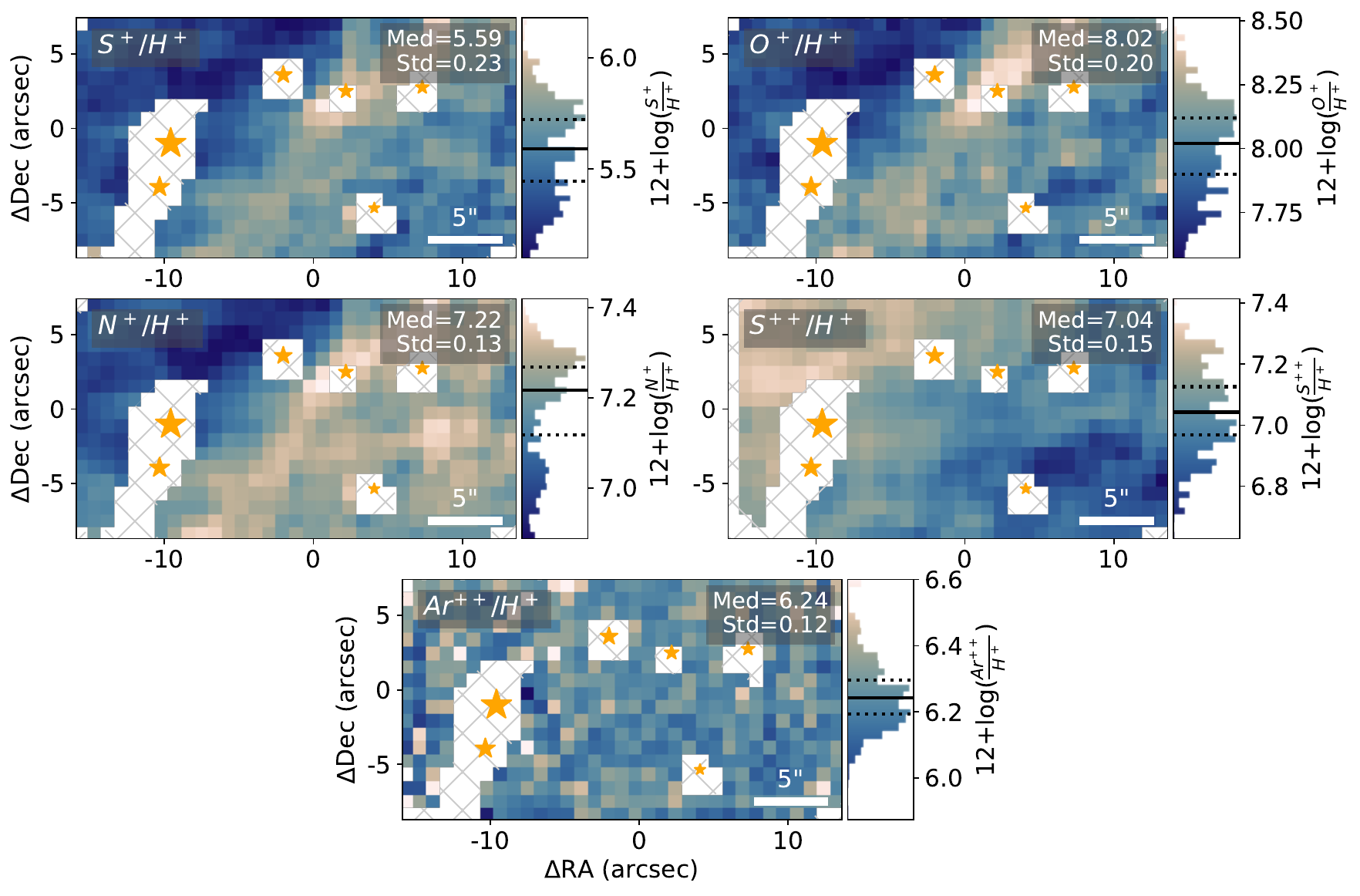}\\
\caption{  \textbf{Spatial distributions of ionic abundances for \Sp, \Op, \Np, \Spp\ and \Arpp.} All abundances are derived using the direct-$T_e$ method with temperatures appropriate to each ion (Table~\ref{tab:phys_method}).}
\label{fig:ion_maps}
\end{figure}

\begin{figure}[htbp]
\centering
\includegraphics[width=0.99\textwidth]{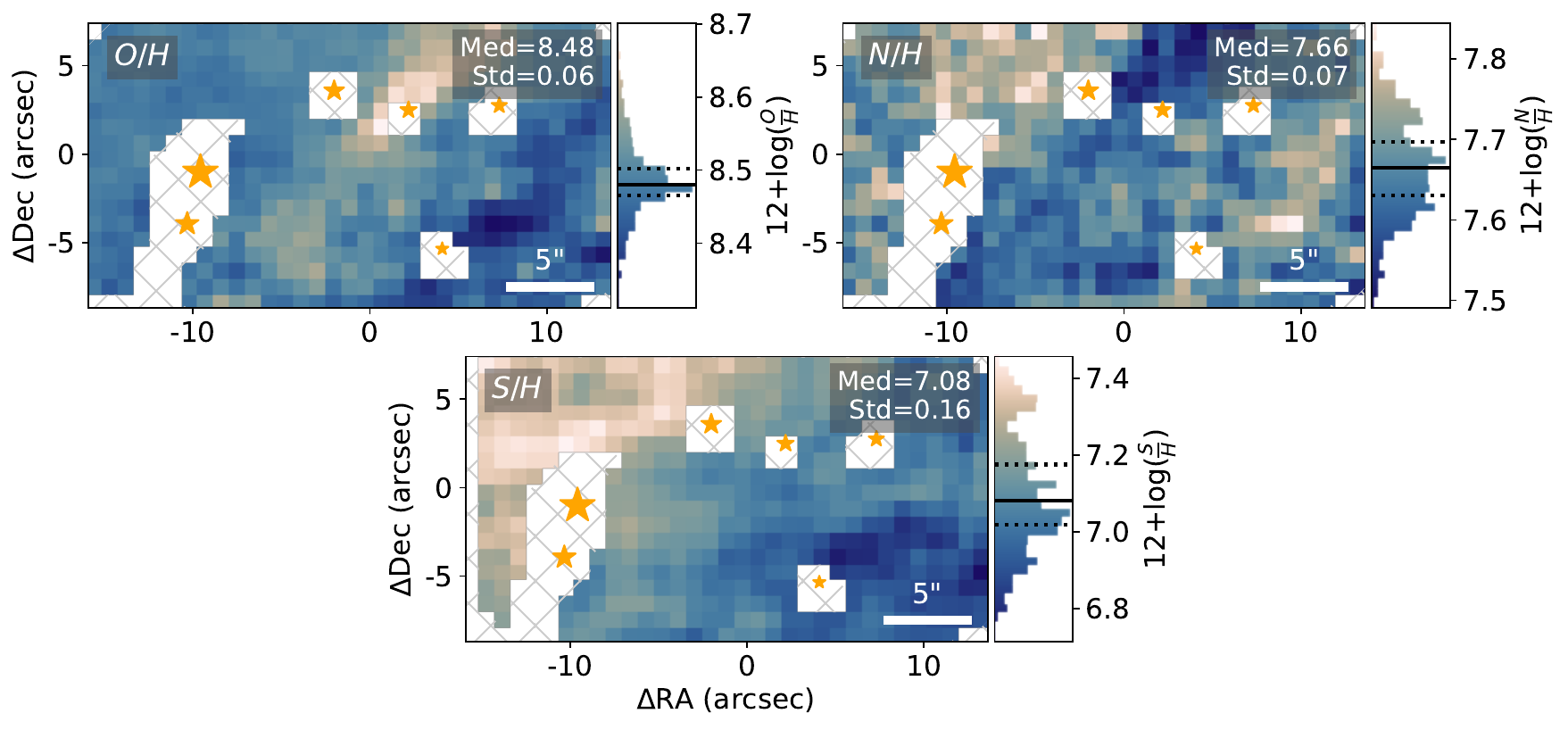}\\
\caption{     \textbf{Maps of total elemental abundances for O, N and S.} Methodological details are provided in Table~\ref{tab:phys_method}.}
\label{fig:totmet_maps}
\end{figure}

\begin{figure}[htbp]
\centering
\includegraphics[width=0.7\textwidth]{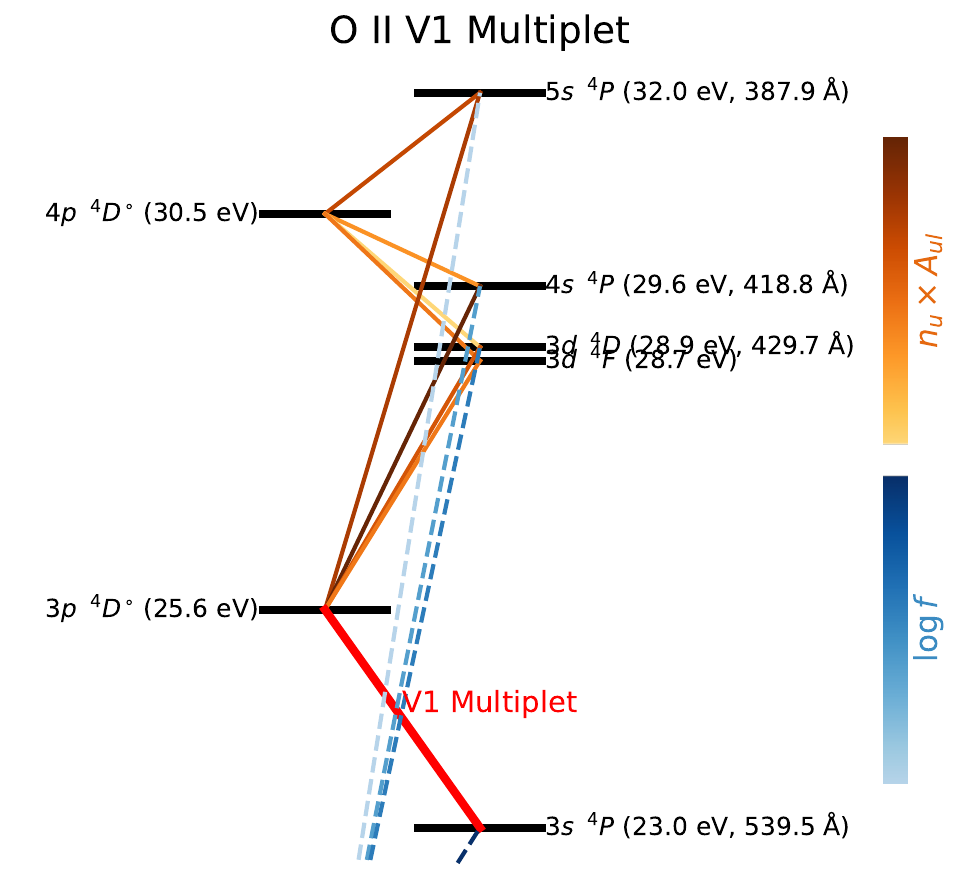}\\
\caption{   \textbf{Grotrian diagram of the O \textsc{II} V1 multiplet used to derive $\Opp/\Hp$ abundances.} The V1 transitions are highlighted in red. Upstream transitions that can influence the V1 population are shown as orange solid lines, with color intensity scaled by $n_u \times A_{ul}$, representing the approximate transition probability. Blue dashed lines connect the relevant levels to the ground state, indicating resonant transitions; their color depth reflects the oscillator strength $\log f$. The $n_u \times A_{ul}$ values are indicative only, as they are computed from a representative layer in a \textsc{Cloudy} simulation.}
\label{fig:v1_grotrian}
\end{figure}

\begin{figure}[htbp]
\centering
\includegraphics[width=0.99\textwidth]{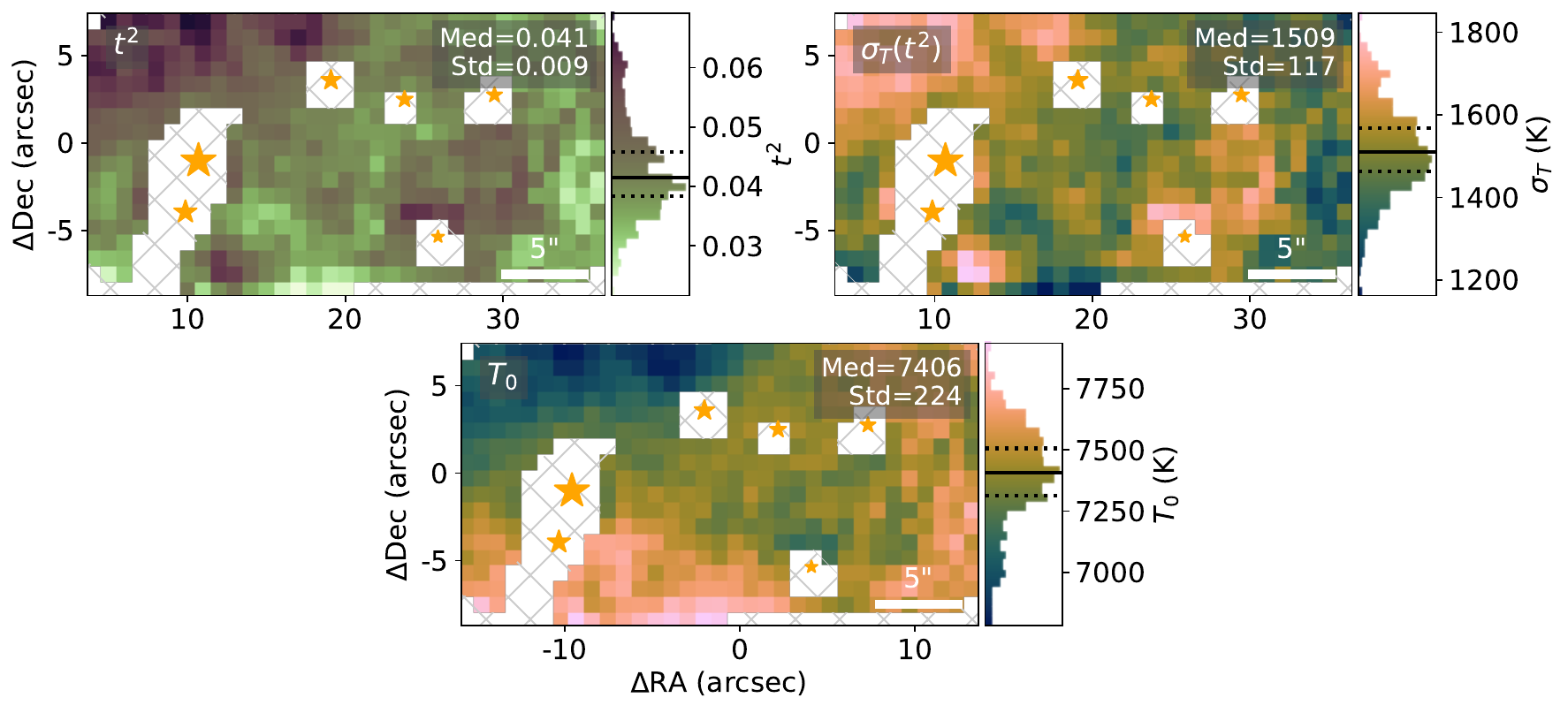}
\caption{\textbf{Maps of the temperature fluctuation parameters in M42.} Spatial distributions of $t^2$, the corresponding temperature dispersion $\sigma_T(t^2)$, and the mean temperature $T_0$, derived following the temperature fluctuation formalism.}
\label{fig:t2_maps}
\end{figure} 

\begin{figure}[htbp!]
\centering
\includegraphics[width=0.8\textwidth]{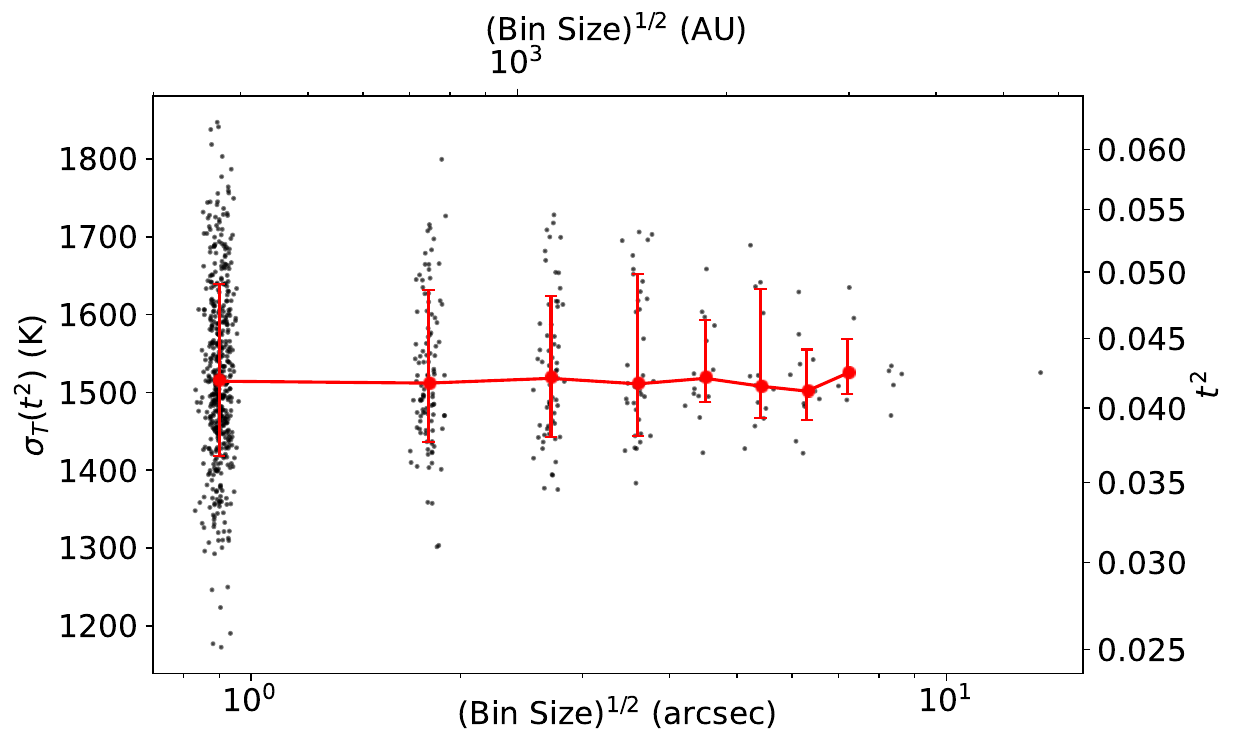}
\caption{   \textbf{Required temperature fluctuations from the ADF, $\sigma_T(t^2)$, as a function of spatial bin size.} Black points represent individual binned spectra, with a small random horizontal offset applied for clarity. Red circles indicate the median $\sigma_T(t^2)$ at each bin size, and error bars denote the standard deviation. The spatial binning does not have impact on the required temperature fluctuations to explain the ADF, consistent with the expectation that the measured variations of $T_e(\Opp)$ is $>1$ dex lower than what is required to explain the ADF.}
\label{fig:t2_cog}
\end{figure} 

\begin{figure}[htbp!]
\centering
\includegraphics[width=0.8\textwidth]{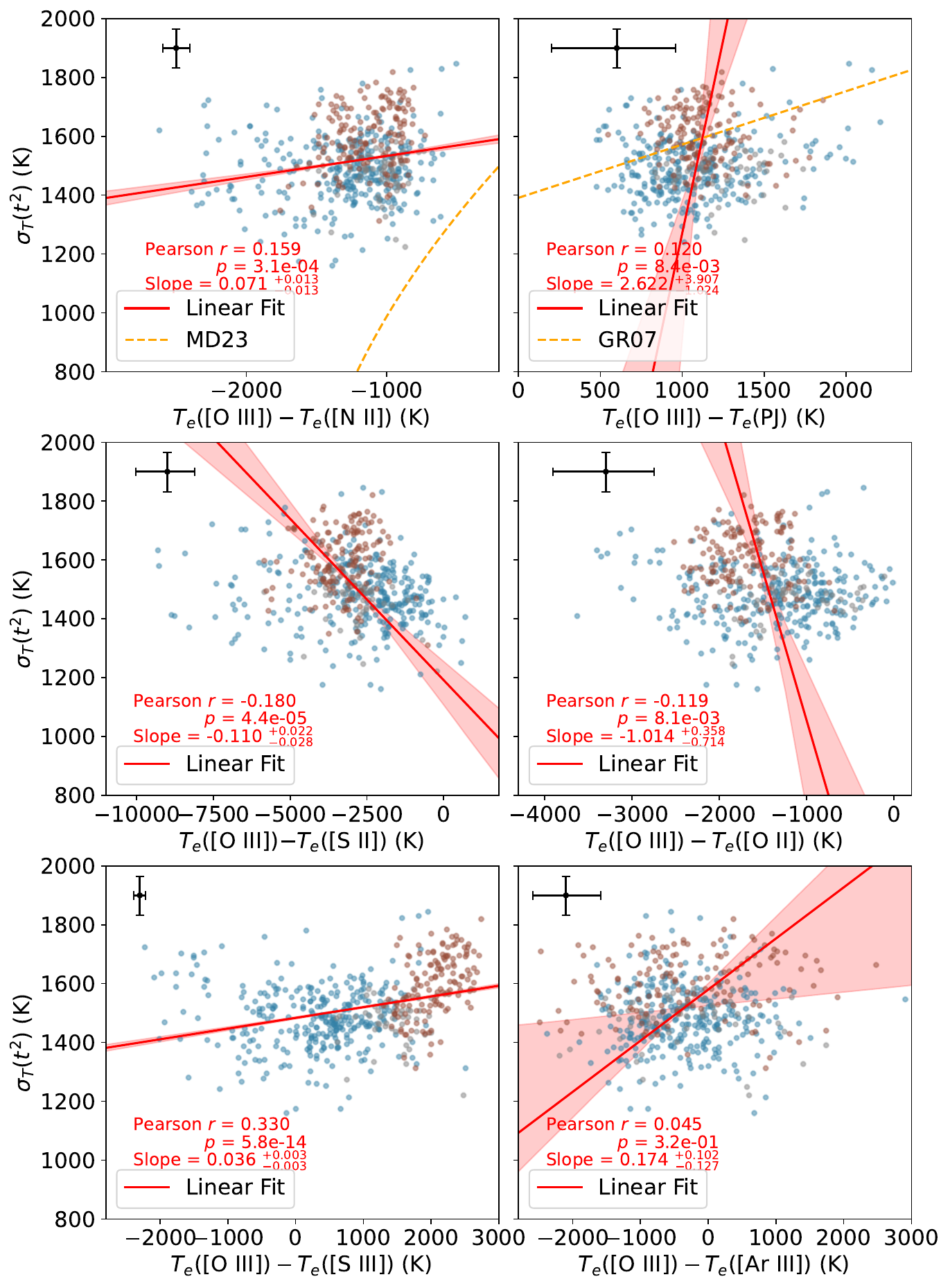}\\
\caption{  \textbf{Relation between $\sigma_T(t^2)$ and the line-of-sight inter-ionic temperature differences ($\Delta T_e = T_e(\textrm{[O \textsc{III}]}) - T_e(X)$) for individual spaxels.} Point colors encode ionization state (Fig.~\ref{fig:o3n2}). Empirical correlations established between different \Hii\ regions \cite{mendez-delgado23, garcia-rojas07b} are shown as the dashed-yellow lines.  Inter-ionic temperature difference do not drive the spatially resolved ADF. }
\label{fig:tsig_to_dt}
\end{figure}

\begin{figure}[htbp!]
\centering
\includegraphics[width=0.8\textwidth]{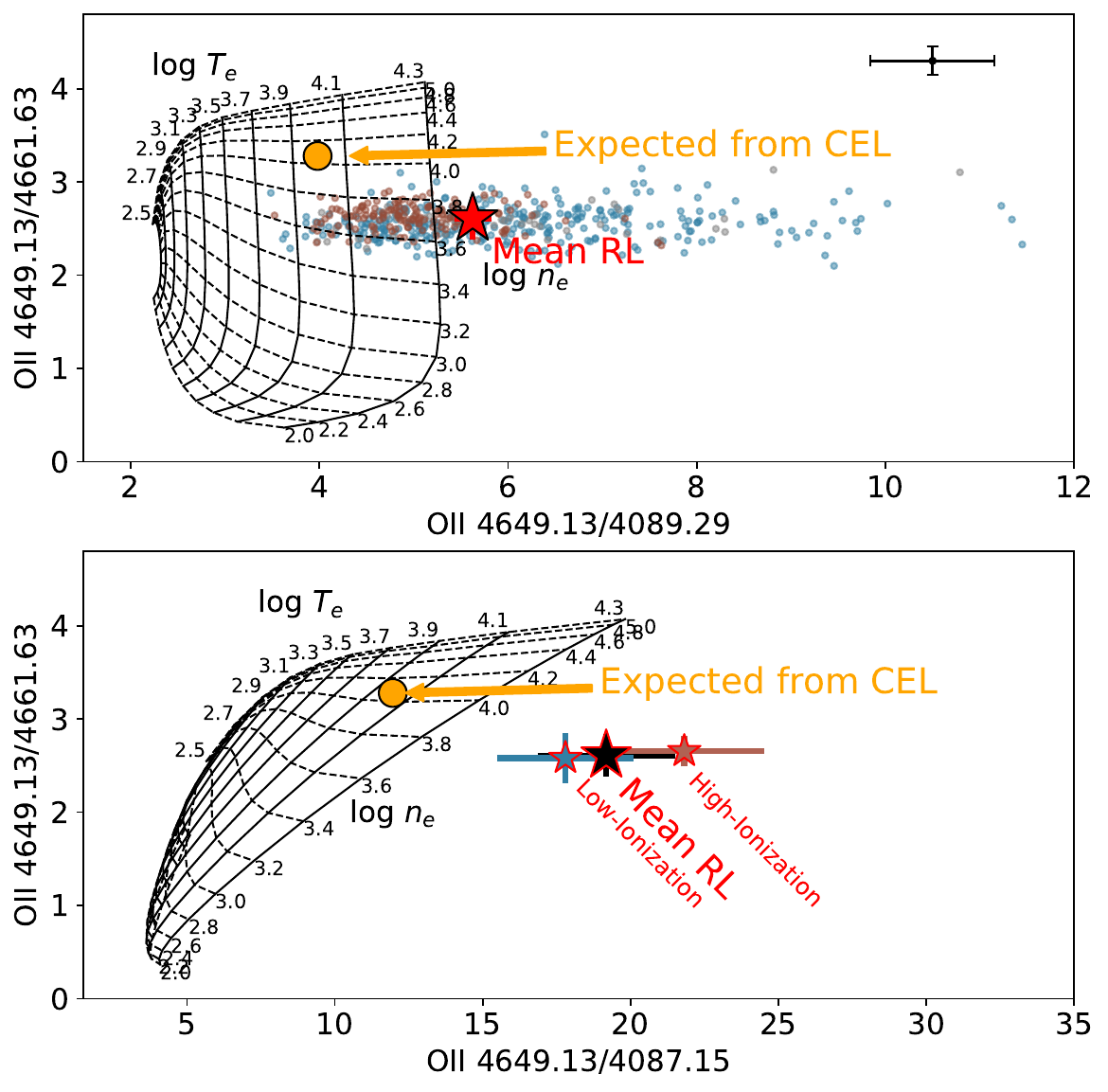}
\caption{  \textbf{Electron temperature and density diagnostics from pure \Oiirec\ recombination-line ratios.} \textbf{Top:} Distribution of the $n_e$-sensitive \Oiirec\ $\lambda\lambda 4649/4089$ ratio vs. the $T_e$-sensitive \Oiirec\ $\lambda\lambda 4649/4662$ ratio. Each colored point corresponds to an individual spaxel, with colors denoting ionization states (see Fig.~\ref{fig:o3n2}). Contamination from scattered \ion{Si}{IV} $\lambda4088.86$ absorption has been corrected. The black grid shows theoretical predictions computed for the corresponding $\log(n_e)$ and $\log(T_e)$ using updated atomic data \cite{storey17}. The red star marks the mean RL ratios, and the orange circle indicates the expectation based on CEL-derived $T_e$(\Oiii) and $n_e$(\Sii). 
\textbf{Bottom:} The $n_e$-sensitive \Oiirec\ $\lambda\lambda 4649/4662$ ratio is plotted against the \Oiirec\ $\lambda\lambda 4649/4087$ ratio, which is sensitive to both $n_e$ and $T_e$, for the stacked spectra. In both diagnostics, the density-sensitive ratio is similar to the CEL-based expectation, whereas the temperature-sensitive ratios, $\lambda\lambda 4649/4089$ and $\lambda\lambda 4649/4087$, are significantly enhanced.
}
\label{fig:o2rl_tene}
\end{figure} 

\begin{figure}[htbp!]
\centering
\includegraphics[width=0.9\textwidth]{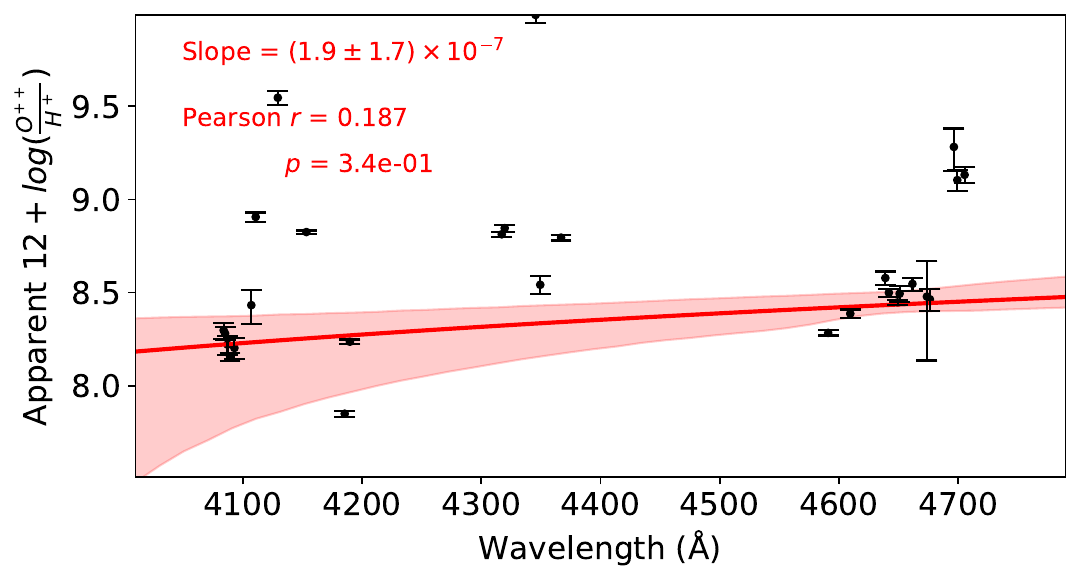}
\caption{ \textbf{Apparent $\Opp/\Hp$ abundances inferred from individual \Oiirec\ RLs as a function of wavelength in the stacked spectrum.} As indicated by the red best-fit linear relation and its shaded error, the absence of a systematic wavelength-dependent trend, and the large scatter, disfavor additional dust obscuration as the origin of the elevated $\lambda\lambda 4649/4089$ and $\lambda\lambda 4649/4087$ ratios.  } 
\label{fig:o2rl_wave}
\end{figure} 

\begin{figure}[htbp]
\centering
\includegraphics[width=0.85\textwidth]{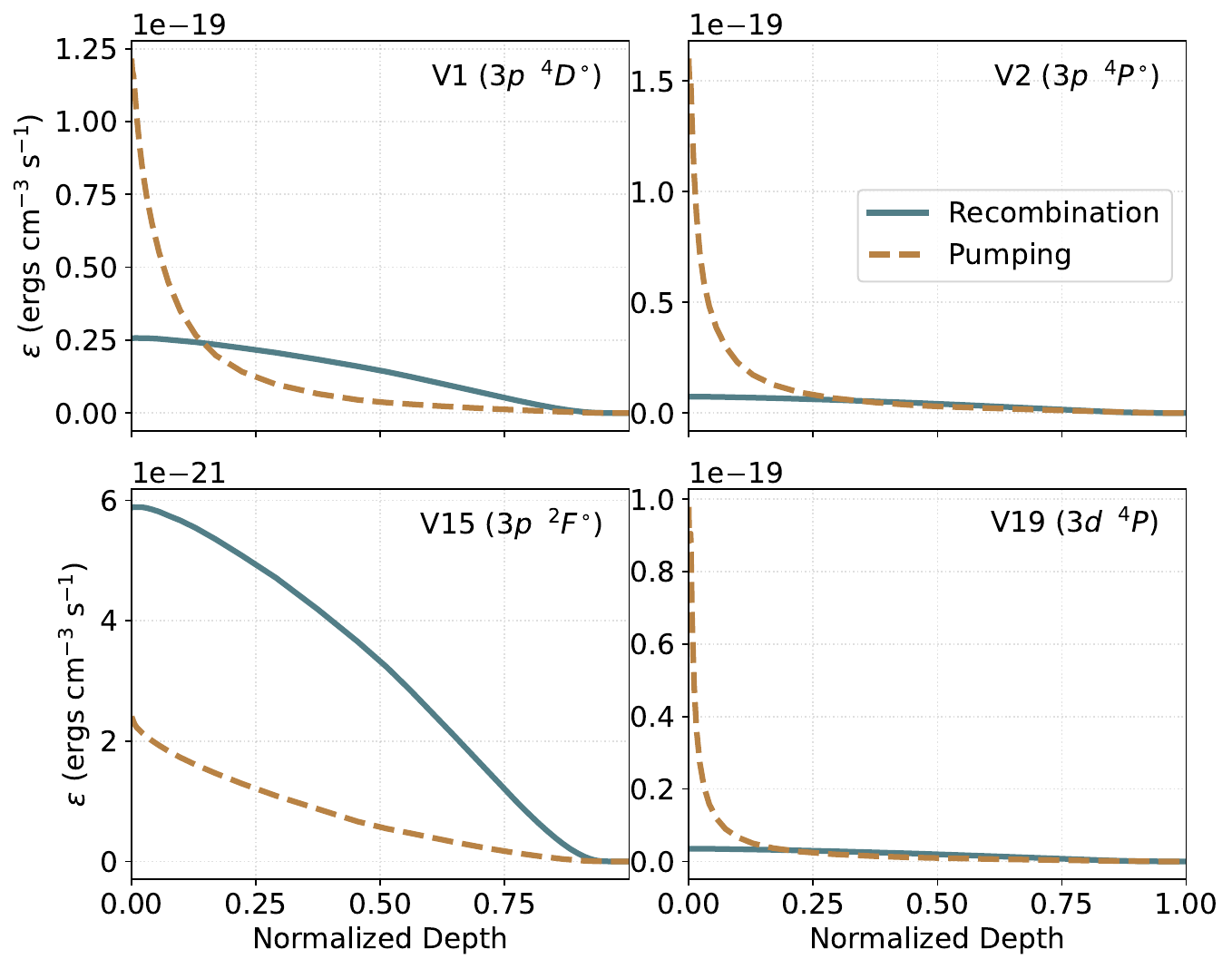}
\caption{   \textbf{Radial emissivity profiles from the \textsc{Cloudy} simulation for selected \Oiirec\ RL multiplets.} Blue solid curves denote the recombination contribution, and brown dashed curves shows the continuum-pumping contribution. Atomic terms in parentheses indicate the upper levels of the transitions. For each multiplet, only lines securely detected in our observations (Table~\ref{tab:oii_rl}) are included.}
\label{fig:pumping_profile}
\end{figure}

\begin{figure}[htbp]
\centering
\includegraphics[width=0.85\textwidth]{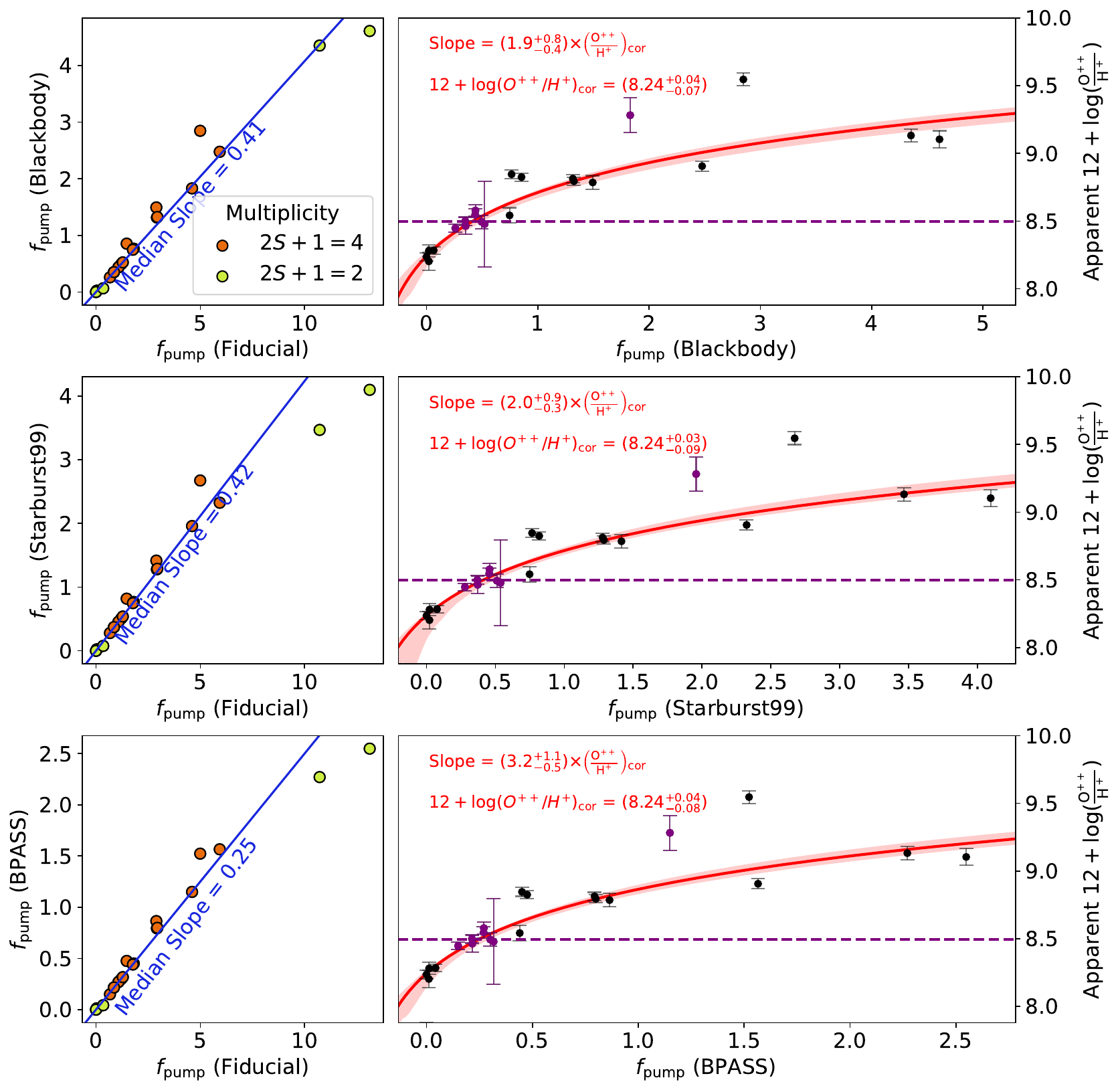}
\caption{    \textbf{Effect of the ionizing spectrum on the predicted pumping fraction ($f_\mathrm{pump}$).} From top to bottom, the panels compare $f_\mathrm{pump}$ predicted using a blackbody with $T = 37{,}500$~K, a Starburst99 model, and a BPASS model. \textbf{Left:} Comparison of the predicted $f_\mathrm{pump}$ values from each alternative ionizing spectrum with those from the fiducial Trapezium model. Each point represents an \Oiirec\ RL, color-coded by multiplicity ($2S + 1$). The blue line indicates the median relation, reflecting the typical scaling of the pumping fraction between the two models. \textbf{Right:} Apparent \Oiirec\ RL abundances as a function of the corresponding $f_\mathrm{pump}$. Purple points denote the V1 multiplet, and the dashed horizontal line marks mean V1 abundance. Although the fitted slopes differ because the absolute values of $f_\mathrm{pump}$ depend on the adopted ionizing spectrum, the extrapolated RL abundance at $f_\mathrm{pump}=0$ is nearly identical to that obtained with the tailored Trapezium simulation (Fig.~\ref{fig:o2rl_fluo}).
 }
\label{fig:fpump_from_sed}
\end{figure}

\begin{figure}[htbp]
\centering
\includegraphics[height=4.2cm]{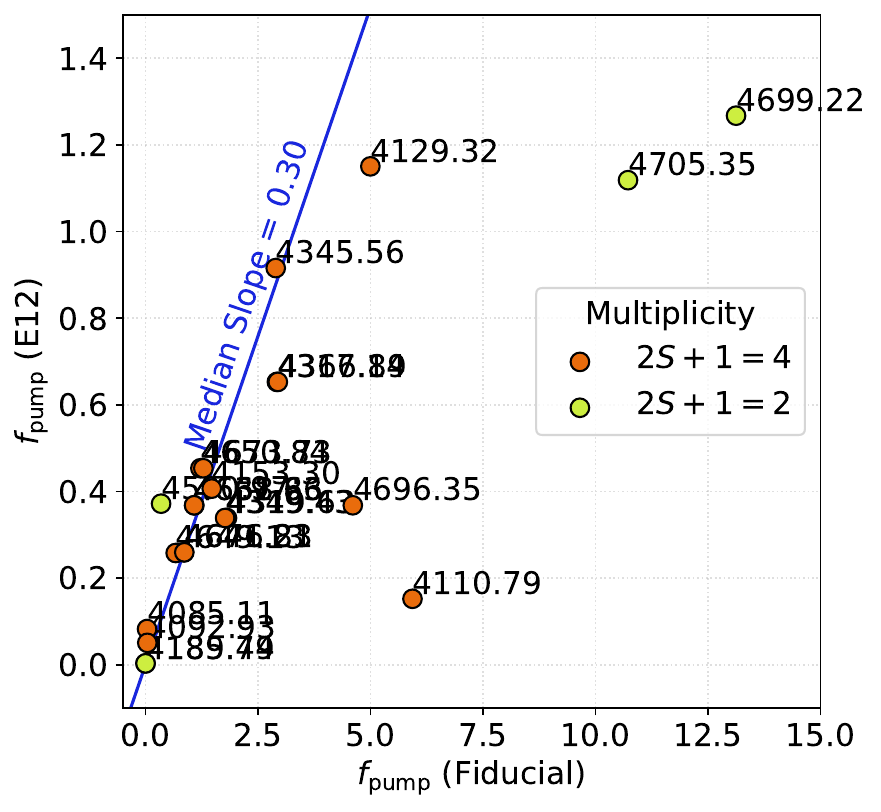}
\includegraphics[height=4.2cm]{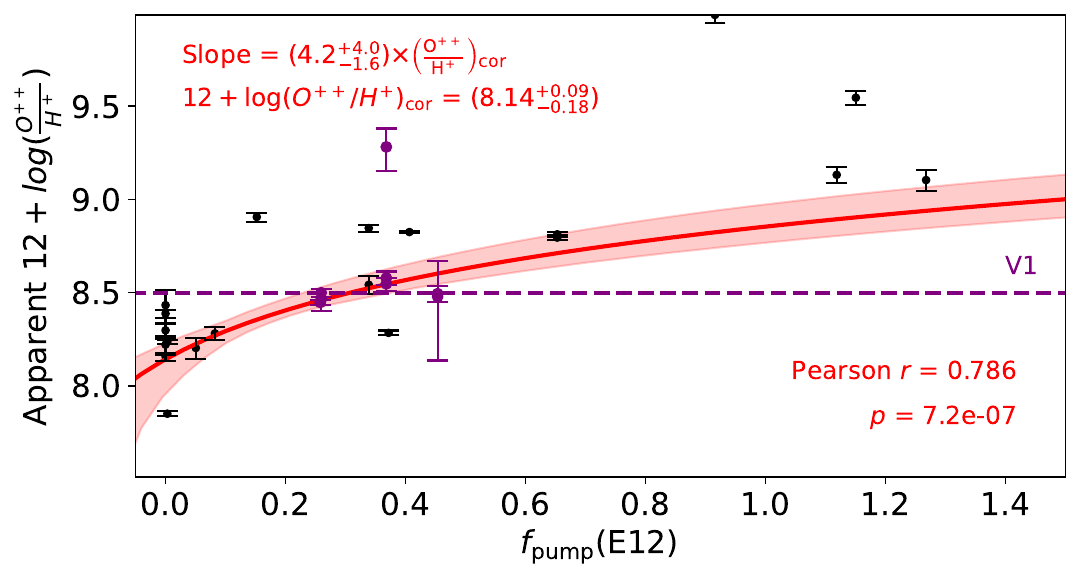}
\caption{\textbf{Comparison between our fiducial continuum pumping model and the literature planetary nebula model for IC~418 (E12).} The figure layout follows Fig.~\ref{fig:fpump_from_sed}.}
\label{fig:model_comparison}
\end{figure}

\begin{figure}[htbp]
\centering
\includegraphics[width=0.8\textwidth]{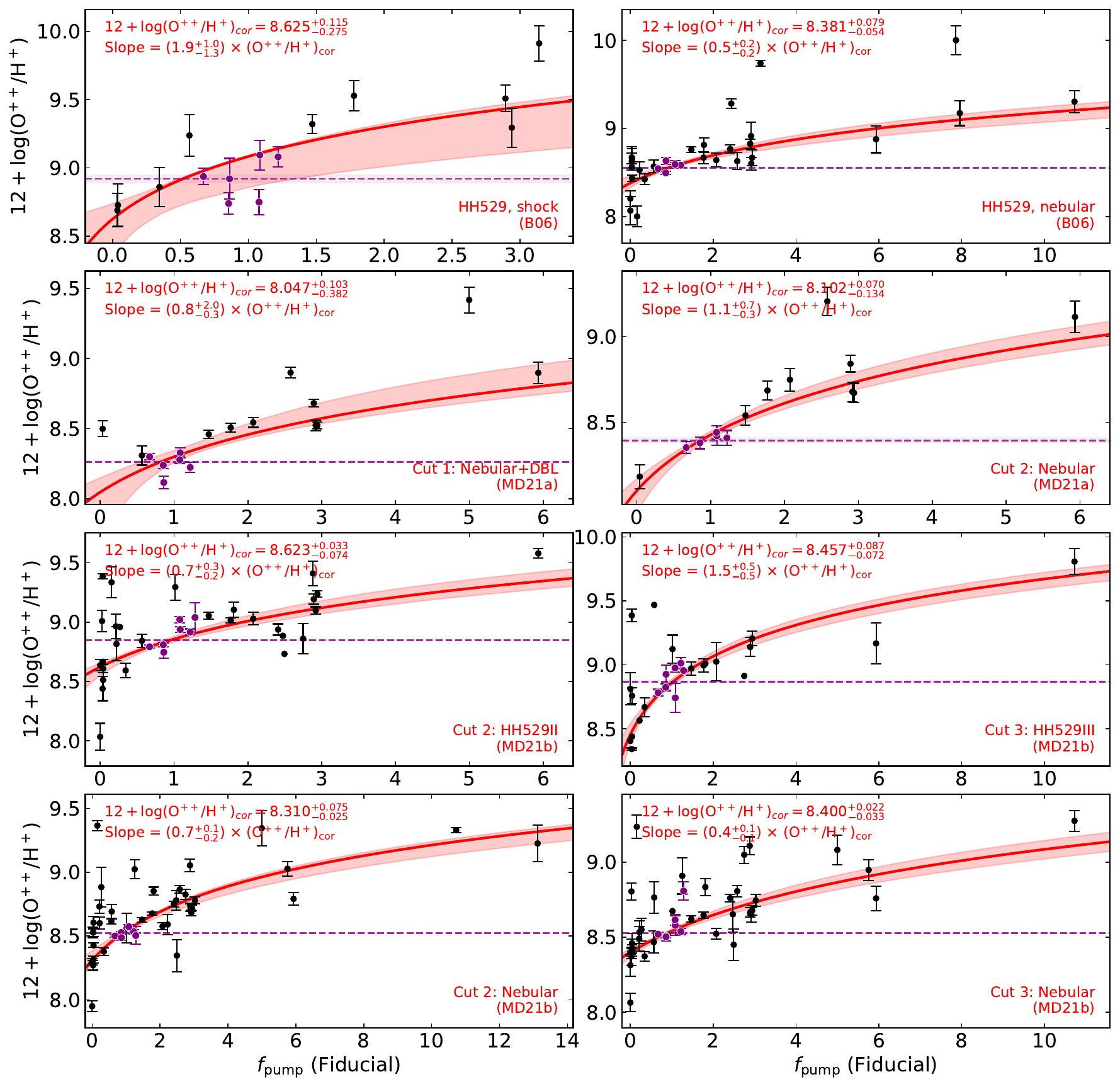}
\caption{\textbf{Apparent \Oiirec\ RL abundances in HH objects and kinematically resolved components of M42 compiled from the literature \cite{blagrave06, mendez-delgado21a, mendez-delgado21b}.} The figure layout follows the right-hand panels of Fig.~\ref{fig:fpump_from_sed}. In all cases, the apparent \Opp/\Hp\ abundance derived from individual \Oiirec\ RLs exhibits a positive correlation with the continuum pumping fraction, $f_\mathrm{pump}$, predicted by our fiducial photoionization model.}
\label{fig:hh_oh_fpump}
\end{figure}

\begin{figure}[htbp]
\centering
\includegraphics[width=0.6\textwidth]{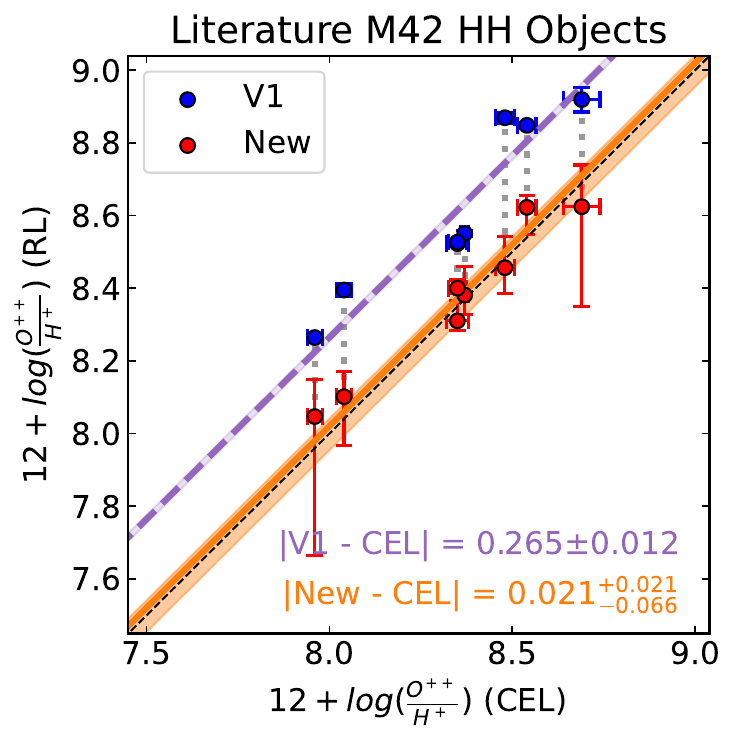}
\caption{\textbf{Comparison of \Oiirec\ RL- and \Oiii\ CEL-derived \Opp/\Hp\ abundances for the literature sample of kinematically resolved HH objects in M42.} The figure layout follows Fig.~\ref{fig:new_adf}. The underlying data are summarized in Table~\ref{tab:hh_oii_rl}. The difference between the \Opp/\Hp\ abundances derived from the V1 multiplet and the CELs is consistent with the classical ADF of $\sim0.25$~dex. After correcting for continuum pumping (Fig.~\ref{fig:hh_oh_fpump}), the RL abundances become nearly identical to the CEL abundances.}
\label{fig:hh_adf}
\end{figure}

\begin{figure}[htbp]
\centering
\includegraphics[width=0.8\textwidth]{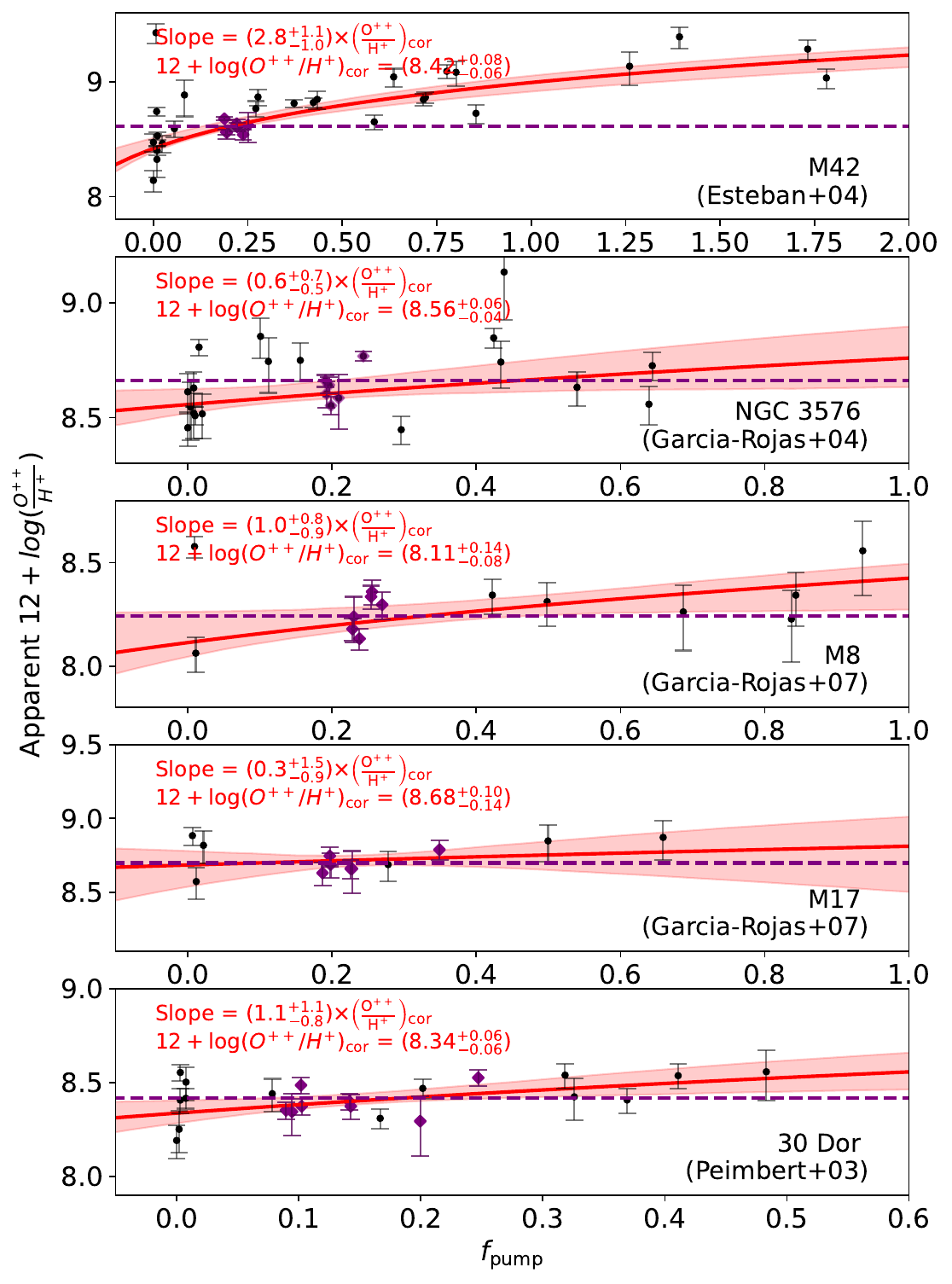}
\caption{   \textbf{Linear fits between the apparent $\Opp/\Hp$ abundance and $f_\mathrm{pump}$ for \Oiirec\ RL measurements compiled from the literature \cite{esteban04, garcia-rojas04, garcia-rojas07, peimbert03}.} The purple points represent the V1 multiplets, with their average V1 abundance marked as the horizontal dashed purple line. In all five cases, the inferred correction reduces the RL-based abundance. The resulting revised ADF values are shown in the right panel of Fig.~\ref{fig:new_adf}.}
\label{fig:recalculate_met}
\end{figure}

\begin{table*}[htbp]
\centering
\caption{\textbf{KCWI observing log for M42.}}
\label{tab:kcwi_observing_log}

\resizebox{\textwidth}{!}{
\begin{tabular}{ll c c c c l ccl }
\hline
Pointing & Slicer & PAs & Setup & \multicolumn{3}{c}{Red channel} & \multicolumn{3}{c}{Blue channel}\\
& & (deg) & & Grating & Exposure Time & Frame no. & Grating & Exposure Time & Frame no. \\
\hline
M42 East & Medium & 0, 90, 180, 270 & deep     & RL7250 & $30\,\mathrm{s}\times12$ & 118--129 & BM4400 & $30\,\mathrm{s}\times8$ & 96--103 \\
Sky      & Medium & 270               & deep     & RL7250 & $30\,\mathrm{s}\times3$  & 130--132 & BM4400 & $30\,\mathrm{s}\times2$ & 104, 105 \\
Sky      & Medium & 270               & snapshot & RL8000 & $1\,\mathrm{s}\times2$   & 133, 134 & BL4500 & $1\,\mathrm{s}\times2$  & 106, 107 \\
M42 East & Medium & 0, 270            & snapshot & RL8000 & $1\,\mathrm{s}\times4$   & 135--138 & BL4500 & $1\,\mathrm{s}\times4$  & 108--111 \\
M42 West & Medium & 0, 90             & snapshot & RL8000 & $1\,\mathrm{s}\times4$   & 139--142 & BL4500 & $1\,\mathrm{s}\times4$  & 112--115 \\
Sky      & Medium & 90                & snapshot & RL8000 & $1\,\mathrm{s}\times2$   & 143, 144 & BL4500 & $1\,\mathrm{s}\times2$  & 116, 117 \\
Sky      & Medium & 90                & deep     & RL7250 & $30\,\mathrm{s}\times2$  & 145, 146 & BM4400 & $30\,\mathrm{s}\times2$ & 118, 119 \\
M42 West & Medium & 90, 180           & deep     & RL7250 & $30\,\mathrm{s}\times7$  & 147--153 & BM4400 & $30\,\mathrm{s}\times4$ & 120--123 \\
\hline
\end{tabular}}

\vspace{2mm}
\footnotesize
\textit{Notes.} Exposure times are given as individual exposure duration multiplied by the number of frames. 
\end{table*}

\begin{table*}
\centering
\caption{\textbf{Flux summary of important nebular emission lines.}}
\label{tab:strong_line_fluxes}
\resizebox{\textwidth}{!}{
\begin{tabular}{lccccccccccc}
\toprule
Line & $\lambda$ & Grating &
\multicolumn{3}{c}{Full FoV} &
\multicolumn{3}{c}{Low-Ionization} &
\multicolumn{3}{c}{High-Ionization} \\
\cmidrule(lr){4-6}\cmidrule(lr){7-9}\cmidrule(lr){10-12}
 & (\AA) & &
$F$ & $I$ &
&
$F$ & $I$ &
&
$F$ & $I$ \\
\midrule
H14 & 3721.94 & BL4500
& $243\pm17$ & $3.46\pm0.30$
& & $243\pm45$ & $3.22\pm0.62$
& & $168\pm22$ & $3.11\pm0.40$ & \\

[\ion{O}{II}] & 3726.03 & BL4500
& $5025\pm225$ & $71.4\pm5.0$
& & $6498\pm87$ & $86.1\pm1.5$
& & $3164\pm42$ & $58.5\pm0.99$ & \\

[\ion{O}{II}] & 3728.82 & BL4500
& $2376\pm125$ & $33.8\pm2.6$
& & $2689\pm59$ & $35.6\pm0.88$
& & $1442\pm29$ & $26.6\pm0.60$ & \\

H13 & 3734.37 & BL4500
& $177\pm12$ & $2.52\pm0.22$
& & $204\pm44$ & $2.70\pm0.59$
& & $142\pm21$ & $2.63\pm0.39$ & \\

H12 & 3750.15 & BL4500
& $228\pm13$ & $3.22\pm0.25$
& & $258\pm7$ & $3.41\pm0.11$
& & $190\pm5.7$ & $3.50\pm0.11$ & \\

H11 & 3770.63 & BL4500
& $279\pm14$ & $3.95\pm0.28$
& & $319\pm7.6$ & $4.20\pm0.11$
& & $229\pm5.8$ & $4.20\pm0.12$ & \\

H10 & 3797.90 & BL4500
& $369\pm15$ & $5.19\pm0.34$
& & $423\pm8.1$ & $5.55\pm0.13$
& & $301\pm6.2$ & $5.50\pm0.13$ & \\

H9 & 3835.39 & BL4500
& $516\pm18$ & $7.21\pm0.43$
& & $583\pm9.1$ & $7.61\pm0.14$
& & $423\pm6.9$ & $7.67\pm0.14$ & \\

H8 & 3889.05 & BL4500
& $1084\pm29$ & $15.0\pm0.83$
& & $1244\pm15$ & $16.1\pm0.26$
& & $837\pm10$ & $15.0\pm0.23$ & \\

[\ion{S}{II}] & 4068.60 & BM4400
& $227\pm5.3$ & $3.03\pm0.14$
& & $306\pm6.7$ & $3.83\pm0.09$
& & $93.2\pm2.6$ & $1.60\pm0.05$ & \\

[\ion{S}{II}] & 4076.35 & BM4400
& $78.8\pm2.3$ & $1.05\pm0.05$
& & $105.6\pm2.7$ & $1.32\pm0.04$
& & $34.1\pm1.4$ & $0.584\pm0.025$ & \\

H6 & 4101.74 & BL4500
& $1946\pm58$ & $25.8\pm1.3$
& & $2135\pm23$ & $26.5\pm0.35$
& & $1582\pm17$ & $26.9\pm0.36$ & \\

H$\gamma$ & 4340.47 & BL4500
& $3771\pm176$ & $47.2\pm2.4$
& & $4060\pm47$ & $47.9\pm0.61$
& & $3049\pm35$ & $48.3\pm0.59$ & \\

[\ion{O}{III}] & 4363.21 & BM4400
& $93.2\pm2.3$ & $1.16\pm0.04$
& & $96.7\pm2.6$ & $1.14\pm0.03$
& & $88.7\pm2.1$ & $1.40\pm0.03$ & \\

H$\beta$ & 4861.33 & BL4500
& $9054\pm484$ & $100\pm5.5$
& & $9486\pm101$ & $100\pm1.1$
& & $7333\pm78$ & $100\pm1.1$ & \\

[\ion{O}{III}] & 4958.91 & BL4500
& $9667\pm580$ & $104.5\pm6.7$
& & $9317\pm105$ & $96.3\pm1.1$
& & $9566\pm108$ & $127.1\pm1.5$ & \\

[\ion{Ar}{III}] & 5191.82 & BL4500
& $5.83\pm0.29$ & $0.0600\pm0.0032$
& & $6.49\pm0.33$ & $0.0641\pm0.0031$
& & $4.97\pm0.25$ & $0.0622\pm0.0030$ & \\

[\ion{N}{II}] & 5754.64 & RL7250
& $142\pm1.7$ & $1.32\pm0.05$
& & $188\pm2.2$ & $1.69\pm0.02$
& & $66.9\pm1.0$ & $0.735\pm0.012$ & \\

[\ion{S}{III}] & 6312.10 & RL7250
& $251\pm3.2$ & $2.10\pm0.13$
& & $284\pm3.7$ & $2.33\pm0.04$
& & $197\pm2.5$ & $1.93\pm0.03$ & \\

[\ion{N}{II}] & 6548.03 & RL8000
& $2434\pm70$ & $19.6\pm1.5$
& & $3044\pm91$ & $24.1\pm0.81$
& & $1377\pm40$ & $12.8\pm0.40$ & \\

H$\alpha$ & 6562.82 & RL8000
& $36024\pm895$ & $289\pm21$
& & $38119\pm977$ & $301\pm8.9$
& & $33134\pm769$ & $307\pm8.1$ & \\

[\ion{N}{II}] & 6583.41 & RL8000
& $8037\pm199$ & $64.3\pm4.8$
& & $10003\pm258$ & $78.7\pm2.4$
& & $4631\pm110$ & $42.8\pm1.2$ & \\

[\ion{S}{II}] & 6716.47 & RL7250
& $346\pm4.2$ & $2.71\pm0.20$
& & $416\pm5.1$ & $3.21\pm0.07$
& & $229\pm2.8$ & $2.06\pm0.04$ & \\

[\ion{S}{II}] & 6730.85 & RL7250
& $677\pm7.9$ & $5.28\pm0.40$
& & $829\pm9.7$ & $6.38\pm0.13$
& & $424\pm4.9$ & $3.80\pm0.07$ & \\

[\ion{Ar}{III}] & 7135.78 & RL7250
& $2245\pm40$ & $16.4\pm1.5$
& & $2341\pm46$ & $16.9\pm0.46$
& & $2125\pm32$ & $17.6\pm0.40$ & \\

[\ion{O}{II}] & 7319 & RL7250
& $1506\pm22$ & $10.6\pm1.0$
& & $1933\pm27$ & $13.6\pm0.34$
& & $797\pm12$ & $6.34\pm0.16$ & \\

[\ion{O}{II}] & 7330 & RL7250
& $1226\pm18$ & $8.65\pm0.84$
& & $1574\pm22$ & $11.1\pm0.28$
& & $650\pm9.7$ & $5.17\pm0.12$ & \\

[\ion{Ar}{III}] & 7751.10 & RL7250
& $584\pm7.8$ & $3.85\pm0.43$
& & $603\pm8.3$ & $3.98\pm0.11$
& & $565\pm7.2$ & $4.13\pm0.10$ & \\

P11 & 8862.79 & RL8000
& $261\pm3.3$ & $1.46\pm0.22$
& & $265\pm3.4$ & $1.50\pm0.05$
& & $263\pm3.5$ & $1.57\pm0.05$ & \\

P10 & 9014.90 & RL8000
& $362\pm6.4$ & $1.97\pm0.31$
& & $364\pm4.3$ & $2.03\pm0.07$
& & $368\pm12$ & $2.14\pm0.09$ & \\

[\ion{S}{III}] & 9068.90 & RL8000
& $7477\pm184$ & $40.5\pm6.3$
& & $7685\pm195$ & $42.5\pm1.8$
& & $7264\pm183$ & $41.9\pm1.7$ & \\

P9 & 9229.01 & RL8000
& $506\pm7.7$ & $2.68\pm0.43$
& & $512\pm8.1$ & $2.78\pm0.11$
& & $511\pm7.7$ & $2.88\pm0.10$ & \\

[\ion{S}{III}] & 9530.60 & RL8000
& $18177\pm988$ & $93.0\pm16$
& & $18811\pm1010$ & $98.7\pm6.2$
& & $17442\pm987$ & $94.0\pm6.3$ & \\

P8 & 9545.97 & RL8000
& $713\pm47$ & $3.64\pm0.67$
& & $718\pm46$ & $3.76\pm0.30$
& & $725\pm49$ & $3.90\pm0.32$ & \\
\bottomrule
\end{tabular}}

\vspace{2mm}
\begin{minipage}{0.95\textwidth}
\footnotesize
\textit{Notes.} $F$ denotes observed fluxes in average surface brightness units: $10^{-16}~\mathrm{ergs}~\mathrm{s}^{-1}~\mathrm{cm}^{-2}~\mathrm{\AA}^{-1}~\mathrm{arcsec}^{-2}$; $I$ denotes reddening-corrected intensities normalized to
$I(\mathrm{H}\beta)=100$. Uncertainties are $1\sigma$. All flux uncertainties quoted here include the 1\% flux calibration uncertainty, while the noralized intensity uncertainties also include the dust-correction uncertainty. The [\ion{O}{II}]~$\lambda\lambda$7319 flux is a sum of the 7318.39 and 7319.99 lines; the [\ion{O}{II}]~$\lambda\lambda$7330 flux is a sum of the 7329.66, and 7330.73 lines.
\end{minipage}
\end{table*}

\begin{table*}[!htbp]
\centering
\caption{\textbf{Summary of the methodology and assumptions used to measure nebular physical properties.}}
\label{tab:phys_method}

\resizebox{\textwidth}{!}{
\begin{tabular}{l l p{10cm} p{3cm}}
\toprule
Ion & Quantity & Method & Atomic \& extinction data \\
\midrule

& \ebv &
Average \ebv\ from the H$\beta$/H$\alpha$ and H$\gamma$/H$\beta$ ratios, assuming
$n_e=10{,}000~\mathrm{cm^{-3}}$, $T_e=10{,}000~\mathrm{K}$, and $R_V=5.5$. &
CCM89; SH95 (Case~B) \\
\addlinespace

\Hp & $T_e(\mathrm{PJ})$ &
$\dfrac{I_\lambda(8157)-I_\lambda(8567)}{I(\mathrm{P11})}$. &
SH95 (Case~B) \\
\addlinespace

\Op &
$n_e(\Oii)$ &
Jointly from \Oii\ $\lambda\lambda 3726/3729$ and
\Oii\ $\lambda\lambda(7319+7330)/(3726+3729)$. &
Z82, WFD96, K09 \\
& $T_e(\Oii)$ &
\Oii\ $\lambda\lambda(7319+7330)/(3726+3729)$. & \\
& $12+\log(\Op/\Hp)$ &
\Oii\ $\lambda\lambda(3726+3729)$/H$\beta$. & \\
\addlinespace

\Opp &
$T_e(\Oiii)$ &
\Oiii\ $\lambda\lambda 4363/4959$. &
\multirow{2}{3.2cm}{FFT04, SZ00, SSB14} \\
& $12+\log(\Opp/\Hp)$ &
\Oiii\ $\lambda 4959$/H$\beta$. & \\
\addlinespace

O & $12+\log(\mathrm{O}/\mathrm{H})$ &
$(\Op+\Opp)/\Hp$. & \\
\addlinespace

\Np &
$T_e(\Nii)$ &
\Nii\ $\lambda\lambda 5755/6583$. &
\multirow{2}{3.2cm}{FFT04, T11} \\
& $12+\log(\Np/\Hp)$ &
\Nii\ $\lambda 6583$/H$\alpha$. & \\
\addlinespace

N & $12+\log(\mathrm{N}/\mathrm{H})$ &
$(\Np+\Npp)/\Hp$, with $\Npp/\Np=\Opp/\Op$. & \\
\addlinespace

\Sp &
$n_e(\Sii)$ \& $T_e(\Sii)$ &
Jointly from \Sii\ $\lambda\lambda 6716/6731$ and
\Sii\ $\lambda\lambda 4068/(6716+6731)$. &
\multirow{2}{3.2cm}{RGJ19, TZ10} \\
& $12+\log(\Sp/\Hp)$ &
\Sii\ $\lambda\lambda(6716+6731)$/H$\alpha$. & \\
\addlinespace

\Spp &
$T_e(\Siii)$ &
\Siii\ $\lambda\lambda 6312/9069$. &
\multirow{2}{3.2cm}{FFTI06, TG99} \\
& $12+\log(\Spp/\Hp)$ &
\Siii\ $\lambda 9069$/H$\alpha$. & \\
\addlinespace

S & $12+\log(\mathrm{S}/\mathrm{H})$ &
$\mathrm{ICF}\times \dfrac{\Sp+\Spp}{\Hp}$, where the ionization correction factor (ICF)
follows the theoretical calibration of Ref. \cite{dors16}. & \\
\addlinespace

\Arpp &
$T_e(\Ariii)$ &
\Ariii\ $\lambda\lambda 5192/7136$. &
\multirow{2}{3.2cm}{MB09} \\
& $12+\log(\Arpp/\Hp)$ &
\Ariii\ $\lambda 7136$/H$\alpha$. & \\

\bottomrule
\end{tabular}}

\begin{tablenotes}[flushleft]
\footnotesize
\item \textit{Notes.} Unless otherwise noted, all measurements assume $n_e = n_e(\Sii)$.
Extinction curve: CCM89 \cite{cardelli89}.
Atomic data references: SH95 \cite{storey95}, Z82 \cite{zeippen82}, WFD96 \cite{wiese96},
K09 \cite{kisielius09}, FFT04 \cite{froese-fischer04}, SZ00 \cite{storey00},
SSB14 \cite{storey14}, T11 \cite{tayal11}, RGJ19 \cite{rynkun19}, TZ10 \cite{tayal10},
FFTI06 \cite{froese-fischer06}, TG99 \cite{tayal99}, MB09 \cite{munoz-burgos09}.
\end{tablenotes}

\end{table*}

\begin{table*}
\centering
\caption{\textbf{Derived physical conditions and chemical abundances.}}
\label{tab:phys_properties}
\begin{tabular}{lccc}
\toprule
Quantity
& Full FoV
& Low-Ionization
& High-Ionization \\
\midrule

$E(B-V)$
& $0.261^{+0.059}_{-0.054}$
& $0.236^{+0.013}_{-0.012}$
& $0.314^{+0.011}_{-0.011}$ \\

$n_e$([\ion{S}{II}]) (cm$^{-3}$)
& $11800^{+820}_{-630}$ 
& $13300^{+890}_{-720}$ 
& $8130^{+360}_{-330}$ \\

$T_e$([\ion{S}{II}]) (K)
& $12580^{+2280}_{-1580}$ 
& $11830^{+840}_{-850}$ 
& $12070^{+640}_{-600}$ \\

$12+\log(\mathrm{S}^{+}/\mathrm{H}^{+})$
& $5.50^{+0.14}_{-0.14}$ 
& $5.654^{+0.095}_{-0.081}$ 
& $5.286^{+0.060}_{-0.057}$ \\

$n_e$([\ion{O}{II}]) (cm$^{-3}$)
& $6820^{+1900}_{-1060}$ 
& $11740^{+1470}_{-1200}$ 
& $7100^{+530}_{-460}$ \\

$T_e$([\ion{O}{II}]) (K)
& $10140^{+1340}_{-1160}$ 
& $10110^{+390}_{-450}$ 
& $10200^{+310}_{-310}$ \\

$12+\log(\mathrm{O}^{+}/\mathrm{H}^{+})$
& $8.00^{+0.27}_{-0.25}$ 
& $8.103^{+0.108}_{-0.088}$ 
& $7.805^{+0.069}_{-0.067}$ \\

$T_e$([\ion{N}{II}]) (K])
& $10100^{+170}_{-150}$
& $10150^{+140}_{-130}$
& $9724^{+101}_{-88}$ \\

$12+\log(\mathrm{N}^{+}/\mathrm{H}^{+})$
& $7.145^{+0.029}_{-0.029}$
& $7.210^{+0.027}_{-0.028}$
& $6.986^{+0.022}_{-0.024}$ \\

$T_e$([\ion{O}{III}]) (K)
& $8430^{+160}_{-150}$
& $8548^{+56}_{-57}$
& $8442^{+49}_{-52}$ \\

$12+\log(\mathrm{O}^{++}/\mathrm{H}^{+})$
& $8.311^{+0.065}_{-0.065}$
& $8.253^{+0.012}_{-0.012}$
& $8.391^{+0.011}_{-0.011}$ \\

$T_e$([\ion{S}{III}]) (K)
& $8840^{+340}_{-310}$
& $9016^{+114}_{-110}$
& $8490^{+96}_{-93}$ \\

$12+\log(\mathrm{S}^{++}/\mathrm{H}^{+})$
& $6.882^{+0.073}_{-0.076}$
& $6.867^{+0.030}_{-0.031}$
& $6.909^{+0.028}_{-0.028}$ \\

$T_e$(PJ) (K)
& $7550^{+730}_{-610}$
& $7550^{+480}_{-410}$
& $7440^{+420}_{-420}$ \\

$T_e$([\ion{Ar}{III}]) (K)
& $8875^{+240}_{-230}$
& $8960^{+150}_{-140}$
& $8800^{+130}_{-140}$ \\

$12+\log(\mathrm{Ar}^{++}/\mathrm{H}^{+})$
& $6.258^{+0.042}_{-0.044}$
& $6.244^{+0.026}_{-0.028}$
& $6.272^{+0.025}_{-0.023}$ \\

$12+\log(\mathrm{O}/\mathrm{H})$
& $8.484^{+0.107}_{-0.060}$ 
& $8.485^{+0.050}_{-0.033}$ 
& $8.491^{+0.018}_{-0.014}$ \\

$12+\log(\mathrm{S}/\mathrm{H})$
& $6.94^{+0.12}_{-0.10}$ 
& $6.918^{+0.036}_{-0.034}$ 
& $7.003^{+0.036}_{-0.032}$ \\

$12+\log(\mathrm{N}/\mathrm{H})$
& $7.63^{+0.22}_{-0.19}$ 
& $7.591^{+0.062}_{-0.067}$ 
& $7.670^{+0.065}_{-0.063}$ \\

\bottomrule
\end{tabular}
\end{table*}

\begin{sidewaystable*}
\centering
\caption{\textbf{\Oiirec\ recombination lines and derived abundances.}}
\label{tab:oii_rl}
\resizebox{\textwidth}{!}{

\begin{tabular}{c c c l l c cc cc cc}
\toprule
$\lambda$ &
$f_{\rm pump}$ &
$f_{\rm pump}$ &
Upper term &
Lower term &
Multiplet &
\multicolumn{2}{c}{Full FoV} &
\multicolumn{2}{c}{Low-Ionization} &
\multicolumn{2}{c}{High-Ionization} \\
\cmidrule(lr){7-8}\cmidrule(lr){9-10}\cmidrule(lr){11-12}
(\AA) & (E12) & (Fiducial) &&&&
Flux & $12+\log({\mathrm{O}^{++}/\mathrm{H}^+})$ &
Flux & $12+\log({\mathrm{O}^{++}/\mathrm{H}^+})$ &
Flux & $12+\log({\mathrm{O}^{++}/\mathrm{H}^+})$ \\
\midrule

4083.90 & 0 & --- &
$4f\;\;G\,{}^{2}[4]_{7/2}^{\circ}$ &
$3d\;\;{}^{4}F_{5/2}$ &
V47 &
$0.557\pm0.054$ & $8.316\pm0.050$ &
$0.618\pm0.058$ & $8.338\pm0.044$ &
$0.469\pm0.056$ & $8.328\pm0.051$ \\

4085.11 & 0.082 & 0.034 &
$3d\;\;{}^{4}F_{5/2}$ &
$3p\;\;{}^{4}D_{5/2}^{\circ}$ &
V10 &
$0.710\pm0.055$ & $8.283\pm0.047$ &
$0.691\pm0.059$ & $8.248\pm0.040$ &
$0.754\pm0.058$ & $8.400\pm0.033$ \\

4087.15 & 0 & --- &
$4f\;\;G\,{}^{2}[3]_{5/2}^{\circ}$ &
$3d\;\;{}^{4}F_{3/2}$ &
V48 &
$0.457\pm0.053$ & $8.239\pm0.060$ &
$0.465\pm0.058$ & $8.225\pm0.055$ &
$0.458\pm0.056$ & $8.324\pm0.055$ \\

4089.29 & 0 & --- &
$4f\;\;G\,{}^{2}[5]_{11/2}^{\circ}$ &
$3d\;\;{}^{4}F_{9/2}$ &
V48 &
$1.187\pm0.060$ & $8.176\pm0.037$ &
$1.055\pm0.063$ & $8.093\pm0.027$ &
$1.502\pm0.065$ & $8.399\pm0.019$ \\

4092.93 & 0.050 & 0.039 &
$3d\;\;{}^{4}F_{7/2}$ &
$3p\;\;{}^{4}D_{7/2}^{\circ}$ &
V10 &
$0.405\pm0.052$ & $8.205\pm0.064$ &
$0.407\pm0.057$ & $8.182\pm0.062$ &
$0.423\pm0.055$ & $8.324\pm0.055$ \\

4107.09 & 0 & --- &
$4f\;\;D\,{}^{2}[3]_{7/2}^{\circ}$ &
$3d\;\;{}^{4}F_{5/2}$ &
V48.01 &
$0.358\pm0.075$ & $8.45\pm0.11$ &
$0.387\pm0.084$ & $8.47\pm0.10$ &
$0.321\pm0.076$ & $8.48\pm0.12$ \\

4110.79 & 0.15 & 5.9 &
$3d\;\;{}^{4}D_{1/2}$ &
$3p\;\;{}^{4}P_{3/2}^{\circ}$ &
V20 &
$1.536\pm0.087$ & $8.911\pm0.038$ &
$1.350\pm0.093$ & $8.826\pm0.031$ &
$1.958\pm0.095$ & $9.132\pm0.022$ \\

4129.32 & 1.2 & 5.0 &
$3d\;\;{}^{4}P_{1/2}$ &
$3p\;\;{}^{4}P_{3/2}^{\circ}$ &
V19 &
$1.31\pm0.11$ & $9.551\pm0.050$ &
$1.36\pm0.10$ & $9.541\pm0.034$ &
$1.22\pm0.12$ & $9.623\pm0.044$ \\

4153.30 & 0.41 & 1.5 &
$3d\;\;{}^{4}P_{5/2}$ &
$3p\;\;{}^{4}P_{3/2}^{\circ}$ &
V19 &
$4.498\pm0.085$ & $8.830\pm0.031$ &
$3.975\pm0.088$ & $8.755\pm0.011$ &
$5.713\pm0.096$ & $9.020\pm0.009$ \\

4185.44 & 0.0030 & $0$ &
$3d\;\;{}^{2}G_{7/2}$ &
$3p\;\;{}^{2}F_{5/2}^{\circ}$ &
V36 &
$1.055\pm0.037$ & $7.850\pm0.032$ &
$1.043\pm0.037$ & $7.814\pm0.016$ &
$1.094\pm0.043$ & $7.988\pm0.018$ \\

4189.79 & 0.0030 & $0$ &
$3d\;\;{}^{2}G_{9/2}$ &
$3p\;\;{}^{2}F_{7/2}^{\circ}$ &
V36 &
$1.446\pm0.040$ & $8.213\pm0.031$ &
$1.399\pm0.040$ & $8.167\pm0.014$ &
$1.577\pm0.047$ & $8.377\pm0.015$ \\

4317.14 & 0.65 & 2.9 &
$3p\;\;{}^{4}P_{3/2}^{\circ}$ &
$3s\;\;{}^{4}P_{1/2}$ &
V2 &
$2.92\pm0.10$ & $8.811\pm0.030$ &
$2.55\pm0.11$ & $8.729\pm0.020$ &
$3.723\pm0.093$ & $9.014\pm0.012$ \\

4319.63 & 0.34 & 1.8 &
$3p\;\;{}^{4}P_{5/2}^{\circ}$ &
$3s\;\;{}^{4}P_{3/2}$ &
V2 &
$2.061\pm0.092$ & $8.841\pm0.034$ &
$2.16\pm0.11$ & $8.835\pm0.022$ &
$1.960\pm0.077$ & $8.926\pm0.018$ \\

4345.55 & 0.92 & 2.9 &
$3p\;\;{}^{4}P_{1/2}^{\circ}$ &
$3s\;\;{}^{4}P_{3/2}$ &
V2 &
$3.47\pm0.34$ & $10.008\pm0.049$ &
$3.11\pm0.41$ & $9.942\pm0.061$ &
$4.38\pm0.24$ & $10.189\pm0.024$ \\

4349.43 & 0.34 & 1.8 &
$3p\;\;{}^{4}P_{5/2}^{\circ}$ &
$3s\;\;{}^{4}P_{5/2}$ &
V2 &
$2.85\pm0.32$ & $8.537\pm0.053$ &
$2.46\pm0.39$ & $8.446\pm0.072$ &
$3.71\pm0.23$ & $8.758\pm0.027$ \\

4366.89 & 0.65 & 2.9 &
$3p\;\;{}^{4}P_{3/2}^{\circ}$ &
$3s\;\;{}^{4}P_{5/2}$ &
V2 &
$2.97\pm0.10$ & $8.794\pm0.030$ &
$2.67\pm0.13$ & $8.724\pm0.022$ &
$3.72\pm0.14$ & $8.987\pm0.017$ \\

4590.97 & 0.37 & 0.35 &
$3p\;\;{}^{2}F_{7/2}^{\circ}$ &
$3s\;\;{}^{2}D_{5/2}$ &
V15 &
$1.757\pm0.055$ & $8.261\pm0.028$ &
$1.727\pm0.060$ & $8.227\pm0.016$ &
$1.875\pm0.054$ & $8.405\pm0.014$ \\

4609.44 & 0 & --- &
$4f\;\;F\,{}^{2}[4]_{7/2}^{\circ}$ &
$3d\;\;{}^{2}D_{5/2}$ &
V93 &
$0.987\pm0.051$ & $8.409\pm0.033$ &
$0.961\pm0.056$ & $8.373\pm0.027$ &
$1.063\pm0.049$ & $8.544\pm0.020$ \\

4638.86 & 0.37 & 1.1 &
$3p\;\;{}^{4}D_{3/2}^{\circ}$ &
$3s\;\;{}^{4}P_{1/2}$ &
V1 &
$3.68\pm0.31$ & $8.574\pm0.046$ &
$3.58\pm0.36$ & $8.544\pm0.044$ &
$4.01\pm0.25$ & $8.692\pm0.028$ \\

4641.81 & 0.26 & 0.85 &
$3p\;\;{}^{4}D_{5/2}^{\circ}$ &
$3s\;\;{}^{4}P_{3/2}$ &
V1 &
$6.92\pm0.34$ & $8.495\pm0.029$ &
$6.54\pm0.39$ & $8.450\pm0.026$ &
$7.92\pm0.29$ & $8.643\pm0.016$ \\

4649.13 & 0.26 & 0.67 &
$3p\;\;{}^{4}D_{7/2}^{\circ}$ &
$3s\;\;{}^{4}P_{5/2}$ &
V1 &
$10.03\pm0.36$ & $8.444\pm0.027$ &
$9.49\pm0.40$ & $8.393\pm0.019$ &
$11.43\pm0.30$ & $8.613\pm0.012$ \\

4650.84 & 0.45 & 1.2 &
$3p\;\;{}^{4}D_{1/2}^{\circ}$ &
$3s\;\;{}^{4}P_{1/2}$ &
V1 &
$3.09\pm0.30$ & $8.491\pm0.049$ &
$2.90\pm0.35$ & $8.445\pm0.050$ &
$3.61\pm0.25$ & $8.636\pm0.033$ \\

4661.63 & 0.37 & 1.1 &
$3p\;\;{}^{4}D_{3/2}^{\circ}$ &
$3s\;\;{}^{4}P_{3/2}$ &
V1 &
$3.86\pm0.31$ & $8.543\pm0.041$ &
$3.69\pm0.36$ & $8.505\pm0.043$ &
$4.32\pm0.25$ & $8.672\pm0.027$ \\

4673.73 & 0.45 & 1.3 &
$3p\;\;{}^{4}D_{1/2}^{\circ}$ &
$3s\;\;{}^{4}P_{3/2}$ &
V1 &
$0.52\pm0.29$ & $8.48\pm0.33$ &
$0.47\pm0.33$ & $8.41\pm0.38$ &
$0.65\pm0.23$ & $8.64\pm0.18$ \\

4676.23 & 0.26 & 0.86 &
$3p\;\;{}^{4}D_{5/2}^{\circ}$ &
$3s\;\;{}^{4}P_{5/2}$ &
V1 &
$2.22\pm0.30$ & $8.460\pm0.064$ &
$2.09\pm0.34$ & $8.413\pm0.071$ &
$2.53\pm0.24$ & $8.604\pm0.042$ \\

4696.36 & 0.37 & 4.6 &
$3p\;\;{}^{4}D_{3/2}^{\circ}$ &
$3s\;\;{}^{4}P_{5/2}$ &
V1 &
$0.39\pm0.10$ & $9.30\pm0.11$ &
$0.38\pm0.12$ & $9.28\pm0.14$ &
$0.414\pm0.093$ & $9.39\pm0.11$ \\

4699.22 & 1.3 & 13 &
$3d\;\;{}^{2}F_{5/2}$ &
$3p\;\;{}^{2}D_{3/2}^{\circ}$ &
V25 &
$0.78\pm0.10$ & $9.109\pm0.064$ &
$0.77\pm0.12$ & $9.082\pm0.074$ &
$0.846\pm0.094$ & $9.236\pm0.048$ \\

4705.35 & 1.1 & 11 &
$3d\;\;{}^{2}F_{7/2}$ &
$3p\;\;{}^{2}D_{5/2}^{\circ}$ &
V25 &
$1.06\pm0.10$ & $9.139\pm0.046$ &
$1.00\pm0.13$ & $9.088\pm0.060$ &
$1.198\pm0.095$ & $9.298\pm0.034$ \\

\midrule

\multicolumn{5}{l}{\it V1 Abundance} &
--- & $8.497\pm0.022$ &
--- & $8.452\pm0.024$ &
--- & $8.645\pm0.021$ \\

\multicolumn{5}{l}{\it Corrected Abundance} &
--- & $8.235_{-0.098}^{+0.064}$ &
--- & $8.213_{-0.152}^{+0.016}$ &
--- & $8.285_{-0.131}^{+0.091}$ \\

\bottomrule
\end{tabular}}

\vspace{2mm}
\begin{minipage}{0.95\textwidth}
\footnotesize
\textit{Notes.} Fluxes are presented in average surface brightness units:
$10^{-16}~\mathrm{ergs}~\mathrm{s}^{-1}~\mathrm{cm}^{-2}~
\mathrm{\AA}^{-1}~\mathrm{arcsec}^{-2}$, the same as in
Table~\ref{tab:strong_line_fluxes}.
\end{minipage}
\end{sidewaystable*}

\begin{table*}
\centering
\caption{\textbf{Key model assumptions to caculate the corrected literature \Oiirec\ RL metallicities.}}
\label{tab:model_assumptions_new_adf}
\begin{tabular}{l c c c c c c}
\toprule
Value & M42 & M42 & NGC~3576 & M8 & M17 & 30~Dor \\
& (Fiducial) & (E04) & (GR04) & (GR07) & (GR07) & (P03)\\
\midrule
\multicolumn{6}{l}{\it \texttt{Cloudy} photoionization model:} \\
Stellar Age (Myr) & --- & 1 & 1 & 1 & 1 & 1\\
$\log(Z_*)$ & --- & $-1.85$ & $-1.80$ & $-1.85$ & $-1.84$ & $-2.03$\\
$\log(U)$ & $-2.5$ & $-2.5$ & $-2.5$ & $-2.5$ & $-2.5$ & $-2.5$\\
$\log(n_H / \mathrm{cm}^{-3})$ & 4 & 3.38 & 2.98 & 3.26 & 2.68 & 2.45\\
$\log(\mathrm{O}/\mathrm{H})_\mathrm{gas}$ & $-3.49$ & $-3.49$ & $-3.44$ & $-3.49$ & $-3.48$ & $-3.67$ \\
\midrule
\multicolumn{6}{l}{\it \texttt{PyNeb} \Oiirec\ RL emissivities:} \\
$n_e$ ($\mathrm{cm}^{-3}$) & 11,790 & 2,400 & 950 & 1,800 & 480 & 280\\
$T_e$ (K) & 8,430 & 8,300 & 8,500 & 8,090 & 8,020 & 9,950\\
\bottomrule
\end{tabular}

\begin{tablenotes}[flushleft]
\footnotesize
\item \textit{Notes.} Reference literature sources: E04 \cite{esteban04}, GR04 \cite{garcia-rojas04}, GR07 \cite{garcia-rojas07}, and P03 \cite{peimbert03}.
\end{tablenotes}
\end{table*}

\begin{sidewaystable*}
\centering
\caption{\textbf{Summary of the \Oiirec\ RLs for HH Objects in M42.}}
\label{tab:hh_oii_rl}
\resizebox{\textwidth}{!}{%
\begin{tabular}{c c l l cc cc cc cc}
\toprule
$\lambda$ & $f_{\rm pump}$ & Upper term & Lower term & \multicolumn{2}{c}{HH529 shock (B06)} & \multicolumn{2}{c}{HH529 neb. (B06)} & \multicolumn{2}{c}{Cut 1 neb.+DBL (MD21a)} & \multicolumn{2}{c}{Cut 2 neb. (MD21a)} \\
\cmidrule(lr){5-6}\cmidrule(lr){7-8}\cmidrule(lr){9-10}\cmidrule(lr){11-12}
(\AA) & & & & $I(\lambda)$ & $12+\log({\rm O}^{++}/{\rm H}^+)$ & $I(\lambda)$ & $12+\log({\rm O}^{++}/{\rm H}^+)$ & $I(\lambda)$ & $12+\log({\rm O}^{++}/{\rm H}^+)$ & $I(\lambda)$ & $12+\log({\rm O}^{++}/{\rm H}^+)$ \\
\midrule
3882.19 & 0.575 & $3d\;\;{}^{4}D_{7/2}$ & $3p\;\;{}^{4}D_{7/2}^{\circ}$ & --- & --- & --- & --- & --- & --- & --- & --- \\
3954.36 & 1.253 & $3p\;\;{}^{2}P_{1/2}^{\circ}$ & $3s\;\;{}^{2}P_{1/2}$ & --- & --- & --- & --- & --- & --- & --- & --- \\
3973.26 & 1.017 & $3p\;\;{}^{2}P_{3/2}^{\circ}$ & $3s\;\;{}^{2}P_{3/2}$ & --- & --- & --- & --- & --- & --- & --- & --- \\
4069.62 & 0.026 & $3d\;\;{}^{4}F_{3/2}$ & $3p\;\;{}^{4}D_{1/2}^{\circ}$ & --- & --- & --- & --- & --- & --- & --- & --- \\
4069.88 & 0.033 & $3d\;\;{}^{4}F_{5/2}$ & $3p\;\;{}^{4}D_{3/2}^{\circ}$ & --- & --- & $0.079\pm0.017$ & $8.625\pm0.095$ & --- & --- & --- & --- \\
4072.15 & 0.039 & $3d\;\;{}^{4}F_{7/2}$ & $3p\;\;{}^{4}D_{5/2}^{\circ}$ & $0.132\pm0.047$ & $8.727\pm0.155$ & $0.070\pm0.006$ & $8.438\pm0.036$ & --- & --- & $0.041\pm0.007$ & $8.184\pm0.069$ \\
4075.86 & 0.036 & $3d\;\;{}^{4}F_{9/2}$ & $3p\;\;{}^{4}D_{7/2}^{\circ}$ & $0.151\pm0.042$ & $8.689\pm0.122$ & $0.105\pm0.009$ & $8.565\pm0.035$ & --- & --- & --- & --- \\
4078.84 & 0.026 & $3d\;\;{}^{4}F_{3/2}$ & $3p\;\;{}^{4}D_{3/2}^{\circ}$ & --- & --- & --- & --- & --- & --- & --- & --- \\
4085.11 & 0.034 & $3d\;\;{}^{4}F_{5/2}$ & $3p\;\;{}^{4}D_{5/2}^{\circ}$ & --- & --- & $0.025\pm0.007$ & $8.673\pm0.121$ & $0.022\pm0.003$ & $8.499\pm0.056$ & --- & --- \\
4092.93 & 0.039 & $3d\;\;{}^{4}F_{7/2}$ & $3p\;\;{}^{4}D_{7/2}^{\circ}$ & --- & --- & $0.016\pm0.004$ & $8.650\pm0.123$ & --- & --- & --- & --- \\
4097.23 & 7.860 & $3d\;\;{}^{4}D_{3/2}$ & $3p\;\;{}^{4}P_{1/2}^{\circ}$ & --- & --- & $0.063\pm0.024$ & $10.003\pm0.166$ & --- & --- & --- & --- \\
4104.99 & 2.495 & $3d\;\;{}^{4}D_{3/2}$ & $3p\;\;{}^{4}P_{3/2}^{\circ}$ & --- & --- & --- & --- & --- & --- & --- & --- \\
4110.79 & 5.932 & $3d\;\;{}^{4}D_{1/2}$ & $3p\;\;{}^{4}P_{3/2}^{\circ}$ & --- & --- & $0.018\pm0.006$ & $8.879\pm0.153$ & $0.016\pm0.003$ & $8.898\pm0.078$ & $0.027\pm0.006$ & $9.115\pm0.091$ \\
4119.22 & 0.566 & $3d\;\;{}^{4}D_{7/2}$ & $3p\;\;{}^{4}P_{5/2}^{\circ}$ & $0.151\pm0.052$ & $9.238\pm0.151$ & $0.032\pm0.005$ & $8.570\pm0.071$ & $0.015\pm0.002$ & $8.309\pm0.069$ & --- & --- \\
4121.46 & 2.578 & $3d\;\;{}^{4}P_{1/2}$ & $3p\;\;{}^{4}P_{1/2}^{\circ}$ & --- & --- & $0.014\pm0.003$ & $8.630\pm0.097$ & $0.026\pm0.002$ & $8.899\pm0.039$ & $0.053\pm0.010$ & $9.206\pm0.083$ \\
4129.32 & 4.996 & $3d\;\;{}^{4}P_{1/2}$ & $3p\;\;{}^{4}P_{3/2}^{\circ}$ & --- & --- & --- & --- & $0.013\pm0.003$ & $9.417\pm0.091$ & --- & --- \\
4132.80 & 2.073 & $3d\;\;{}^{4}P_{3/2}$ & $3p\;\;{}^{4}P_{1/2}^{\circ}$ & --- & --- & $0.029\pm0.005$ & $8.642\pm0.078$ & $0.028\pm0.002$ & $8.544\pm0.035$ & $0.044\pm0.007$ & $8.748\pm0.065$ \\
4153.30 & 1.470 & $3d\;\;{}^{4}P_{5/2}$ & $3p\;\;{}^{4}P_{3/2}^{\circ}$ & $0.187\pm0.030$ & $9.321\pm0.069$ & $0.055\pm0.004$ & $8.760\pm0.033$ & $0.041\pm0.003$ & $8.460\pm0.030$ & $0.047\pm0.006$ & $8.540\pm0.056$ \\
4156.53 & 2.436 & $3d\;\;{}^{4}P_{3/2}$ & $3p\;\;{}^{4}P_{5/2}^{\circ}$ & --- & --- & $0.025\pm0.003$ & $9.284\pm0.054$ & --- & --- & --- & --- \\
4185.44 & 0 & $3d\;\;{}^{2}G_{7/2}$ & $3p\;\;{}^{2}F_{5/2}^{\circ}$ & --- & --- & $0.029\pm0.006$ & $8.206\pm0.089$ & --- & --- & --- & --- \\
4189.79 & 0 & $3d\;\;{}^{2}G_{9/2}$ & $3p\;\;{}^{2}F_{7/2}^{\circ}$ & --- & --- & $0.012\pm0.004$ & $8.069\pm0.161$ & --- & --- & --- & --- \\
4317.14 & 2.919 & $3p\;\;{}^{4}P_{3/2}^{\circ}$ & $3s\;\;{}^{4}P_{1/2}$ & --- & --- & $0.023\pm0.004$ & $8.602\pm0.072$ & $0.023\pm0.002$ & $8.523\pm0.039$ & $0.032\pm0.004$ & $8.678\pm0.056$ \\
4319.63 & 1.810 & $3p\;\;{}^{4}P_{5/2}^{\circ}$ & $3s\;\;{}^{4}P_{3/2}$ & --- & --- & --- & --- & --- & --- & --- & --- \\
4325.76 & 2.881 & $3p\;\;{}^{4}P_{1/2}^{\circ}$ & $3s\;\;{}^{4}P_{1/2}$ & --- & --- & --- & --- & --- & --- & --- & --- \\
4336.86 & 2.914 & $3p\;\;{}^{4}P_{3/2}^{\circ}$ & $3s\;\;{}^{4}P_{3/2}$ & --- & --- & $0.020\pm0.007$ & $8.916\pm0.158$ & --- & --- & --- & --- \\
4345.56 & 2.893 & $3p\;\;{}^{4}P_{1/2}^{\circ}$ & $3s\;\;{}^{4}P_{3/2}$ & $0.231\pm0.051$ & $9.509\pm0.096$ & $0.049\pm0.006$ & $8.827\pm0.052$ & $0.039\pm0.002$ & $8.682\pm0.026$ & $0.056\pm0.006$ & $8.842\pm0.048$ \\
4347.41 & 2.225 & $3p\;\;{}^{2}D_{3/2}^{\circ}$ & $3s\;\;{}^{2}D_{3/2}$ & --- & --- & --- & --- & --- & --- & --- & --- \\
4349.43 & 1.768 & $3p\;\;{}^{4}P_{5/2}^{\circ}$ & $3s\;\;{}^{4}P_{5/2}$ & --- & --- & $0.047\pm0.007$ & $8.670\pm0.067$ & $0.031\pm0.002$ & $8.507\pm0.030$ & $0.047\pm0.006$ & $8.686\pm0.056$ \\
4351.26 & 2.748 & $3p\;\;{}^{2}D_{5/2}^{\circ}$ & $3s\;\;{}^{2}D_{5/2}$ & --- & --- & --- & --- & --- & --- & --- & --- \\
4366.89 & 2.940 & $3p\;\;{}^{4}P_{3/2}^{\circ}$ & $3s\;\;{}^{4}P_{5/2}$ & --- & --- & --- & --- & $0.024\pm0.001$ & $8.523\pm0.026$ & $0.033\pm0.004$ & $8.672\pm0.056$ \\
4366.90 & 2.940 & $3p\;\;{}^{4}P_{3/2}^{\circ}$ & $3s\;\;{}^{4}P_{5/2}$ & $0.117\pm0.039$ & $9.294\pm0.144$ & $0.028\pm0.006$ & $8.670\pm0.091$ & --- & --- & --- & --- \\
4414.90 & 2.409 & $3p\;\;{}^{2}D_{5/2}^{\circ}$ & $3s\;\;{}^{2}P_{3/2}$ & --- & --- & $0.032\pm0.003$ & $8.767\pm0.045$ & --- & --- & --- & --- \\
4416.97 & 3.028 & $3p\;\;{}^{2}D_{3/2}^{\circ}$ & $3s\;\;{}^{2}P_{1/2}$ & --- & --- & --- & --- & --- & --- & --- & --- \\
4452.38 & 3.141 & $3p\;\;{}^{2}D_{3/2}^{\circ}$ & $3s\;\;{}^{2}P_{3/2}$ & $0.062\pm0.019$ & $9.913\pm0.131$ & $0.044\pm0.003$ & $9.741\pm0.034$ & --- & --- & --- & --- \\
4590.97 & 0.345 & $3p\;\;{}^{2}F_{7/2}^{\circ}$ & $3s\;\;{}^{2}D_{5/2}$ & $0.077\pm0.025$ & $8.860\pm0.143$ & $0.027\pm0.004$ & $8.424\pm0.059$ & --- & --- & --- & --- \\
4595.96 & 0.154 & $3p\;\;{}^{2}F_{5/2}^{\circ}$ & $3s\;\;{}^{2}D_{5/2}$ & --- & --- & --- & --- & --- & --- & --- & --- \\
4596.18 & 0.162 & $3p\;\;{}^{2}F_{5/2}^{\circ}$ & $3s\;\;{}^{2}D_{3/2}$ & --- & --- & $0.014\pm0.004$ & $8.002\pm0.118$ & --- & --- & --- & --- \\
4638.86 & 1.084 & $3p\;\;{}^{4}D_{3/2}^{\circ}$ & $3s\;\;{}^{4}P_{1/2}$ & $0.142\pm0.036$ & $9.094\pm0.109$ & $0.048\pm0.005$ & $8.593\pm0.045$ & $0.034\pm0.003$ & $8.328\pm0.035$ & $0.041\pm0.005$ & $8.422\pm0.056$ \\
4641.81 & 0.855 & $3p\;\;{}^{4}D_{5/2}^{\circ}$ & $3s\;\;{}^{4}P_{3/2}$ & $0.141\pm0.026$ & $8.739\pm0.080$ & $0.083\pm0.003$ & $8.496\pm0.017$ & $0.050\pm0.003$ & $8.240\pm0.026$ & $0.069\pm0.006$ & $8.381\pm0.035$ \\
4649.13 & 0.669 & $3p\;\;{}^{4}D_{7/2}^{\circ}$ & $3s\;\;{}^{4}P_{5/2}$ & --- & --- & --- & --- & $0.054\pm0.003$ & $8.300\pm0.022$ & $0.064\pm0.005$ & $8.353\pm0.035$ \\
4649.14 & 0.669 & $3p\;\;{}^{4}D_{7/2}^{\circ}$ & $3s\;\;{}^{4}P_{5/2}$ & $0.355\pm0.049$ & $8.936\pm0.061$ & $0.136\pm0.004$ & $8.544\pm0.014$ & --- & --- & --- & --- \\
4650.84 & 1.219 & $3p\;\;{}^{4}D_{1/2}^{\circ}$ & $3s\;\;{}^{4}P_{1/2}$ & $0.141\pm0.024$ & $9.081\pm0.075$ & $0.048\pm0.003$ & $8.585\pm0.026$ & $0.030\pm0.002$ & $8.224\pm0.035$ & $0.044\pm0.004$ & $8.409\pm0.043$ \\
4661.63 & 1.079 & $3p\;\;{}^{4}D_{3/2}^{\circ}$ & $3s\;\;{}^{4}P_{3/2}$ & $0.072\pm0.015$ & $8.749\pm0.093$ & $0.053\pm0.002$ & $8.593\pm0.019$ & $0.034\pm0.002$ & $8.279\pm0.026$ & $0.048\pm0.004$ & $8.442\pm0.039$ \\
4673.73 & 1.285 & $3p\;\;{}^{4}D_{1/2}^{\circ}$ & $3s\;\;{}^{4}P_{3/2}$ & --- & --- & --- & --- & --- & --- & --- & --- \\
4676.23 & 0.861 & $3p\;\;{}^{4}D_{5/2}^{\circ}$ & $3s\;\;{}^{4}P_{5/2}$ & $0.074\pm0.025$ & $8.920\pm0.149$ & $0.039\pm0.004$ & $8.635\pm0.040$ & $0.013\pm0.001$ & $8.117\pm0.043$ & --- & --- \\
4699.01 & 7.953 & $3d\;\;{}^{2}F_{7/2}$ & $3p\;\;{}^{2}D_{5/2}^{\circ}$ & --- & --- & $0.012\pm0.004$ & $9.174\pm0.142$ & --- & --- & --- & --- \\
4699.22 & 13.122 & $3d\;\;{}^{2}F_{5/2}$ & $3p\;\;{}^{2}D_{3/2}^{\circ}$ & --- & --- & --- & --- & --- & --- & --- & --- \\
4705.35 & 10.722 & $3d\;\;{}^{2}F_{7/2}$ & $3p\;\;{}^{2}D_{5/2}^{\circ}$ & --- & --- & $0.017\pm0.005$ & $9.306\pm0.125$ & --- & --- & --- & --- \\
4890.86 & 5.752 & $3d\;\;{}^{4}P_{1/2}$ & $3p\;\;{}^{4}S_{3/2}^{\circ}$ & --- & --- & --- & --- & --- & --- & --- & --- \\
4924.53 & 1.778 & $3d\;\;{}^{4}P_{5/2}$ & $3p\;\;{}^{4}S_{3/2}^{\circ}$ & $0.157\pm0.041$ & $9.528\pm0.113$ & $0.033\pm0.006$ & $8.814\pm0.083$ & --- & --- & --- & --- \\
6721.39 & 2.474 & $3p\;\;{}^{2}S_{1/2}^{\circ}$ & $3s\;\;{}^{2}P_{3/2}$ & --- & --- & --- & --- & --- & --- & --- & --- \\
6895.10 & 0.268 & $4p\;\;{}^{4}D_{7/2}^{\circ}$ & $3d\;\;{}^{4}F_{9/2}$ & --- & --- & --- & --- & --- & --- & --- & --- \\
6906.44 & 0.221 & $4p\;\;{}^{4}D_{5/2}^{\circ}$ & $3d\;\;{}^{4}F_{7/2}$ & --- & --- & $0.006\pm0.001$ & $8.529\pm0.094$ & --- & --- & --- & --- \\
\multicolumn{4}{l}{\it V1 Abundance} & --- & $8.919 \pm 0.034$ & --- & $8.551 \pm  0.008$ & --- & $8.264 \pm 0.012$ & --- & $8.395 \pm 0.018$ \\
\multicolumn{4}{l}{\it Corrected Abundance} & --- & $8.62_{-0.28}^{+0.11}$ & --- & $8.381_{-0.054}^{+0.079}$ & --- & $8.05_{-0.38}^{+0.10}$ & --- & $8.10_{-0.13}^{+0.07}$ \\
\bottomrule
\end{tabular}%
}

\vspace{2mm}
\begin{minipage}{0.95\textwidth}
\footnotesize
\textit{Notes.} Literature dereddened intensities ($I$) are normalized to $I({\rm H}\beta)=100$. Lines without reliable $f_\mathrm{pump}$ calculations are omitted. Literature sources: B06 \cite{blagrave06}, MD21a \cite{mendez-delgado21a}, MD21b \cite{mendez-delgado21b}.
\end{minipage}
\end{sidewaystable*}

\begin{sidewaystable*}
\centering
\caption{\textbf{Summary of the \Oiirec\ RLs for HH Objects in M42 (continued).}}
\label{tab:hh_oii_rl_continued}
\resizebox{\textwidth}{!}{%
\begin{tabular}{c c l l cc cc cc cc}
\toprule
$\lambda$ & $f_{\rm pump}$ & Upper term & Lower term & \multicolumn{2}{c}{Cut 2 HH529II (MD21b)} & \multicolumn{2}{c}{Cut 3 HH529III (MD21b)} & \multicolumn{2}{c}{Cut 2 neb. (MD21b)} & \multicolumn{2}{c}{Cut 3 neb. (MD21b)} \\
\cmidrule(lr){5-6}\cmidrule(lr){7-8}\cmidrule(lr){9-10}\cmidrule(lr){11-12}
(\AA) & & & & $I(\lambda)$ & $12+\log({\rm O}^{++}/{\rm H}^+)$ & $I(\lambda)$ & $12+\log({\rm O}^{++}/{\rm H}^+)$ & $I(\lambda)$ & $12+\log({\rm O}^{++}/{\rm H}^+)$ & $I(\lambda)$ & $12+\log({\rm O}^{++}/{\rm H}^+)$ \\
\midrule
3882.19 & 0.575 & $3d\;\;{}^{4}D_{7/2}$ & $3p\;\;{}^{4}D_{7/2}^{\circ}$ & --- & --- & $0.100$ & $9.468$ & $0.016\pm0.002$ & $8.691\pm0.056$ & $0.019\pm0.005$ & $8.765\pm0.104$ \\
3954.36 & 1.253 & $3p\;\;{}^{2}P_{1/2}^{\circ}$ & $3s\;\;{}^{2}P_{1/2}$ & --- & --- & --- & --- & $0.013\pm0.002$ & $9.023\pm0.074$ & $0.010\pm0.003$ & $8.909\pm0.122$ \\
3973.26 & 1.017 & $3p\;\;{}^{2}P_{3/2}^{\circ}$ & $3s\;\;{}^{2}P_{3/2}$ & $0.055\pm0.014$ & $9.296\pm0.109$ & $0.038\pm0.009$ & $9.124\pm0.109$ & $0.010\pm0.003$ & $8.563\pm0.117$ & $0.013$ & $8.674$ \\
4069.62 & 0.026 & $3d\;\;{}^{4}F_{3/2}$ & $3p\;\;{}^{4}D_{1/2}^{\circ}$ & --- & --- & --- & --- & $0.024\pm0.002$ & $8.270\pm0.039$ & $0.030\pm0.004$ & $8.372\pm0.061$ \\
4069.88 & 0.033 & $3d\;\;{}^{4}F_{5/2}$ & $3p\;\;{}^{4}D_{3/2}^{\circ}$ & $0.048\pm0.012$ & $8.440\pm0.104$ & $0.036\pm0.001$ & $8.343\pm0.009$ & $0.039\pm0.002$ & $8.316\pm0.026$ & $0.047\pm0.003$ & $8.400\pm0.026$ \\
4072.15 & 0.039 & $3d\;\;{}^{4}F_{7/2}$ & $3p\;\;{}^{4}D_{5/2}^{\circ}$ & $0.100\pm0.008$ & $8.608\pm0.035$ & $0.137\pm0.019$ & $8.759\pm0.061$ & $0.068\pm0.003$ & $8.427\pm0.017$ & $0.068\pm0.003$ & $8.428\pm0.022$ \\
4075.86 & 0.036 & $3d\;\;{}^{4}F_{9/2}$ & $3p\;\;{}^{4}D_{7/2}^{\circ}$ & $0.145\pm0.007$ & $8.666\pm0.022$ & --- & --- & --- & --- & --- & --- \\
4078.84 & 0.026 & $3d\;\;{}^{4}F_{3/2}$ & $3p\;\;{}^{4}D_{3/2}^{\circ}$ & $0.044\pm0.009$ & $9.009\pm0.091$ & --- & --- & $0.016\pm0.002$ & $8.528\pm0.043$ & $0.030\pm0.004$ & $8.805\pm0.056$ \\
4085.11 & 0.034 & $3d\;\;{}^{4}F_{5/2}$ & $3p\;\;{}^{4}D_{5/2}^{\circ}$ & $0.120\pm0.006$ & $9.385\pm0.022$ & $0.113\pm0.012$ & $9.386\pm0.048$ & $0.018\pm0.001$ & $8.527\pm0.030$ & $0.014\pm0.002$ & $8.422\pm0.065$ \\
4092.93 & 0.039 & $3d\;\;{}^{4}F_{7/2}$ & $3p\;\;{}^{4}D_{7/2}^{\circ}$ & $0.011\pm0.004$ & $8.513\pm0.174$ & $0.009$ & $8.440$ & $0.014\pm0.002$ & $8.604\pm0.048$ & $0.010\pm0.002$ & $8.459\pm0.065$ \\
4097.23 & 7.860 & $3d\;\;{}^{4}D_{3/2}$ & $3p\;\;{}^{4}P_{1/2}^{\circ}$ & --- & --- & --- & --- & --- & --- & --- & --- \\
4104.99 & 2.495 & $3d\;\;{}^{4}D_{3/2}$ & $3p\;\;{}^{4}P_{3/2}^{\circ}$ & $0.028$ & $8.732$ & --- & --- & $0.011\pm0.003$ & $8.347\pm0.126$ & $0.014\pm0.003$ & $8.450\pm0.104$ \\
4110.79 & 5.932 & $3d\;\;{}^{4}D_{1/2}$ & $3p\;\;{}^{4}P_{3/2}^{\circ}$ & $0.096\pm0.009$ & $9.579\pm0.039$ & $0.038\pm0.014$ & $9.167\pm0.161$ & $0.015\pm0.002$ & $8.791\pm0.048$ & $0.014\pm0.003$ & $8.759\pm0.083$ \\
4119.22 & 0.566 & $3d\;\;{}^{4}D_{7/2}$ & $3p\;\;{}^{4}P_{5/2}^{\circ}$ & $0.061\pm0.008$ & $8.843\pm0.056$ & --- & --- & $0.035\pm0.002$ & $8.615\pm0.022$ & $0.025\pm0.004$ & $8.468\pm0.074$ \\
4121.46 & 2.578 & $3d\;\;{}^{4}P_{1/2}$ & $3p\;\;{}^{4}P_{1/2}^{\circ}$ & --- & --- & --- & --- & $0.024\pm0.002$ & $8.864\pm0.030$ & $0.021\pm0.002$ & $8.807\pm0.039$ \\
4129.32 & 4.996 & $3d\;\;{}^{4}P_{1/2}$ & $3p\;\;{}^{4}P_{3/2}^{\circ}$ & --- & --- & --- & --- & $0.011\pm0.004$ & $9.344\pm0.139$ & $0.006\pm0.001$ & $9.082\pm0.100$ \\
4132.80 & 2.073 & $3d\;\;{}^{4}P_{3/2}$ & $3p\;\;{}^{4}P_{1/2}^{\circ}$ & $0.067\pm0.008$ & $9.030\pm0.052$ & $0.063\pm0.022$ & $9.025\pm0.152$ & $0.025\pm0.002$ & $8.576\pm0.026$ & $0.022\pm0.002$ & $8.523\pm0.035$ \\
4153.30 & 1.470 & $3d\;\;{}^{4}P_{5/2}$ & $3p\;\;{}^{4}P_{3/2}^{\circ}$ & $0.100\pm0.007$ & $9.055\pm0.030$ & $0.077\pm0.009$ & $8.972\pm0.052$ & $0.041\pm0.002$ & $8.627\pm0.017$ & $0.040\pm0.002$ & $8.621\pm0.022$ \\
4156.53 & 2.436 & $3d\;\;{}^{4}P_{3/2}$ & $3p\;\;{}^{4}P_{5/2}^{\circ}$ & --- & --- & --- & --- & --- & --- & --- & --- \\
4185.44 & 0 & $3d\;\;{}^{2}G_{7/2}$ & $3p\;\;{}^{2}F_{5/2}^{\circ}$ & $0.021\pm0.005$ & $8.034\pm0.113$ & $0.052$ & $8.406$ & $0.016\pm0.002$ & $7.949\pm0.043$ & $0.021\pm0.003$ & $8.064\pm0.061$ \\
4189.79 & 0 & $3d\;\;{}^{2}G_{9/2}$ & $3p\;\;{}^{2}F_{7/2}^{\circ}$ & $0.047\pm0.005$ & $8.637\pm0.048$ & $0.075\pm0.022$ & $8.813\pm0.126$ & $0.020\pm0.001$ & $8.297\pm0.026$ & $0.021\pm0.004$ & $8.313\pm0.074$ \\
4317.14 & 2.919 & $3p\;\;{}^{4}P_{3/2}^{\circ}$ & $3s\;\;{}^{4}P_{1/2}$ & $0.071\pm0.005$ & $9.099\pm0.030$ & --- & --- & $0.028\pm0.001$ & $8.681\pm0.022$ & $0.026\pm0.003$ & $8.650\pm0.048$ \\
4319.63 & 1.810 & $3p\;\;{}^{4}P_{5/2}^{\circ}$ & $3s\;\;{}^{4}P_{3/2}$ & $0.047\pm0.007$ & $9.105\pm0.065$ & $0.038$ & $9.008$ & $0.026\pm0.002$ & $8.852\pm0.030$ & $0.025\pm0.003$ & $8.834\pm0.056$ \\
4325.76 & 2.881 & $3p\;\;{}^{4}P_{1/2}^{\circ}$ & $3s\;\;{}^{4}P_{1/2}$ & $0.033\pm0.008$ & $9.411\pm0.104$ & --- & --- & $0.015\pm0.002$ & $9.053\pm0.048$ & $0.017\pm0.002$ & $9.109\pm0.061$ \\
4336.86 & 2.914 & $3p\;\;{}^{4}P_{3/2}^{\circ}$ & $3s\;\;{}^{4}P_{3/2}$ & --- & --- & --- & --- & --- & --- & --- & --- \\
4345.56 & 2.893 & $3p\;\;{}^{4}P_{1/2}^{\circ}$ & $3s\;\;{}^{4}P_{3/2}$ & $0.111\pm0.011$ & $9.194\pm0.043$ & $0.095\pm0.016$ & $9.139\pm0.074$ & $0.039\pm0.003$ & $8.724\pm0.030$ & $0.034\pm0.002$ & $8.666\pm0.030$ \\
4347.41 & 2.225 & $3p\;\;{}^{2}D_{3/2}^{\circ}$ & $3s\;\;{}^{2}D_{3/2}$ & --- & --- & --- & --- & $0.007\pm0.001$ & $8.590\pm0.078$ & --- & --- \\
4349.43 & 1.768 & $3p\;\;{}^{4}P_{5/2}^{\circ}$ & $3s\;\;{}^{4}P_{5/2}$ & $0.106\pm0.005$ & $9.016\pm0.022$ & $0.102\pm0.012$ & $8.995\pm0.052$ & $0.048\pm0.001$ & $8.676\pm0.013$ & $0.045\pm0.002$ & $8.648\pm0.022$ \\
4351.26 & 2.748 & $3p\;\;{}^{2}D_{5/2}^{\circ}$ & $3s\;\;{}^{2}D_{5/2}$ & $0.018\pm0.005$ & $8.861\pm0.126$ & $0.021$ & $8.914$ & $0.016\pm0.001$ & $8.825\pm0.039$ & $0.027\pm0.004$ & $9.049\pm0.056$ \\
4366.89 & 2.940 & $3p\;\;{}^{4}P_{3/2}^{\circ}$ & $3s\;\;{}^{4}P_{5/2}$ & $0.102\pm0.006$ & $9.237\pm0.026$ & $0.093\pm0.012$ & $9.206\pm0.056$ & $0.032\pm0.002$ & $8.720\pm0.022$ & $0.029\pm0.002$ & $8.678\pm0.030$ \\
4366.90 & 2.940 & $3p\;\;{}^{4}P_{3/2}^{\circ}$ & $3s\;\;{}^{4}P_{5/2}$ & --- & --- & --- & --- & --- & --- & --- & --- \\
4414.90 & 2.409 & $3p\;\;{}^{2}D_{5/2}^{\circ}$ & $3s\;\;{}^{2}P_{3/2}$ & $0.050\pm0.005$ & $8.938\pm0.048$ & --- & --- & $0.031\pm0.002$ & $8.751\pm0.022$ & $0.032\pm0.002$ & $8.761\pm0.026$ \\
4416.97 & 3.028 & $3p\;\;{}^{2}D_{3/2}^{\circ}$ & $3s\;\;{}^{2}P_{1/2}$ & --- & --- & --- & --- & $0.026\pm0.002$ & $8.778\pm0.026$ & $0.024\pm0.002$ & $8.745\pm0.039$ \\
4452.38 & 3.141 & $3p\;\;{}^{2}D_{3/2}^{\circ}$ & $3s\;\;{}^{2}P_{3/2}$ & --- & --- & --- & --- & --- & --- & --- & --- \\
4590.97 & 0.345 & $3p\;\;{}^{2}F_{7/2}^{\circ}$ & $3s\;\;{}^{2}D_{5/2}$ & $0.042\pm0.006$ & $8.594\pm0.061$ & $0.053\pm0.009$ & $8.671\pm0.074$ & $0.024\pm0.002$ & $8.377\pm0.030$ & $0.024\pm0.002$ & $8.373\pm0.030$ \\
4595.96 & 0.154 & $3p\;\;{}^{2}F_{5/2}^{\circ}$ & $3s\;\;{}^{2}D_{5/2}$ & $0.020\pm0.006$ & $9.336\pm0.130$ & --- & --- & $0.020\pm0.002$ & $9.364\pm0.039$ & $0.015\pm0.003$ & $9.236\pm0.078$ \\
4596.18 & 0.162 & $3p\;\;{}^{2}F_{5/2}^{\circ}$ & $3s\;\;{}^{2}D_{3/2}$ & --- & --- & --- & --- & --- & --- & --- & --- \\
4638.86 & 1.084 & $3p\;\;{}^{4}D_{3/2}^{\circ}$ & $3s\;\;{}^{4}P_{1/2}$ & $0.098\pm0.006$ & $8.938\pm0.026$ & $0.059\pm0.015$ & $8.742\pm0.113$ & $0.045\pm0.002$ & $8.567\pm0.017$ & $0.046\pm0.007$ & $8.580\pm0.065$ \\
4641.81 & 0.855 & $3p\;\;{}^{4}D_{5/2}^{\circ}$ & $3s\;\;{}^{4}P_{3/2}$ & $0.164\pm0.007$ & $8.809\pm0.017$ & $0.165\pm0.010$ & $8.825\pm0.026$ & $0.089\pm0.002$ & $8.527\pm0.009$ & $0.085\pm0.003$ & $8.508\pm0.013$ \\
4649.13 & 0.669 & $3p\;\;{}^{4}D_{7/2}^{\circ}$ & $3s\;\;{}^{4}P_{5/2}$ & $0.258\pm0.008$ & $8.793\pm0.013$ & $0.265\pm0.016$ & $8.783\pm0.026$ & $0.122\pm0.002$ & $8.500\pm0.009$ & $0.129\pm0.004$ & $8.520\pm0.013$ \\
4649.14 & 0.669 & $3p\;\;{}^{4}D_{7/2}^{\circ}$ & $3s\;\;{}^{4}P_{5/2}$ & --- & --- & --- & --- & --- & --- & --- & --- \\
4650.84 & 1.219 & $3p\;\;{}^{4}D_{1/2}^{\circ}$ & $3s\;\;{}^{4}P_{1/2}$ & $0.095\pm0.006$ & $8.918\pm0.026$ & $0.111\pm0.011$ & $9.013\pm0.043$ & $0.043\pm0.002$ & $8.534\pm0.017$ & $0.043\pm0.002$ & $8.538\pm0.022$ \\
4661.63 & 1.079 & $3p\;\;{}^{4}D_{3/2}^{\circ}$ & $3s\;\;{}^{4}P_{3/2}$ & $0.133\pm0.008$ & $9.022\pm0.026$ & $0.113\pm0.008$ & $8.976\pm0.030$ & $0.051\pm0.002$ & $8.572\pm0.017$ & $0.056\pm0.004$ & $8.616\pm0.035$ \\
4673.73 & 1.285 & $3p\;\;{}^{4}D_{1/2}^{\circ}$ & $3s\;\;{}^{4}P_{3/2}$ & $0.022\pm0.006$ & $9.040\pm0.126$ & $0.017$ & $8.956$ & $0.007\pm0.001$ & $8.503\pm0.069$ & $0.014\pm0.002$ & $8.808\pm0.061$ \\
4676.23 & 0.861 & $3p\;\;{}^{4}D_{5/2}^{\circ}$ & $3s\;\;{}^{4}P_{5/2}$ & $0.049\pm0.005$ & $8.747\pm0.048$ & $0.072\pm0.012$ & $8.927\pm0.069$ & $0.028\pm0.001$ & $8.487\pm0.022$ & $0.029\pm0.002$ & $8.503\pm0.026$ \\
4699.01 & 7.953 & $3d\;\;{}^{2}F_{7/2}$ & $3p\;\;{}^{2}D_{5/2}^{\circ}$ & --- & --- & --- & --- & --- & --- & --- & --- \\
4699.22 & 13.122 & $3d\;\;{}^{2}F_{5/2}$ & $3p\;\;{}^{2}D_{3/2}^{\circ}$ & --- & --- & --- & --- & $0.012\pm0.004$ & $9.225\pm0.143$ & --- & --- \\
4705.35 & 10.722 & $3d\;\;{}^{2}F_{7/2}$ & $3p\;\;{}^{2}D_{5/2}^{\circ}$ & --- & --- & $0.060\pm0.014$ & $9.807\pm0.100$ & $0.018\pm0.001$ & $9.329\pm0.022$ & $0.016\pm0.003$ & $9.276\pm0.069$ \\
4890.86 & 5.752 & $3d\;\;{}^{4}P_{1/2}$ & $3p\;\;{}^{4}S_{3/2}^{\circ}$ & --- & --- & --- & --- & $0.012\pm0.002$ & $9.027\pm0.056$ & $0.010\pm0.002$ & $8.948\pm0.069$ \\
4924.53 & 1.778 & $3d\;\;{}^{4}P_{5/2}$ & $3p\;\;{}^{4}S_{3/2}^{\circ}$ & --- & --- & --- & --- & --- & --- & --- & --- \\
6721.39 & 2.474 & $3p\;\;{}^{2}S_{1/2}^{\circ}$ & $3s\;\;{}^{2}P_{3/2}$ & $0.005$ & $8.885$ & --- & --- & $0.004\pm0.001$ & $8.779\pm0.078$ & $0.003\pm0.001$ & $8.653\pm0.117$ \\
6895.10 & 0.268 & $4p\;\;{}^{4}D_{7/2}^{\circ}$ & $3d\;\;{}^{4}F_{9/2}$ & $0.022$ & $8.959$ & --- & --- & $0.017\pm0.006$ & $8.882\pm0.156$ & $0.008\pm0.001$ & $8.551\pm0.074$ \\
6906.44 & 0.221 & $4p\;\;{}^{4}D_{5/2}^{\circ}$ & $3d\;\;{}^{4}F_{7/2}$ & $0.011\pm0.004$ & $8.817\pm0.143$ & $0.006$ & $8.566$ & $0.007\pm0.001$ & $8.600\pm0.048$ & $0.006\pm0.001$ & $8.534\pm0.074$ \\
\multicolumn{4}{l}{\it V1 Abundance} & --- & $8.848 \pm 0.008$ & --- & $8.869 \pm 0.014$ & --- & $8.523 \pm 0.005$ & --- & $8.527 \pm 0.008$ \\
\multicolumn{4}{l}{\it Corrected Abundance} & --- & $8.622_{-0.074}^{+0.033}$ & --- & $8.457_{-0.072}^{+0.087}$ & --- & $8.310_{-0.025}^{+0.075}$ & --- & $8.400_{-0.033}^{+0.022}$ \\
\bottomrule
\end{tabular}%
}

\vspace{2mm}
\begin{minipage}{0.95\textwidth}
\footnotesize
\textit{Notes.} Literature dereddened intensities ($I$) are normalized to $I({\rm H}\beta)=100$. Lines without reliable $f_\mathrm{pump}$ calculations are omitted. Literature sources: B06 \cite{blagrave06}, MD21a \cite{mendez-delgado21a}, MD21b \cite{mendez-delgado21b}.
\end{minipage}
\end{sidewaystable*}

\begin{table*}
\centering\caption{\textbf{Recalculating M42 (E04) \Oiirec\ RL abundances.}}
\label{tab:new_lit_m42}
\resizebox{0.7\textwidth}{!}{

\begin{tabular}{c c l l c c}
\toprule
$\lambda$ (\AA) & $f_{\mathrm{pump}}$ & Upper term & Lower term & $I$ & $12+\log(\mathrm{\Opp/\Hp})$ \\
\midrule
3749.48 & 0.64 & $3p\;\;{}^{4}S_{3/2}^{\circ}$ & $3s\;\;{}^{4}P_{5/2}$ & $0.115 \pm 0.021$ & $9.042 \pm 0.079$ \\
3973.26 & 0.082 & $3p\;\;{}^{2}P_{3/2}^{\circ}$ & $3s\;\;{}^{2}P_{3/2}$ & $0.020 \pm 0.007$ & $8.884 \pm 0.16$ \\
4069.62 & 0.0086 & $3d\;\;{}^{4}F_{3/2}$ & $3p\;\;{}^{4}D_{1/2}^{\circ}$ & $0.086 \pm 0.007$ & $8.741 \pm 0.035$ \\
4072.15 & 0.010  & $3d\;\;{}^{4}F_{7/2}$ & $3p\;\;{}^{4}D_{5/2}^{\circ}$ & $0.067 \pm 0.006$ & $8.401 \pm 0.039$ \\
4075.86 & 0.011  & $3d\;\;{}^{4}F_{9/2}$ & $3p\;\;{}^{4}D_{7/2}^{\circ}$ & $0.079 \pm 0.006$ & $8.528 \pm 0.035$ \\
4085.11 & 0.0094 & $3d\;\;{}^{4}F_{5/2}$ & $3p\;\;{}^{4}D_{5/2}^{\circ}$ & $0.013 \pm 0.004$ & $8.32 \pm 0.13$ \\
4104.99 & 0.85   & $3d\;\;{}^{4}D_{3/2}$ & $3p\;\;{}^{4}P_{3/2}^{\circ}$ & $0.024 \pm 0.005$ & $8.725 \pm 0.084$ \\
4110.79 & 1.8    & $3d\;\;{}^{4}D_{1/2}$ & $3p\;\;{}^{4}P_{3/2}^{\circ}$ & $0.024 \pm 0.005$ & $9.033 \pm 0.084$ \\
4119.22 & 0.056  & $3d\;\;{}^{4}D_{7/2}$ & $3p\;\;{}^{4}P_{5/2}^{\circ}$ & $0.031 \pm 0.005$ & $8.591 \pm 0.070$ \\
4121.46 & 0.78   & $3d\;\;{}^{4}P_{1/2}$ & $3p\;\;{}^{4}P_{1/2}^{\circ}$ & $0.041 \pm 0.005$ & $9.091 \pm 0.057$ \\
4132.80 & 0.58   & $3d\;\;{}^{4}P_{3/2}$ & $3p\;\;{}^{4}P_{1/2}^{\circ}$ & $0.033 \pm 0.005$ & $8.651 \pm 0.066$ \\
4153.30 & 0.37   & $3d\;\;{}^{4}P_{5/2}$ & $3p\;\;{}^{4}P_{3/2}^{\circ}$ & $0.076 \pm 0.006$ & $8.812 \pm 0.035$ \\
4185.44 & $0$ & $3d\;\;{}^{2}G_{7/2}$ & $3p\;\;{}^{2}F_{5/2}^{\circ}$ & $0.021 \pm 0.004$ & $8.140 \pm 0.093$ \\
4189.79 & $0$ & $3d\;\;{}^{2}G_{9/2}$ & $3p\;\;{}^{2}F_{7/2}^{\circ}$ & $0.025 \pm 0.005$ & $8.472 \pm 0.079$ \\
4317.14 & 0.71   & $3p\;\;{}^{4}P_{3/2}^{\circ}$ & $3s\;\;{}^{4}P_{1/2}$ & $0.044 \pm 0.005$ & $8.845 \pm 0.052$ \\
4319.63 & 0.43   & $3p\;\;{}^{4}P_{5/2}^{\circ}$ & $3s\;\;{}^{4}P_{3/2}$ & $0.025 \pm 0.005$ & $8.846 \pm 0.079$ \\
4325.76 & 0.80   & $3p\;\;{}^{4}P_{1/2}^{\circ}$ & $3s\;\;{}^{4}P_{1/2}$ & $0.017 \pm 0.004$ & $9.08 \pm 0.11$ \\
4349.43 & 0.42   & $3p\;\;{}^{4}P_{5/2}^{\circ}$ & $3s\;\;{}^{4}P_{5/2}$ & $0.065 \pm 0.006$ & $8.819 \pm 0.039$ \\
4366.89 & 0.72   & $3p\;\;{}^{4}P_{3/2}^{\circ}$ & $3s\;\;{}^{4}P_{5/2}$ & $0.048 \pm 0.005$ & $8.864 \pm 0.048$ \\
4414.90 & 0.28   & $3p\;\;{}^{2}D_{5/2}^{\circ}$ & $3s\;\;{}^{2}P_{3/2}$ & $0.036 \pm 0.006$ & $8.867 \pm 0.070$ \\
4416.97 & 0.27   & $3p\;\;{}^{2}D_{3/2}^{\circ}$ & $3s\;\;{}^{2}P_{1/2}$ & $0.028 \pm 0.005$ & $8.766 \pm 0.070$ \\
4590.97 & 0.024  & $3p\;\;{}^{2}F_{7/2}^{\circ}$ & $3s\;\;{}^{2}D_{5/2}$ & $0.025 \pm 0.004$ & $8.462 \pm 0.075$ \\
4595.96 & 0.0068 & $3p\;\;{}^{2}F_{5/2}^{\circ}$ & $3s\;\;{}^{2}D_{5/2}$ & $0.020 \pm 0.004$ & $9.426 \pm 0.088$ \\
4638.86 & 0.22   & $3p\;\;{}^{4}D_{3/2}^{\circ}$ & $3s\;\;{}^{4}P_{1/2}$ & $0.057 \pm 0.005$ & $8.607 \pm 0.039$ \\
4641.81 & 0.19   & $3p\;\;{}^{4}D_{5/2}^{\circ}$ & $3s\;\;{}^{4}P_{3/2}$ & $0.102 \pm 0.005$ & $8.561 \pm 0.022$ \\
4649.13 & 0.19   & $3p\;\;{}^{4}D_{7/2}^{\circ}$ & $3s\;\;{}^{4}P_{5/2}$ & $0.155 \pm 0.005$ & $8.677 \pm 0.013$ \\
4650.84 & 0.24   & $3p\;\;{}^{4}D_{1/2}^{\circ}$ & $3s\;\;{}^{4}P_{1/2}$ & $0.052 \pm 0.005$ & $8.539 \pm 0.044$ \\
4661.63 & 0.22   & $3p\;\;{}^{4}D_{3/2}^{\circ}$ & $3s\;\;{}^{4}P_{3/2}$ & $0.068 \pm 0.005$ & $8.635 \pm 0.035$ \\
4673.73 & 0.25   & $3p\;\;{}^{4}D_{1/2}^{\circ}$ & $3s\;\;{}^{4}P_{3/2}$ & $0.011 \pm 0.003$ & $8.62 \pm 0.13$ \\
4676.23 & 0.19   & $3p\;\;{}^{4}D_{5/2}^{\circ}$ & $3s\;\;{}^{4}P_{5/2}$ & $0.035 \pm 0.005$ & $8.558 \pm 0.057$ \\
4699.22 & 1.3    & $3d\;\;{}^{2}F_{5/2}$ & $3p\;\;{}^{2}D_{3/2}^{\circ}$ & $0.010 \pm 0.003$ & $9.14 \pm 0.14$ \\
4705.35 & 1.4    & $3d\;\;{}^{2}F_{7/2}$ & $3p\;\;{}^{2}D_{5/2}^{\circ}$ & $0.018 \pm 0.004$ & $9.390 \pm 0.093$ \\
4752.69 & 0.59   & $3d\;\;{}^{4}D_{5/2}$ & $3p\;\;{}^{2}D_{5/2}^{\circ}$ & $0.010 \pm 0.003$ & $10.42 \pm 0.14$ \\
4890.86 & 1.7    & $3d\;\;{}^{4}P_{1/2}$ & $3p\;\;{}^{4}S_{3/2}^{\circ}$ & $0.022 \pm 0.004$ & $9.284 \pm 0.084$ \\
\midrule
\multicolumn{4}{l}{\it V1 Abundance} & --- & $8.610 \pm 0.011$ \\
\multicolumn{4}{l}{\it Corrected Abundance} & --- & $8.418_{-0.036}^{+0.084}$ \\
\bottomrule
\end{tabular}}
\end{table*}

\begin{table*}
\centering
\caption{\textbf{Recalculating NGC~3576 \Oiirec\ RL abundances.}}
\label{tab:new_lit_ngc3576}
\begin{tabular}{c c l l c c}
\toprule
$\lambda$ (\AA) & $f_{\mathrm{pump}}$ & Upper term & Lower term & $I$ & $12+\log(\mathrm{\Opp/\Hp})$ \\
\midrule
4071.93 & 0.010 & $3d\;\;{}^{4}F_{7/2}$ & $3p\;\;{}^{4}D_{5/2}^{\circ}$ & $0.085 \pm 0.008$ & $8.508 \pm 0.039$ \\
4075.62 & 0.016 & $3d\;\;{}^{4}F_{9/2}$ & $3p\;\;{}^{4}D_{7/2}^{\circ}$ & $0.108 \pm 0.008$ & $8.806 \pm 0.035$ \\
4083.52 & 0.0040 & $4f\;\;G\,{}^{2}[4]_{7/2}^{\circ}$ & $3d\;\;{}^{4}F_{5/2}$ & $0.017 \pm 0.005$ & $8.55 \pm 0.12$ \\
4084.90 & 0.0085 & $3d\;\;{}^{4}F_{5/2}$ & $3p\;\;{}^{4}D_{5/2}^{\circ}$ & $0.024 \pm 0.005$ & $8.518 \pm 0.097$ \\
4119.05 & 0.020 & $3d\;\;{}^{4}D_{7/2}$ & $3p\;\;{}^{4}P_{5/2}^{\circ}$ & $0.023 \pm 0.005$ & $8.516 \pm 0.097$ \\
4132.57 & 0.54  & $3d\;\;{}^{4}P_{3/2}$ & $3p\;\;{}^{4}P_{1/2}^{\circ}$ & $0.035 \pm 0.006$ & $8.632 \pm 0.075$ \\
4153.05 & 0.30  & $3d\;\;{}^{4}P_{5/2}$ & $3p\;\;{}^{4}P_{3/2}^{\circ}$ & $0.043 \pm 0.006$ & $8.447 \pm 0.061$ \\
4185.22 & $0$ & $3d\;\;{}^{2}G_{7/2}$ & $3p\;\;{}^{2}F_{5/2}^{\circ}$ & $0.034 \pm 0.006$ & $8.455 \pm 0.075$ \\
4189.49 & $0$ & $3d\;\;{}^{2}G_{9/2}$ & $3p\;\;{}^{2}F_{7/2}^{\circ}$ & $0.026 \pm 0.005$ & $8.612 \pm 0.088$ \\
4316.86 & 0.64  & $3p\;\;{}^{4}P_{3/2}^{\circ}$ & $3s\;\;{}^{4}P_{1/2}$ & $0.026 \pm 0.005$ & $8.558 \pm 0.084$ \\
4319.35 & 0.43  & $3p\;\;{}^{4}P_{5/2}^{\circ}$ & $3s\;\;{}^{4}P_{3/2}$ & $0.019 \pm 0.004$ & $8.74 \pm 0.10$ \\
4349.18 & 0.42  & $3p\;\;{}^{4}P_{5/2}^{\circ}$ & $3s\;\;{}^{4}P_{5/2}$ & $0.067 \pm 0.007$ & $8.847 \pm 0.044$ \\
4366.62 & 0.64  & $3p\;\;{}^{4}P_{3/2}^{\circ}$ & $3s\;\;{}^{4}P_{5/2}$ & $0.040 \pm 0.006$ & $8.726 \pm 0.061$ \\
4414.64 & 0.10  & $3p\;\;{}^{2}D_{5/2}^{\circ}$ & $3s\;\;{}^{2}P_{3/2}$ & $0.029 \pm 0.006$ & $8.853 \pm 0.088$ \\
4416.71 & 0.16  & $3p\;\;{}^{2}D_{3/2}^{\circ}$ & $3s\;\;{}^{2}P_{1/2}$ & $0.031 \pm 0.006$ & $8.749 \pm 0.084$ \\
4590.67 & 0.0083 & $3p\;\;{}^{2}F_{7/2}^{\circ}$ & $3s\;\;{}^{2}D_{5/2}$ & $0.029 \pm 0.005$ & $8.629 \pm 0.075$ \\
4638.57 & 0.20  & $3p\;\;{}^{4}D_{3/2}^{\circ}$ & $3s\;\;{}^{4}P_{1/2}$ & $0.074 \pm 0.006$ & $8.642 \pm 0.035$ \\
4641.53 & 0.19  & $3p\;\;{}^{4}D_{5/2}^{\circ}$ & $3s\;\;{}^{4}P_{3/2}$ & $0.132 \pm 0.008$ & $8.660 \pm 0.026$ \\
4648.86 & 0.24  & $3p\;\;{}^{4}D_{7/2}^{\circ}$ & $3s\;\;{}^{4}P_{5/2}$ & $0.145 \pm 0.007$ & $8.767 \pm 0.022$ \\
4650.54 & 0.20  & $3p\;\;{}^{4}D_{1/2}^{\circ}$ & $3s\;\;{}^{4}P_{1/2}$ & $0.069 \pm 0.006$ & $8.552 \pm 0.039$ \\
4673.39 & 0.21  & $3p\;\;{}^{4}D_{1/2}^{\circ}$ & $3s\;\;{}^{4}P_{3/2}$ & $0.013 \pm 0.004$ & $8.58 \pm 0.12$ \\
4675.92 & 0.19  & $3p\;\;{}^{4}D_{5/2}^{\circ}$ & $3s\;\;{}^{4}P_{5/2}$ & $0.040 \pm 0.005$ & $8.604 \pm 0.057$ \\
4705.09 & 0.44  & $3d\;\;{}^{2}F_{7/2}$ & $3p\;\;{}^{2}D_{5/2}^{\circ}$ & $0.008 \pm 0.003$ & $9.13 \pm 0.17$ \\
4752.60 & 0.42  & $3d\;\;{}^{4}D_{5/2}$ & $3p\;\;{}^{2}D_{5/2}^{\circ}$ & $0.008 \pm 0.003$ & $10.28 \pm 0.16$ \\
6720.92 & 0.11  & $3p\;\;{}^{2}S_{1/2}^{\circ}$ & $3s\;\;{}^{2}P_{3/2}$ & $0.004 \pm 0.001$ & $8.74 \pm 0.12$ \\
\midrule
\multicolumn{4}{l}{\it V1 Abundance} & --- & $8.661 \pm 0.014$ \\
\multicolumn{4}{l}{\it Corrected Abundance} & --- & $8.557_{-0.045}^{+0.067}$ \\
\bottomrule
\end{tabular}
\end{table*}

\begin{table*}
\centering
\caption{\textbf{Recalculating M8 \Oiirec\ RL abundances.}}
\label{tab:new_lit_m8}
\begin{tabular}{c c l l c c}
\toprule
$\lambda$ (\AA) & $f_{\mathrm{pump}}$ & Upper term & Lower term & $I$ & $12+\log(\mathrm{\Opp/\Hp})$ \\
\midrule
4069.62 & 0.0094 & $3d\;\;{}^{4}F_{3/2}$ & $3p\;\;{}^{4}D_{1/2}^{\circ}$ & $0.064 \pm 0.008$ & $8.578 \pm 0.052$ \\
4072.15 & 0.011  & $3d\;\;{}^{4}F_{7/2}$ & $3p\;\;{}^{4}D_{5/2}^{\circ}$ & $0.031 \pm 0.006$ & $8.062 \pm 0.084$ \\
4121.46 & 0.94   & $3d\;\;{}^{4}P_{1/2}$ & $3p\;\;{}^{4}P_{1/2}^{\circ}$ & $0.012 \pm 0.005$ & $8.56 \pm 0.18$ \\
4132.80 & 0.69   & $3d\;\;{}^{4}P_{3/2}$ & $3p\;\;{}^{4}P_{1/2}^{\circ}$ & $0.014 \pm 0.005$ & $8.26 \pm 0.16$ \\
4153.30 & 0.42   & $3d\;\;{}^{4}P_{5/2}$ & $3p\;\;{}^{4}P_{3/2}^{\circ}$ & $0.028 \pm 0.005$ & $8.343 \pm 0.084$ \\
4317.14 & 0.84   & $3p\;\;{}^{4}P_{3/2}^{\circ}$ & $3s\;\;{}^{4}P_{1/2}$ & $0.011 \pm 0.004$ & $8.23 \pm 0.17$ \\
4349.43 & 0.50   & $3p\;\;{}^{4}P_{5/2}^{\circ}$ & $3s\;\;{}^{4}P_{5/2}$ & $0.020 \pm 0.005$ & $8.31 \pm 0.11$ \\
4366.89 & 0.84   & $3p\;\;{}^{4}P_{3/2}^{\circ}$ & $3s\;\;{}^{4}P_{5/2}$ & $0.015 \pm 0.004$ & $8.34 \pm 0.13$ \\
4638.86 & 0.26   & $3p\;\;{}^{4}D_{3/2}^{\circ}$ & $3s\;\;{}^{4}P_{1/2}$ & $0.034 \pm 0.005$ & $8.360 \pm 0.061$ \\
4641.81 & 0.23   & $3p\;\;{}^{4}D_{5/2}^{\circ}$ & $3s\;\;{}^{4}P_{3/2}$ & $0.043 \pm 0.005$ & $8.179 \pm 0.052$ \\
4649.13 & 0.24   & $3p\;\;{}^{4}D_{7/2}^{\circ}$ & $3s\;\;{}^{4}P_{5/2}$ & $0.041 \pm 0.005$ & $8.132 \pm 0.052$ \\
4650.84 & 0.27   & $3p\;\;{}^{4}D_{1/2}^{\circ}$ & $3s\;\;{}^{4}P_{1/2}$ & $0.032 \pm 0.005$ & $8.297 \pm 0.066$ \\
4661.63 & 0.25   & $3p\;\;{}^{4}D_{3/2}^{\circ}$ & $3s\;\;{}^{4}P_{3/2}$ & $0.036 \pm 0.005$ & $8.336 \pm 0.057$ \\
4676.23 & 0.23   & $3p\;\;{}^{4}D_{5/2}^{\circ}$ & $3s\;\;{}^{4}P_{5/2}$ & $0.017 \pm 0.004$ & $8.24 \pm 0.11$ \\
\midrule
\multicolumn{4}{l}{\it V1 Abundance} & --- & $8.243 \pm 0.025$ \\
\multicolumn{4}{l}{\it Corrected Abundance} & --- & $8.11_{-0.06}^{+0.16}$ \\
\bottomrule
\end{tabular}
\end{table*}

\begin{table*}
\centering
\caption{\textbf{Recalculating M17 \Oiirec\ RL abundances.}}
\label{tab:new_lit_m17}
\begin{tabular}{c c l l c c}
\toprule
$\lambda$ (\AA) & $f_{\mathrm{pump}}$ & Upper term & Lower term & $I$ & $12+\log(\mathrm{\Opp/\Hp})$ \\
\midrule
4069.62 & 0.0065 & $3d\;\;{}^{4}F_{3/2}$ & $3p\;\;{}^{4}D_{1/2}^{\circ}$ & $0.190 \pm 0.027$ & $8.884 \pm 0.061$ \\
4072.15 & 0.012  & $3d\;\;{}^{4}F_{7/2}$ & $3p\;\;{}^{4}D_{5/2}^{\circ}$ & $0.091 \pm 0.022$ & $8.57 \pm 0.11$ \\
4075.86 & 0.022  & $3d\;\;{}^{4}F_{9/2}$ & $3p\;\;{}^{4}D_{7/2}^{\circ}$ & $0.087 \pm 0.022$ & $8.82 \pm 0.11$ \\
4153.30 & 0.28   & $3d\;\;{}^{4}P_{5/2}$ & $3p\;\;{}^{4}P_{3/2}^{\circ}$ & $0.093 \pm 0.021$ & $8.69 \pm 0.10$ \\
4317.14 & 0.66   & $3p\;\;{}^{4}P_{3/2}^{\circ}$ & $3s\;\;{}^{4}P_{1/2}$ & $0.061 \pm 0.018$ & $8.87 \pm 0.13$ \\
4349.43 & 0.50   & $3p\;\;{}^{4}P_{5/2}^{\circ}$ & $3s\;\;{}^{4}P_{5/2}$ & $0.066 \pm 0.019$ & $8.85 \pm 0.13$ \\
4638.86 & 0.20   & $3p\;\;{}^{4}D_{3/2}^{\circ}$ & $3s\;\;{}^{4}P_{1/2}$ & $0.093 \pm 0.018$ & $8.688 \pm 0.084$ \\
4641.81 & 0.23   & $3p\;\;{}^{4}D_{5/2}^{\circ}$ & $3s\;\;{}^{4}P_{3/2}$ & $0.128 \pm 0.019$ & $8.661 \pm 0.066$ \\
4649.13 & 0.35   & $3p\;\;{}^{4}D_{7/2}^{\circ}$ & $3s\;\;{}^{4}P_{5/2}$ & $0.123 \pm 0.019$ & $8.788 \pm 0.066$ \\
4650.84 & 0.19   & $3p\;\;{}^{4}D_{1/2}^{\circ}$ & $3s\;\;{}^{4}P_{1/2}$ & $0.100 \pm 0.018$ & $8.631 \pm 0.079$ \\
4661.63 & 0.20   & $3p\;\;{}^{4}D_{3/2}^{\circ}$ & $3s\;\;{}^{4}P_{3/2}$ & $0.119 \pm 0.018$ & $8.746 \pm 0.066$ \\
4676.23 & 0.23   & $3p\;\;{}^{4}D_{5/2}^{\circ}$ & $3s\;\;{}^{4}P_{5/2}$ & $0.044 \pm 0.014$ & $8.66 \pm 0.14$ \\
\midrule
\multicolumn{4}{l}{\it V1 Abundance} & --- & $8.699 \pm 0.031$ \\
\multicolumn{4}{l}{\it Corrected Abundance} & --- & $8.68_{-0.15}^{+0.09}$ \\
\bottomrule
\end{tabular}
\end{table*}

\begin{table*}
\centering
\caption{\textbf{Recalculating 30 Dor \Oiirec\ RL abundances.}}
\label{tab:new_lit_30dor}
\resizebox{0.8\textwidth}{!}{

\begin{tabular}{c c l l c c}
\toprule
$\lambda$ (\AA) & $f_{\mathrm{pump}}$ & Upper term & Lower term & $I$ & $12+\log(\mathrm{\Opp/\Hp})$ \\
\midrule
3875.80 & 0.17  & $3d\;\;{}^{2}F_{5/2}$ & $3p\;\;{}^{4}D_{7/2}^{\circ}$ & $0.0187 \pm 0.0037$ & $10.462 \pm 0.088$ \\
4069.62 & 0.0030 & $3d\;\;{}^{4}F_{3/2}$ & $3p\;\;{}^{4}D_{1/2}^{\circ}$ & $0.104 \pm 0.010$ & $8.553 \pm 0.044$ \\
4072.15 & 0.0075 & $3d\;\;{}^{4}F_{7/2}$ & $3p\;\;{}^{4}D_{5/2}^{\circ}$ & $0.0534 \pm 0.0064$ & $8.416 \pm 0.052$ \\
4078.84 & 0.0030 & $3d\;\;{}^{4}F_{3/2}$ & $3p\;\;{}^{4}D_{3/2}^{\circ}$ & $0.0272 \pm 0.0041$ & $8.406 \pm 0.066$ \\
4110.79 & 1.7    & $3d\;\;{}^{4}D_{1/2}$ & $3p\;\;{}^{4}P_{3/2}^{\circ}$ & $0.0261 \pm 0.0039$ & $9.258 \pm 0.066$ \\
4112.02 & 0.14   & $3d\;\;{}^{2}F_{5/2}$ & $3p\;\;{}^{4}P_{5/2}^{\circ}$ & $0.0143 \pm 0.0029$ & $9.551 \pm 0.088$ \\
4119.22 & 0.0077 & $3d\;\;{}^{4}D_{7/2}$ & $3p\;\;{}^{4}P_{5/2}^{\circ}$ & $0.0164 \pm 0.0033$ & $8.502 \pm 0.088$ \\
4132.80 & 0.41   & $3d\;\;{}^{4}P_{3/2}$ & $3p\;\;{}^{4}P_{1/2}^{\circ}$ & $0.0293 \pm 0.0044$ & $8.537 \pm 0.066$ \\
4153.30 & 0.17   & $3d\;\;{}^{4}P_{5/2}$ & $3p\;\;{}^{4}P_{3/2}^{\circ}$ & $0.0454 \pm 0.0054$ & $8.309 \pm 0.052$ \\
4156.53 & 0.48   & $3d\;\;{}^{4}P_{3/2}$ & $3p\;\;{}^{4}P_{5/2}^{\circ}$ & $0.0060 \pm 0.0018$ & $8.56 \pm 0.13$ \\
4185.44 & 0.00   & $3d\;\;{}^{2}G_{7/2}$ & $3p\;\;{}^{2}F_{5/2}^{\circ}$ & $0.0147 \pm 0.0029$ & $8.192 \pm 0.088$ \\
4319.63 & 0.33   & $3p\;\;{}^{4}P_{5/2}^{\circ}$ & $3s\;\;{}^{4}P_{3/2}$ & $0.0087 \pm 0.0022$ & $8.42 \pm 0.11$ \\
4349.43 & 0.32   & $3p\;\;{}^{4}P_{5/2}^{\circ}$ & $3s\;\;{}^{4}P_{5/2}$ & $0.0314 \pm 0.0047$ & $8.539 \pm 0.066$ \\
4366.89 & 0.37   & $3p\;\;{}^{4}P_{3/2}^{\circ}$ & $3s\;\;{}^{4}P_{5/2}$ & $0.0245 \pm 0.0037$ & $8.407 \pm 0.066$ \\
4416.97 & 0.078  & $3p\;\;{}^{2}D_{3/2}^{\circ}$ & $3s\;\;{}^{2}P_{1/2}$ & $0.0188 \pm 0.0038$ & $8.440 \pm 0.088$ \\
4452.38 & 0.081  & $3p\;\;{}^{2}D_{3/2}^{\circ}$ & $3s\;\;{}^{2}P_{3/2}$ & $0.0195 \pm 0.0039$ & $9.194 \pm 0.088$ \\
4590.97 & 0.0020 & $3p\;\;{}^{2}F_{7/2}^{\circ}$ & $3s\;\;{}^{2}D_{5/2}$ & $0.0089 \pm 0.0022$ & $8.25 \pm 0.11$ \\
4595.96 & 0.00030 & $3p\;\;{}^{2}F_{5/2}^{\circ}$ & $3s\;\;{}^{2}D_{5/2}$ & $0.0133 \pm 0.0027$ & $9.401 \pm 0.088$ \\
4638.86 & 0.10   & $3p\;\;{}^{4}D_{3/2}^{\circ}$ & $3s\;\;{}^{4}P_{1/2}$ & $0.0500 \pm 0.0050$ & $8.373 \pm 0.044$ \\
4641.81 & 0.14   & $3p\;\;{}^{4}D_{5/2}^{\circ}$ & $3s\;\;{}^{4}P_{3/2}$ & $0.0648 \pm 0.0065$ & $8.399 \pm 0.044$ \\
4649.13 & 0.25   & $3p\;\;{}^{4}D_{7/2}^{\circ}$ & $3s\;\;{}^{4}P_{5/2}$ & $0.0536 \pm 0.0054$ & $8.525 \pm 0.044$ \\
4650.84 & 0.090  & $3p\;\;{}^{4}D_{1/2}^{\circ}$ & $3s\;\;{}^{4}P_{1/2}$ & $0.0623 \pm 0.0062$ & $8.352 \pm 0.044$ \\
4661.63 & 0.10   & $3p\;\;{}^{4}D_{3/2}^{\circ}$ & $3s\;\;{}^{4}P_{3/2}$ & $0.0725 \pm 0.0073$ & $8.486 \pm 0.044$ \\
4673.73 & 0.094  & $3p\;\;{}^{4}D_{1/2}^{\circ}$ & $3s\;\;{}^{4}P_{3/2}$ & $0.0107 \pm 0.0027$ & $8.34 \pm 0.11$ \\
4676.23 & 0.14   & $3p\;\;{}^{4}D_{5/2}^{\circ}$ & $3s\;\;{}^{4}P_{5/2}$ & $0.0211 \pm 0.0032$ & $8.374 \pm 0.066$ \\
4696.35 & 0.20   & $3p\;\;{}^{4}D_{3/2}^{\circ}$ & $3s\;\;{}^{4}P_{5/2}$ & $0.0035 \pm 0.0012$ & $8.29 \pm 0.16$ \\
4703.16 & 0.0045 & $3d\;\;{}^{2}F_{5/2}$ & $3p\;\;{}^{2}D_{3/2}^{\circ}$ & $0.0425 \pm 0.0051$ & $9.691 \pm 0.052$ \\
4751.28 & 0.0061 & $3d\;\;{}^{4}D_{7/2}$ & $3p\;\;{}^{2}D_{5/2}^{\circ}$ & $0.0456 \pm 0.0055$ & $10.248 \pm 0.052$ \\
4924.53 & 0.20   & $3d\;\;{}^{4}P_{5/2}$ & $3p\;\;{}^{4}S_{3/2}^{\circ}$ & $0.0341 \pm 0.0041$ & $8.469 \pm 0.052$ \\
\midrule
\multicolumn{4}{l}{\it V1 Abundance} & --- & $8.417 \pm 0.019$ \\
\multicolumn{4}{l}{\it Corrected Abundance} & --- & $8.337_{-0.060}^{+0.063}$ \\
\bottomrule
\end{tabular}}

\end{table*}

\clearpage

\bibliography{references}

\end{document}